\documentclass[twocolumn]{aastex631}
\usepackage{xspace}
\usepackage[hyphens]{url}
\usepackage{comment}
\pdfoutput=1 
\usepackage[figure,figure*]{hypcap}
\usepackage{footnote}
\usepackage{hyperref}
\usepackage{subfigure}
\usepackage{amsmath}
\usepackage{rotating}

\usepackage{longtable, threeparttablex}

\usepackage{epsfig}
\usepackage{natbib}
\usepackage{afterpage}

\begin{document}

\def\Msun{\hbox{M$_{\odot}$}}
\def\Lsun{\hbox{L$_{\odot}$}}
\def\kms{km~s$^{\rm -1}$\xspace}
\def\hcop{HCO$^{+}$}
\def\n2hp{N$_{2}$H$^{+}$}
\def\micron{$\mu$m\xspace}
\def\13CO{$^{13}$CO}
\def\etamb{$\eta_{\rm mb}$}
\def\Inu{I$_{\nu}$}
\def\kapnu{$\kappa _{\nu}$}
\def\ffore{f$_{\rm{fore}}$}
\def\tastar{T$_{A}^{*}$}
\def\sgra{SgrA$^{\star}$\xspace}
\def\nh3{NH$_{3}$}
\def\deg{$^{\circ}$\xspace}
\def\arcsec{$^{\prime\prime}$\xspace}
\def\arcmin{$^{\prime}$\xspace}
\def\Vlsr{\hbox{V$_{LSR}$}}
\def\redchisq{$\chi_{\mathrm{red}}^{2}$\xspace}
\newcommand\paperI{\textit{Paper I}\xspace}
\newcommand\paperII{\textit{Paper II}\xspace}
\newcommand\paperIII{\textit{Paper III}\xspace}
\newcommand\paperIV{\textit{Paper IV}\xspace}

\title{3D CMZ V: A new orbital model of our Galaxy's Center, informed by data across the electromagnetic spectrum}

\author[0000-0002-5776-9473]{Dani R. Lipman}
\affiliation{University of Connecticut, Department of Physics, 196A Hillside Road, Unit 3046
Storrs, CT 06269-3046, USA}

\author[0000-0002-6073-9320]{Cara Battersby}
\affiliation{University of Connecticut, Department of Physics, 196A Hillside Road, Unit 3046
Storrs, CT 06269-3046, USA}
\affiliation{Center for Astrophysics $|$ Harvard \& Smithsonian, MS-78, 60 Garden St., Cambridge, MA 02138 USA}

\author[0000-0001-7330-8856]{Daniel Walker}
\affiliation{UK ALMA Regional Centre Node, Jodrell Bank Centre for Astrophysics, Oxford Road, The University of Manchester, Manchester M13 9PL, United Kingdom}
\affiliation{University of Connecticut, Department of Physics, 196A Hillside Road, Unit 3046
Storrs, CT 06269-3046, USA}

\author[0000-0003-0724-2742]{Ma\"ica Clavel}
\affiliation{Univ. Grenoble Alpes, CNRS, IPAG, 38000 Grenoble, France}

\author[0009-0006-1435-2439]{B.L. DuBois}
\affiliation{Center for Astrophysics $|$ Harvard \& Smithsonian, MS-78, 60 Garden St., Cambridge, MA 02138 USA}

\author[0000-0001-6431-9633]{Adam Ginsburg}
\affiliation{Department of Astronomy, University of Florida, P.O. Box 112055, Gainesville, FL 32611, USA}

\author[0000-0001-9656-7682]{Jonathan~D.~Henshaw}
\affiliation{Max-Planck-Institut f\"{u}r Astronomie, K\"{o}nigstuhl 17, D-69117, Heidelberg, Germany}

\author[0000-0002-0560-3172]{Ralf S.\ Klessen}
\affiliation{Universit\"{a}t Heidelberg, Zentrum f\"{u}r Astronomie, Institut f\"{u}r Theoretische Astrophysik, Albert-Ueberle-Str.\ 2, 69120 Heidelberg, Germany}
\affiliation{Universit\"{a}t Heidelberg, Interdisziplin\"{a}res Zentrum f\"{u}r Wissenschaftliches Rechnen, Im Neuenheimer Feld 225, 69120 Heidelberg, Germany}

\author[0000-0001-8782-1992]{Elisabeth A.C. Mills}
\affiliation{Department of Physics and Astronomy, University of Kansas, 1251 Wescoe Hall Drive, Lawrence, KS 66045, USA}

\author[0000-0002-6379-7593]{Francisco Nogueras-Lara}
\affiliation{Instituto de Astrof\'isica de Andaluc\'ia (CSIC), Glorieta de la Astronom\'ia s/n, 18008 Granada, Spain}

\author[0000-0001-6113-6241]{Mattia~C.~Sormani}
\affiliation{Como Lake centre for AstroPhysics (CLAP), DiSAT, Universit{\`a} dell’Insubria, via Valleggio 11, 22100 Como, Italy}

\author[0000-0002-9483-7164]{Robin~G.~Tress}
\affiliation{Institute of Physics, Laboratory for Galaxy Evolution and Spectral Modelling, EPFL, Observatoire de Sauverny, Chemin Pegasi 51, 1290 Versoix, Switzerland}

\begin{abstract}

The 3D structure of The Milky Way's Central Molecular Zone (CMZ) informs our understanding of star formation cycles, black hole accretion, and the evolution of galactic nuclei. However, a comprehensive 3D model has remained elusive, as no singular dataset nor theory contains the requisite information to describe the orbital motion of the gas. We implement a Bayesian framework to flexibly combine datasets across the electromagnetic spectrum for molecular clouds in our CMZ catalog. We develop near/far metrics for each dataset, including dust extinction, absorption, stellar densities, X-ray echoes, and proper motions; and report a posterior positional probability density function (PPDF) for each cloud. We then use the posterior PPDF distributions for all CMZ clouds to search for a best fitting x$_2$ orbit. We find that no single orbit is a perfect fit, but the structure can overall be represented by nested x$_2$ orbits, with major axes ranging from about $72 < a < 146$~pc. We also present projected line of sight distance estimates for all 31 clouds in the catalog. Our results highlight asymmetries along the line of sight, with most clouds lying on the near side of the Galactic Center, and agree overall with current near/far assumptions for most CMZ clouds, including those in the Sgr A region, which may be much closer to the center. We conclude that the CMZ can be well-described by x$_2$ orbital families, and that the overall gas distribution is more complex than a single closed or open elliptical orbit.

\end{abstract}


\section{Introduction}

The inner 300~pc of the Milky Way, (the Central Molecular Zone, CMZ) is the central depot of the Galaxy, where gas from the disk is funneled inward via Galactic bar lanes, connecting to elliptical x$_2$ orbits aligned perpendicular to the bar. The CMZ exhibits extreme physical properties compared to the Galactic disk, including typical volume densities $> 10^{4}$~cm$^{-3}$, as well as temperatures, degrees of turbulence, and magnetic fields that tend to be 1-2 orders of magnitude higher compared to the solar neighborhood \citep[e.g.][see \citet{Henshaw2023a} for extensive review]{Pillai2015, Krieger2017, Ginsburg2016, Mills2018a, Butterfield2024}. The CMZ's extreme properties also share some similarities with high redshift galaxies \citep{Kruijssen2013, Henshaw2023a}, making it an ideal local laboratory for testing whether star formation varies as a function of environment. As our closest galaxy center at a distance of $\sim8.2$~kpc \citep{Gravity2021}, the Milky Way's CMZ offers an opportunity to study the complicated interplay between gas inflows, star formation, and outflows in the centers of galaxies.


Most of the dense gas in the CMZ is thought to lie on an orbit of $\sim100$~pc (i.e. the 100~pc stream or 100~pc ring) \citep[e.g.][]{Binney1991,Molinari2011,Kruijssen2015,Sormani(2015)_gasflow_barred_potl}. Simulations suggest that trace amounts of gas continue to fall inward onto the circumnuclear disk (CND), a ring of gas filling an inner $1\mathrm{pc}<\mathrm{R_{gal}}<10\mathrm{pc}$ annulus of the Galactic Center (GC), at a rate of $\sim$0.01 -- 0.1~\Msun yr$^{-1}$ \citep{Tress2020,Tress2024}. The low amount of inflow towards the inner few parsecs near the supermassive black hole, \sgra, is supported by the relatively low extinction in this region \citep{Nogueras2022}. However, the mechanisms for transporting gas from the 100~pc ring to the CND are still debated \citep[e.g.][]{Shlosman1989,BalbusHawley1998,Davies2007, KimElmegreen2017, Tress2020,Tress2024}. 
The 3D geometry and orbital motions of CMZ gas is integral to modeling nuclear inflow, and thus impacts our understanding of star formation cycles (whether suppressed or periodic), black hole feeding, and the evolution of galactic nuclei.   


The top-down shape of the CMZ has been described by four main theories: (i) two inner spiral arms \citep{Sofue1995, Sofue2022,Sofue2025}, (ii) a closed ellipse assuming a constant orbital velocity \citep{Molinari2011}, (iii) an open stream of orbits \citep{Kruijssen2015}, or (iv) a closed x$_2$ ellipse assuming constant angular momentum \citep{Binney1991,Sormani(2015)_gasflow_barred_potl,Tress2020,Chaves-Velasquez2025}. Distinguishing between orbital models requires information to differentiate objects on either the near side (in front of \sgra) or far side (behind), which can be used to determine the accuracy of each model \citep[e.g.][]{Henshaw2016_GasKin_250pc, Walker_2025,Lipman_2025}. 

Several studies have sought to determine approximate near/far locations of CMZ structures. For example, Sgr B2 has been strongly constrained to reside in the foreground via parallax measurements \citep{Reid2009} and relative strengths of CO emission and OH absorption features \citep{Yan2017}. The Brick cloud also lies on the near side based on color-magnitude diagrams \citep{Nogueras2021} and proper motion measurements \citep{Martinez-Arranz2022}. Most of these previous studies confirm that clouds reside within the CMZ, but are unable to quantify their relative positions. On the other hand, time delays of reflected X-ray radiation due to past flaring events from \sgra have further constrained near-side positions of Sgr B2, as well as clouds in the Sgr A region \citep[e.g.][]{Ponti2010,Clavel(2013),Chuard2018}. More recent studies of stellar kinematics have also produced promising distance estimates for the 20 and 50 \kms~clouds \citep{Nogueras-Lara2026}. However, these methods only provide distance estimates for a handful of clouds and, in the case of X-ray echoes, involve large systematic uncertainties. Lastly, \citet{Henshaw2016_GasKin_250pc} describes the overall distribution of dense molecular gas in the CMZ in position-position-velocity (PPV) space, but line of sight distances and relative positions of prominent structures have not been obtainable through these methods alone. 


In a previous paper of this series, \citet{Lipman_2025}, we combined dust extinction techniques and molecular line absorption analysis from \citet{Walker_2025} to obtain the near/far positions of clouds in our CMZ catalog and compare to proposed orbital models. We found that an x$_2$ model (as described in point iv above) generally fit the dense gas in the CMZ in both position-position (PP) and position-velocity (PV) space. Simulations of Milky Way-like galaxies produce similar star-forming rings and suggest that, on average, the gas can be approximated by a simple x$_2$ ellipse \citep{Armillotta(2019), Salas(2020), Sormani2020,Tress2020,Tress2024}. Nuclear rings are also seen in many nearby galaxy centers \citep[e.g.][]{Sun2020,Stuber2023,Sun2024} and modeling of external CMZs has shown reasonable fittings to x$_2$ orbits \citep{Levy2022}.  



\begin{figure*}[t!]
    \includegraphics[width=\textwidth]{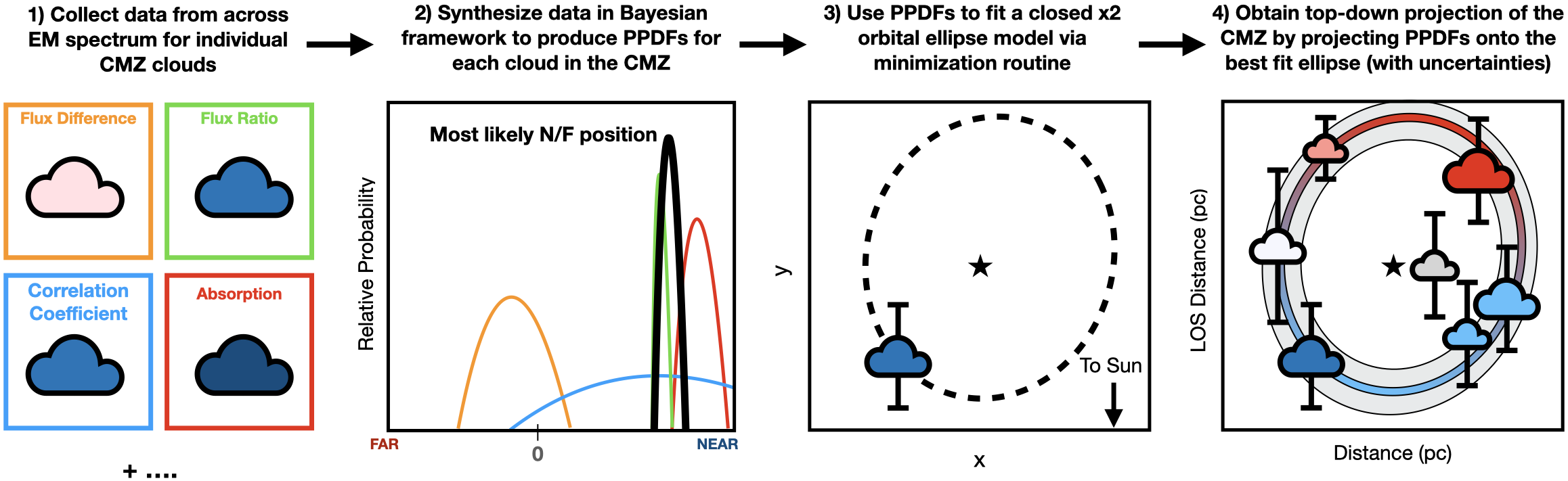}
    \caption{We present a method to synthesize multi wavelength datasets to obtain most likely near/far positions of clouds in the CMZ, and find a best-fitting elliptical orbit. Our methodology follows the above schematic from left to right: 1) we collect both near/far and line of sight distance constraints from a variety of methods across the electromagnetic spectrum (Section~\ref{sec:data_and_methods_EM}), 2) we build a Bayesian framework to combine these different data on a normalized near-to-far scale, producing positional probability density functions which peak at the most likely position of a given cloud (Section~\ref{subsec:bayseian_posteriors}), 3) Find a best fitting closed elliptical x$_2$ orbit using each cloud's PPV location and NF distribution (Section~\ref{subsec:fitting_procedures}), 4) Use the best fit orbital model to present a preliminary top-down projection of the CMZ with line-of-sight distance estimates for all CMZ clouds (Section~\ref{sec:disc_relative_positions}). The black star in panels 3 and 4 indicates the location of \sgra. The blue and red colors of clouds in panels 3 and 4 indicate likely near or far positions, respectively.}
    \label{fig:overview}
\end{figure*}

This paper builds on the analysis of \citet{Battersby_2025_3DCMZI, Battersby_2025_3DCMZII, Walker_2025} and \citet{Lipman_2025} (hereafter \paperI, \paperII, \paperIII, and \paperIV). In this paper, we present a Bayesian framework to synthesize available near vs. far (NF) position data for molecular clouds in the CMZ, and use the posterior NF distributions to find a best fitting x$_2$ orbital model, as summarized in the schematic in Figure~\ref{fig:overview}. In Section ~\ref{sec:data_and_methods_EM} we describe the data used for this work, including brief descriptions of relevant results and data products from papers I-IV. In Section~\ref{sec:Bayesian_synth_and_orbital_fitting} we design a Bayesian framework to determine posterior NF likelihoods for the cataloged clouds (Figure~\ref{fig:overview} panels 1 and 2), and introduce a minimization method for fitting a 4D ($\ell$, b, v, NF) x$_2$ orbital model of the CMZ (panel 3). We present our results in Section~\ref{sec:results}. In Section~\ref{sec:discussion}, we discuss the relative positions of CMZ clouds based on our best fits (panel 4). We also discuss the morphology of the CMZ based on our projected model and interpretations for various structures in PPV space. We address uncertainties and limitations of our methods in Section~\ref{sec:limitations}, and summarize our conclusions in Section~\ref{sec_conclusions}. 
\section{Using datasets across the Electromagnetic spectrum to derive near/far likelihoods}\label{sec:data_and_methods_EM}

We combine datasets from multiple wavelengths and synthesize them to generate posterior positional probability density functions for the locations of each molecular cloud in our CMZ catalog. NF positions of clouds were determined from molecular line and MIR  datasets in \paperIII and \paperIV. In addition to these methods, we incorporate NF positions from star counts, and distance constraints for a handful of clouds based on X-ray data and stellar kinematics. In the following sections, we describe the datasets and products used for deriving NF likelihoods. The datasets available for each cloud are listed in the second to last column of Table~\ref{tab:los_synth_table}.

\subsection{CMZ column density map and cloud catalog}\label{subsec:CMZ catalogs} 

The catalog used for the analysis in this work identifies the densest molecular clouds in the CMZ based on integrated column density maps created using observations from the Herschel Infrared Galactic Plane (HiGAL) Survey \citep{Molinari2010, Molinari2016}. Details for the creation of the column density map can be found in the previous papers of this series. Described briefly: \paperI performed a modified blackbody fit to HiGAL data covering the inner 40\deg of the Milky Way, and produced integrated column density and temperature maps. \paperII produced a hierarchical catalog of structures in the column density map using the \texttt{astrodendro} Python package \citep{astrodendro}. We then calculated and tabulated properties and SFRs for the hierarchy of CMZ structures, and placed the CMZ in a global context with respect to the Milky Way and other galaxies. \paperIII builds upon the catalog by identifying a consistent sample of clouds in PPV space. Nine clouds in the catalog contained multiple velocity components in the HNCO emission not seen in H$_{2}$CO. Those clouds were divided into separate spatial masks based on the projected peak intensity of the velocity components. Lastly, \paperIII presents a detailed kinematic study of each cloud and produced masks corresponding to each dendrogram leaf, which we use in this paper. The column density map from \paperI and cloud masks and ID numbers from \paperIII are shown in the top panel of Figure~\ref{fig:starcount_method}. The first five columns of Table 1 from \paperIII (cloud IDs, cloud names, and central ($\ell, b, v$) positions are summarized in Table~\ref{tab:los_synth_table}.

\vspace{-0.3mm}
\subsection{Near/far distinctions from absorption and extinction methods}\label{subsec:NF_pos_paper34} 

The previous papers in this series distinguished between proposed CMZ models by determining the likely NF positions of all cataloged molecular clouds, and comparing them to different orbital geometries. The analyses leverage the strong absorption at mid-infrared wavelengths, as cold dust absorbs the bright emission from the CMZ, making the material in front appear as dark extinction features. The difference in the bright Galactic background and observed emission are large enough to estimate NF positions of clouds in the CMZ based on molecular line absorption (\paperIII) or dust extinction (\paperIV). We briefly summarize the general concepts of the methods here, and refer the reader to the previous papers in the series for more detailed explanations.

\paperIII quantifies the NF positions of CMZ clouds by assuming a uniform background of diffuse free-free and synchrotron emission in the CMZ. This assumption implies that clouds in front of the continuum will show strong molecular absorption features, while clouds behind show little to no absorption. NF positions of clouds are obtained by comparing molecular line absorption against the radio continuum measured toward each cloud. For this study, we utilize the masked fractional absorption values reported in Table 2 of \paperIII as a NF indicator from absorption. 

\paperIV introduces three dust extinction techniques to determine NF positions of CMZ clouds by using dark MIR features, seen in Spitzer 8\micron \citep{Benjamin2003, Churchwell2009} and HiGAL 70\micron images \citep{Molinari2010}. The 8\micron dust extinction analysis assumes a constant foreground and background, offset by inhomogenous emission from the bright CMZ. Clouds in the CMZ absorb and re-emit this background at different wavelengths. Thus, clouds on the near side appear as dark extinction features, while clouds on the far side may appear similar to or brighter than the background. The analysis of \paperIV presents: 1) a flux difference between 8\micron emission and a smoothed background, and 2) a flux ratio between the 8\micron emission and smoothed map. Additionally, \paperIV uses optically thin 70\micron data to determine the correlation between the cloud column density maps and the amount of material associated with extinction. 
In this paper, we use the flux difference, flux ratio, and Pearson correlation coefficient of emission and extinction column densities for each CMZ catalog cloud, as reported in Table 1 of \paperIV.



\subsection{Near/far distances from star counts}\label{subsec:starcount_data} 
We introduce another method for NF distinction of CMZ clouds based on stellar density maps from \citet{Nishiyama2013}, which report the number of non-foreground stars with extinction-corrected K$_{s,0} > 10.5$. The authors combine data from the NIR camera SIRIUS \citep{Nagayama2003} on the 1.4m telescope IRSF, and 2.25\micron images from ISAAC at the ESO VLT \citep{Nishiyama_Schodel_2013} to supplement the crowded inner 1\arcmin. The SIRIUS data cover the central 840~pc$\times$ 280~pc of the Galaxy, and have $10\sigma$ limiting magnitudes of $(\mathrm{H},\mathrm{K}_{s}) \sim (16.6,15.6)$ and extinction $\mathrm{A}(\mathrm{K}_{s}) \sim
2-3.5$ mag. 
Extinction-corrected stellar density maps were created for the central 20~\arcmin$\times$ 20~\arcmin field, achieving completeness for K$_{s}>10.5$ of $\sim 97\%$ \citep{Hatano2013,Yasui2015}. The data are divided into 2\arcmin~bins to create a map of the number of stars per arcmin$^{2}$. 

\begin{figure*}[t!]
    \includegraphics[width=\textwidth]{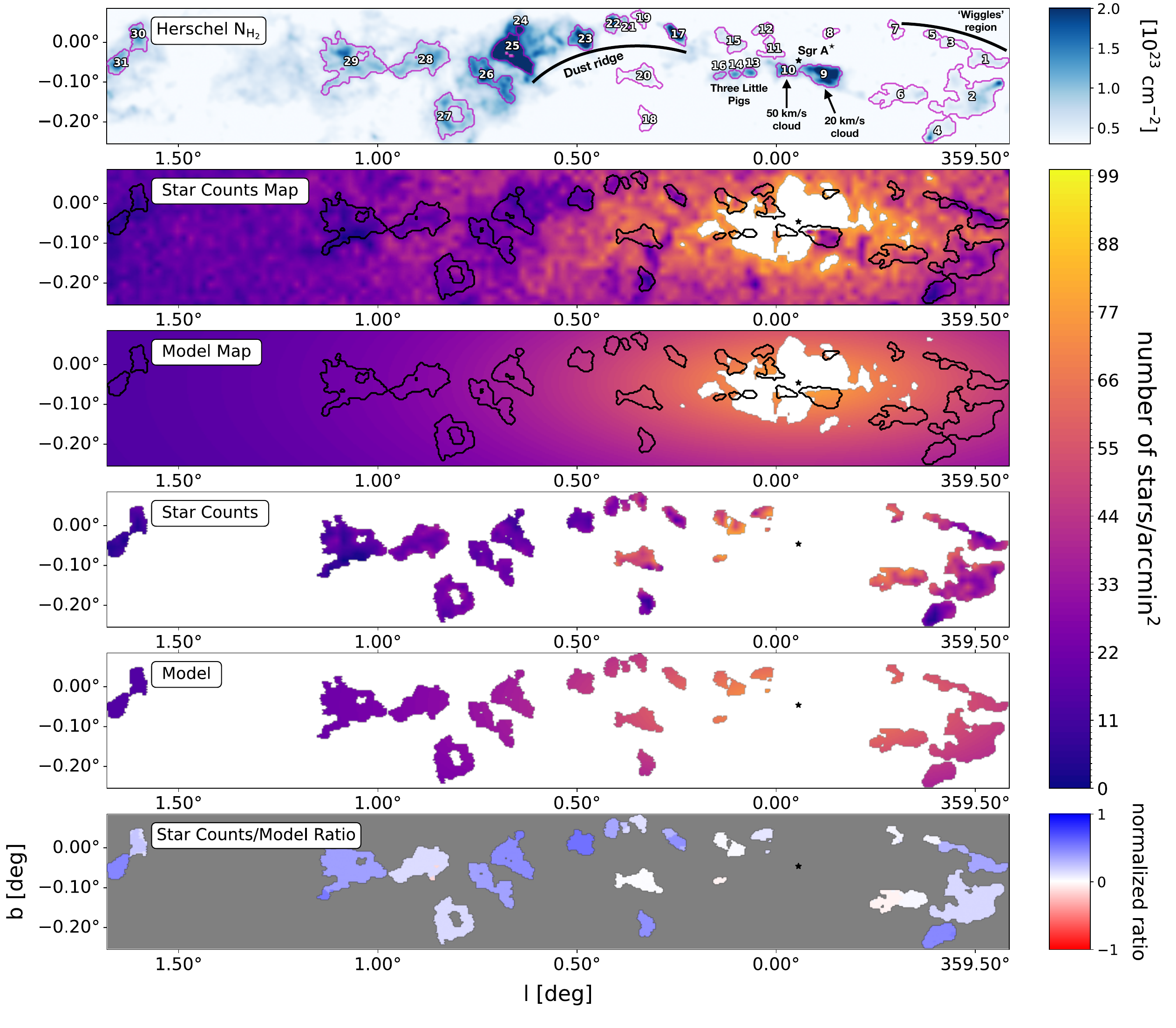}
    \caption{Stellar density maps of the CMZ show dark extinction features that can be used to calculate a star count ratio to determine NF positions of molecular clouds in the catalog. From top to bottom, the panels show: (i) Herschel column density map with black contours and ID numbers correspond to cloud masks summarized in Table~\ref{tab:los_synth_table}, (ii) stellar density map from \citet{Nishiyama2013} and (iii) modeled star count map from \citet{Gallego-Cano2020}, both of which we convolve and regrid to match the 36\arcsec Herschel column density map. White areas are masked pixels where the star count is greater than the maximum of the model. Panel (iv) is a cutout of star counts for the molecular clouds; panel (v) is a cutout of modeled star counts; and panel (vi) is a ratio of the star counts to the model map. The blue and red color scales denote more likely near or far side positions, respectively. Clouds containing $> 20\%$ of masked pixels are excluded from analysis. The black star indicates the location of \sgra.}
    \label{fig:starcount_method}
\end{figure*}

The stellar density map of the CMZ from \citet{Nishiyama2013} (top panel of  Figure~\ref{fig:starcount_method}) shows areas of lower star counts compared to the expected background, where dense molecular clouds in front of the GC block the stars behind. The stellar distribution in this region is dominated by the Galactic bulge and nuclear stellar disk (NSD), a flat disk of stars with a total mass of $\sim 1.05\times10^9 ~\Msun$, which overlaps with the CMZ between a range of $30 < \mathrm{R_{gal}}<300~\mathrm{pc}$ \citep{Launhardt2002,Sormani2020_NSD,Sormani2022_NSD}. We use a ratio of the observed star counts to a modeled stellar distribution from the Galactic bulge and NSD for a quantitative measure of how many stars are being extincted by the dense molecular clouds. For a near-sided cloud, we expect the median ratio to be $<1$, as the cloud blocks the star light expected from the model, while far-sided clouds would have ratios $>1$. A ratio of $1$ implies an exact match to the expected bulge and NSD distribution without a molecular cloud present.

To model the expected star counts, we use the stellar distribution model from \citet{Gallego-Cano2020}. The authors present a best fit to the \citet{Nishiyama2013} data after masking out regions of low density. They also mask out the Nuclear Stellar Cluster (NSC), a massive ($M \simeq 2.5 \times 10^{7}~\Msun$) and dense disk of stars in the central 10~pc \citep{Genzel2010,Schodel2014}. The model is a combination of a Galactic bulge component using a triaxial ellipsoidal bar and an NSD component described by a Sérsic model. We use the best fitting model parameters presented in Table 1 of \citet{Gallego-Cano2020} and refer the reader to Section 4 of that paper for a detailed explanation of the fitting. 

We convolve and regrid both the star count data and model to match the 36\arcsec resolution and pixel scaling of the cloud catalog from \paperIII, as shown in the top three panels of Figure~\ref{fig:starcount_method}. While the model recreates the outer region (i.e. the bulge) quite well, it underestimates the center of the stellar density distribution, as it does not account for the NSC in the central 10~pc. To avoid skewing the NF positions in underestimated modeled regions, we mask out areas where the star count data is greater than the maximum of the model. 

\begingroup 
\setlength{\LTleft}{0pt minus 1000pt}
\setlength{\LTright}{0pt minus 1000pt}

\setlength\LTcapwidth{\textwidth}
\startlongtable

\begin{longtable*}{llcccccccccccc}
\caption{Summary of posterior fitting results from the PPDFs and LOS distance estimates based on the best-fitting ellipses. The first five columns are taken from Table 1 of \citet{Walker_2025}. We report the leaf IDs of the cataloged clouds from (clouds with multiple peak velocity features in HNCO emission are denoted by letters in the leaf ids), the associated cloud names, central coordinates in degrees ($\ell, b$), and central velocity from HNCO (v). From the PPDF fittings (described in Section~\ref{subsec:bayseian_posteriors}) we report the peak relative probability ($\mathrm{A}$), the normalized NF position (d$_{\mathrm{NF}}$), and the 68\% confidence interval for the posterior distribution (CI68). We report the estimated LOS distances in front of (-) or behind (+) the GC (LOS pos) obtained from de-projecting the normalized NF positions onto the median x2 orbit, and associated uncertainties (LOS $\sigma$) obtained from calculating the standard deviation for the LOS positions of a cloud from each of the three ellipses, as described in Section \ref{subsec:results_bestfit_x2}. Lastly, we report the estimated galactocentric radii ($\mathrm{R}_\mathrm{gal}$) and estimated uncertainties based on the projected LOS distances. We report the estimated distances as a qualitative benchmark for the currently available data. The NF distinction methods used for each cloud are noted with a single letter for flux difference (d), flux ratio (r), correlation coefficient (c), star count ratio (s), absorption (a), x-ray distances (x), and estimates from \citet{Nogueras-Lara2026} (k).  Colloquial names for catalog clouds a given in the final column. IDs marked with an asterisk denote the largest mask used for clouds with multiple velocity components.}\\
\hline 
Leaf  & Cloud Name & $\ell$ & $b$ & v & $\mathrm{A}$ & d$_{\mathrm{NF}}$ & $\mathrm{CI68}$ & LOS & LOS & $\mathrm{R}_{\mathrm{gal}}$ & $\mathrm{R}_{\mathrm{gal}}$& NF  &Colloquial\\
\newline
ID&&&&&&&&pos&$\sigma$&&$\sigma$&Methods& Name\\
\hline
 & & $\mathrm{{}^{\circ}}$ & $\mathrm{{}^{\circ}}$ & $\mathrm{km\,s^{-1}}$ &  &  &  & $\mathrm{pc}$ & $\mathrm{pc}$&$\mathrm{pc}$&$\mathrm{pc}$ & &  \\ \hline
1 & G359.475-0.044 & -0.525 & -0.044 & -102.0 & 2.747 & 0.404 & 0.289 & -44.1 & 25.6 & 72.3 & 39.8 & d,r,c,s &  \\
2 & G359.508-0.135 & -0.492 & -0.135 & -56.0 & 2.433 & 0.122 & 0.326 & -13.6 & 17.5 & 51.1 & 39.0 & d,r,c,s & SgrC \\
3 & G359.561-0.001 & -0.439 & -0.001 & -90.0 & 3.723 & 0.168 & 0.213 & -18.6 & 14.5 & 49.0 & 35.1 & d,r,c,a,s &  \\
4a$^*$ & G359.595-0.223 & -0.405 & -0.223 & -27.0 & 2.662 & 0.28 & 0.298 & -30.7 & 21.7 & 53.5 & 30.2 & d,r,c,s &  \\
4b & G359.595-0.223 & -0.405 & -0.223 & -20.0 & 2.314 & 0.292 & 0.343 & -32.0 & 23.9 & 54.4 & 29.7 & d,r,c,s &  \\
5 & G359.608+0.018 & -0.392 & 0.018 & -78.0 & 40.092 & 0.477 & 0.02 & -52.0 & 17.2 & 68.1 & 30.1 & d,r,c,a,s &  \\
6a & G359.688-0.132 & -0.312 & -0.132 & -29.0 & 4.172 & 0.232 & 0.19 & -25.5 & 15.7 & 42.5 & 23.5 & d,r,c,a,s &  \\
6b$^*$ & G359.688-0.132 & -0.312 & -0.132 & -21.0 & 2.451 & -0.053 & 0.348 & 5.3 & 15.7 & 26.5 & 23.5 & d,r,c,a,s &  \\
7a$^*$ & G359.701+0.032 & -0.299 & 0.032 & -73.0 & 11.478 & 0.601 & 0.058 & -65.4 & 23.0 & 73.8 & 20.8 & d,r,c,a,s &  \\
7b & G359.701+0.032 & -0.299 & 0.032 & -37.0 & 3.396 & -0.223 & 0.234 & 23.7 & 16.9 & 20.7 & 22.1 & d,r,c,a,s &  \\
8a$^*$ & G359.865+0.023 & -0.135 & 0.023 & -54.0 & 2.046 & -0.494 & 0.388 & 53.0 & 32.4 & 7.0 & 7.5 & d,r,c,a &  \\
8b & G359.865+0.023 & -0.135 & 0.023 & 15.0 & 2.504 & -0.453 & 0.317 & 48.5 & 28.1 & 7.0 & 7.6 & d,r,a &  \\
8c & G359.865+0.023 & -0.135 & 0.023 & 62.0 & 2.982 & -0.624 & 0.266 & 67.0 & 31.9 & 6.8 & 7.5 & d,r,c,a &  \\
9 & G359.88-0.081 & -0.12 & -0.081 & 15.0 & 6.111 & 0.389 & 0.13 & -42.5 & 18.7 & 44.7 & 7.2 & d,r,c,a,k & 20 km/s \\
10 & G359.979-0.071 & -0.021 & -0.071 & 48.0 & 7.834 & 0.285 & 0.101 & -31.3 & 14.0 & 31.4 & 1.0 & d,r,c,a,k & 50 km/s \\
11a$^*$ & G0.014-0.016 & 0.014 & -0.016 & -11.0 & 1.576 & -0.649 & 0.503 & 69.7 & 42.3 & 0.6 & 0.7 & d,r,c,a &  \\
11b & G0.014-0.016 & 0.014 & -0.016 & 45.0 & 1.625 & -0.658 & 0.488 & 70.7 & 42.0 & 0.6 & 0.7 & d,r,c,a &  \\
11c & G0.014-0.016 & 0.014 & -0.016 & 14.0 & 2.988 & -0.508 & 0.266 & 54.5 & 27.9 & 0.6 & 0.7 & d,r,c,a,s &  \\
12 & G0.035+0.032 & 0.035 & 0.032 & 86.0 & 1.562 & -0.689 & 0.505 & 74.0 & 43.7 & 1.6 & 1.7 & d,r,c,a,s &  \\
13 & G0.068-0.076 & 0.068 & -0.076 & 50.0 & 6.596 & -0.049 & 0.126 & 4.9 & 6.6 & 4.8 & 4.5 & d,r,c,a,x & Stone \\
14 & G0.105-0.08 & 0.105 & -0.08 & 53.0 & 8.681 & -0.049 & 0.091 & 4.9 & 5.2 & 8.0 & 7.9 & d,r,c,a,x & Sticks \\
15 & G0.116+0.003 & 0.116 & 0.003 & 52.0 & 1.0 & -1.107 & 0.422 & 119.2 & 54.6 & 5.6 & 5.9 & d,r,c,a,s &  \\
16a & G0.143-0.083 & 0.143 & -0.083 & -15.0 & nan & nan & nan & nan & nan & nan & nan &  & Straw \\
16b$^*$ & G0.143-0.083 & 0.143 & -0.083 & 57.0 & 6.421 & 0.367 & 0.123 & -40.1 & 17.7 & 43.4 & 9.0 & d,r,c,a,s & Straw \\
17a & G0.255+0.02 & 0.255 & 0.02 & 18.0 & 10.368 & 0.724 & 0.076 & -78.6 & 27.9 & 84.0 & 16.5 & d,r,c,a,s & Brick \\
17b$^*$ & G0.255+0.02 & 0.255 & 0.02 & 37.0 & 13.585 & 0.694 & 0.058 & -75.4 & 26.2 & 81.0 & 16.7 & d,r,c,a,s & Brick \\
17c & G0.255+0.02 & 0.255 & 0.02 & 70.0 & 4.512 & 0.638 & 0.175 & -69.4 & 29.0 & 75.3 & 16.4 & d,r,c,a,s & Brick \\
18 & G0.327-0.195 & 0.327 & -0.195 & 16.0 & 1.983 & 0.215 & 0.4 & -23.6 & 23.6 & 42.5 & 23.1 & d,r,c,s &  \\
19 & G0.342+0.06 & 0.342 & 0.06 & -2.0 & 39.958 & 0.769 & 0.02 & -83.5 & 27.2 & 92.3 & 23.7 & d,r,c,a,s & B \\
20 & G0.342-0.085 & 0.342 & -0.085 & 90.0 & 2.021 & -0.326 & 0.393 & 34.8 & 26.8 & 22.5 & 23.7 & d,r,c,a,s & Sailfish \\
21a$^*$ & G0.379+0.05 & 0.379 & 0.05 & 8.0 & 20.412 & 0.616 & 0.039 & -67.0 & 22.7 & 79.7 & 27.7 & d,r,c,a,s & C \\
21b & G0.379+0.05 & 0.379 & 0.05 & 39.0 & 4.059 & 0.575 & 0.186 & -62.5 & 27.3 & 75.9 & 26.8 & d,r,c,a,s & D \\
22 & G0.413+0.048 & 0.413 & 0.048 & 19.0 & 40.109 & 0.845 & 0.02 & -91.7 & 29.8 & 103.2 & 29.2 & d,r,c,a,s & E/F \\
23 & G0.488+0.008 & 0.488 & 0.008 & 28.0 & 6.223 & 0.684 & 0.127 & -74.3 & 28.6 & 92.5 & 35.9 & d,r,c,a,s &  \\
24 & G0.645+0.03 & 0.645 & 0.03 & 53.0 & 11.963 & 0.471 & 0.059 & -51.3 & 18.6 & 86.8 & 52.5 & d,r,c,a,s &  \\
25 & G0.666-0.028 & 0.666 & -0.028 & 62.0 & 4.328 & 0.213 & 0.183 & -23.4 & 14.8 & 71.8 & 55.6 & d,r,c,a,s & Sgr B2 \\
26a$^*$ & G0.716-0.09 & 0.716 & -0.09 & 28.0 & 3.631 & 0.161 & 0.224 & -17.8 & 14.7 & 73.5 & 60.2 & d,r,c,s & G0.714 \\
26b & G0.716-0.09 & 0.716 & -0.09 & 58.0 & 9.443 & 0.025 & 0.083 & -3.1 & 4.4 & 66.8 & 63.8 & d,r,c,s &  \\
27 & G0.816-0.185 & 0.816 & -0.185 & 39.0 & 9.698 & -0.06 & 0.082 & 6.0 & 5.2 & 72.4 & 72.7 & d,r,c,s &  \\
28a & G0.888-0.044 & 0.888 & -0.044 & 14.0 & 14.373 & -0.049 & 0.055 & 4.9 & 3.7 & 79.4 & 79.8 & d,r,c,s &  \\
28b & G0.888-0.044 & 0.888 & -0.044 & 26.0 & 14.373 & -0.049 & 0.055 & 4.9 & 3.7 & 79.4 & 79.8 & d,r,c,s &  \\
28c$^*$ & G0.888-0.044 & 0.888 & -0.044 & 84.0 & 14.907 & -0.037 & 0.053 & 3.6 & 3.3 & 79.9 & 80.0 & d,r,c,s &  \\
29a & G1.075-0.049 & 1.075 & -0.049 & 74.0 & 1.698 & 0.02 & 0.467 & -2.6 & 19.7 & 99.6 & 91.3 & d,r,c,s &  \\
29b$^*$ & G1.075-0.049 & 1.075 & -0.049 & 85.0 & 6.49 & -0.002 & 0.122 & -0.3 & 5.0 & 98.6 & 96.6 & d,r,c,s &  \\
30a$^*$ & G1.601+0.012 & 1.601 & 0.012 & 48.0 & 24.036 & -0.065 & 0.033 & 6.6 & 3.4 & 144.6 & 145.8 & d,r,c,s & G1.602 \\
30b & G1.601+0.012 & 1.601 & 0.012 & 58.0 & 18.226 & -0.072 & 0.043 & 7.4 & 4.1 & 144.3 & 145.6 & d,r,c,s & G1.602 \\
31 & G1.652-0.052 & 1.652 & -0.052 & 50.0 & 19.485 & -0.039 & 0.041 & 3.8 & 2.8 & 150.4 & 150.8 & d,r,c,s & G1.651 \\
\hline
\hline
\label{tab:los_synth_table}

\end{longtable*}
\endgroup

The catalog cloud masks are then applied to the data and model maps. For each mask, we take the pixel-wise ratio of the star count data and the model. The cloud masks contain anywhere between $\sim $8 -- 400 pixels. To ensure at least 3 pixels per mask, we exclude cloud masks which contain $> 20\%$ of masked pixels from the analysis, resulting in 9 excluded cloud IDs: 8a, 8b, 9: 20 km/s, 10: 50 km/s, 11a, 11b, 13: Stone, 14: Sticks.

A visual summary of the star count ratio method is shown in Figure~\ref{fig:starcount_method}. From top to bottom the panels show: (i) Herschel column density map with contours and ID numbers corresponding to the catalog presented in \paperIII and summarized in Table~\ref{tab:los_synth_table}, (ii) stellar density map from \citet{Nishiyama2013}, (iii) modeled star count map from \citet{Gallego-Cano2020} convolved and regrid to match the 36\arcsec column density map, (iv) cutout of star counts for the molecular clouds, (v) cutout of modeled star counts, and (vi) ratio of the star count map to the model map, normalized to the near/far scaling used in the Bayesian synthesis (See Section~\ref{subsec:bayseian_posteriors}). White areas in panels (i)--(iii) indicate masked pixels where the star count is greater than the maximum of the model. The blue and red color scale in panel (iv) denotes likely near or far side positions, respectively. The black star denotes the location of \sgra.  

The median star count ratio within each mask is used as the overall NF result for a given cloud. The star count ratios are ultimately normalized on a 1 (near) to -1 (far) scale, and folded in to the Bayesian synthesis method described in Section~\ref{subsec:bayseian_posteriors}. 

\subsection{Distance Constraints from X-rays}\label{subsec:xray_data} 
Two objects in the catalog, the Stone (ID 13) and Sticks (ID 14) clouds, have distance constraints determined by detections of  Fe K$_{\alpha}$ emission from reflected X-rays   \citep{Clavel(2013),Chuard2018,Marin2023, Stel2025}. 
\citet{Sunyaev_and_Churazov1998} presents a method for calculating projected line-of-sight distance estimates from the Fe line emission based on the age of a flaring event responsible for the illumination. These calculations provide tight distance constraints on illuminated objects, as it is impossible for illumination to occur without intersecting with the path of the event sweeping out from the central source. \sgra is often assumed as the source for both a single and two-flaring event scenario due to the energy required to explain the observed emission \citep{Clavel(2013)}. \citet{Clavel(2013), Chuard2018} and \citet{Stel2025} use this method to determine the distances of the clouds to be between 5-25~pc behind \sgra. Additionally, \citet{Marin2023} perform polarization measurements of the reflected X-ray emission to constrain the age of a single flaring event to be $205^{+50}_{-30}~\mathrm{yr}$, and assumptions of two flaring events imply ages of 100~yr and 200~yr. Assuming a 200 year old event results in a possible range in line-of-sight distances for both clouds within 30~pc behind the GC and 20~pc (10~pc) in front for the Stone (Sticks) cloud. We use this range as an added constraint on the NF distance estimates for the Sticks and Stone clouds.

\subsection{Distance constraints from Stellar Kinematics}

Distance estimates from stellar density maps are possible to obtain for some clouds in the CMZ. \citet{Nogueras-Lara2026} report distance estimates for the 20\kms~and 50\kms~Clouds (IDs 9 and 10, respectively), based on proper motions estimated from their observed color magnitude. The proper motion measurements were compared with the Galactocentric velocity profile to obtain a distance. This analysis places the 50\kms and 20\kms clouds at $43\pm8$~pc and  $56\pm11$~pc in front of \sgra, respectively. We include these distance estimates in our analysis.


\section{Bayesian synthesis and orbital fitting}\label{sec:Bayesian_synth_and_orbital_fitting}

\subsection{Bayesian synthesis of near/far methods}\label{subsec:bayseian_posteriors}

We present an automated Bayesian framework to combine information from multiple datasets into a posterior distribution of a cloud's likely NF position in the CMZ. We set up the method to synthesize data with the goal of adding future datasets and results, in order to further constrain the posteriors and subsequent orbital fitting. 

Our approach is inspired by \citet{Ellsworth-Bowers2013, Ellsworth-Bowers2015A},who employed a Bayesian approach to produce distance probability density functions (DPDFs) for \\
\vspace{3mm}

\setlength\LTcapwidth{\textwidth}
\begin{longtable*}[ht!]{lcccc}
\caption{Summary of normalization parameters and weights for each near/far distinction method.}\\ \hline
Method & max & min & Gaussian $\sigma$ & weight\\ 
\hline
Flux Difference         &150~MJy/sr & -150~MJy/sr   & normalized $\sigma_{\mathrm{MAD}}$ for each source  & 1\\
Flux Ratio              & 1         & 0             & normalized $\sigma_{\mathrm{MAD}}$ for each source                & 1 \\
Correlation Coefficient & 1         & -1            & based on stdv of dataset                            &2\\
Absorption              & 2         & 0             & normalized error for each source &2\\
Star Count Ratio        & 2         & 0             & 3 $\times$ median $\sigma_{\mathrm{MAD}}$ for the entire sample &1\\
X-ray (IDs 13, 14)      & 150~pc    & -150~pc        &  -                                          &2\\
Nogueras et al. (2026) (IDs 9, 10) & 150~pc & -150~pc&       normalized uncertainties for each source                      &2\\
\hline
\hline

\label{tab:z-scaling parameters}
\end{longtable*}

\noindent constraining distances to infrared dark clouds in the Galactic disk. The authors introduce Galactic rotation curves as a likelihood function, which is subject to ambiguity in projected orbital motions, and use various ancillary datasets as priors to constrain distance estimates \citep[see][]{Ellsworth-Bowers2013}. Similarly, the dense molecular clouds in the CMZ are subject to ambiguity in their line-of-sight locations relative to the GC, making the DPDF approach a promising starting point for developing constraints on cloud positions.

Distinguishing line of sight distances to infrared dark clouds in the Galactic disk has previously been achieved by comparing emission column densities to column densities calculated via MIR extinction \citep{Ellsworth-Bowers2013, Ellsworth-Bowers2015A}, or from stellar extinction using precise estimates from Gaia \citep[e.g][]{Edenhofer2024,Zhang2025}. However, \paperIV showed that MIR extinction in the CMZ is quickly saturated due to the extreme column densities, making obtaining line of sight distances to molecular clouds difficult with currently available data. While recent studies of stellar kinematics can distinguish between near and far positions in the CMZ \citep[e.g.][]{Nogueras-Lara2026}, this has only been obtained for a couple targets. Thus, rather than extracting distances for CMZ clouds, we use Bayes' Theorem to obtain a \textit{position} PDF (hereafter PPDF), which gives the distribution of a cloud's NF position relative to \sgra. We define the PPDF as a function of near-far position ($p_{NF}$) by
\begin{equation}\label{eqn:PPDF}
    \mathrm{PPDF}(p_{NF} )= \mathcal{L}(p_{NF})\prod_{i} P_{i}(p_{NF})\;,
\end{equation}
where $\mathcal{L}(p_{NF})$ is a likelihood function and $P_{i}(p_{NF})$ are prior PPDFs based on ancillary datasets. We use the median 8\micron flux difference as the basis for our likelihood function, as it provides the simplest comparison between the Spitzer 8\micron emission and model background. PDF distributions from the other methods are treated as priors to constrain the NF estimates. The posterior PPDF is then defined as the product of the likelihood function with the priors, normalized to unit total probability, i.e. $\int_{0}^{\inf} PPDF(p_{NF})~~d(p_{NF}) = 1$. 


Before constructing the likelihood function and prior positional distributions, we map each of the NF position measures to the same quantitative scale of -1 (far) to +1 (near) using a min-max normalization procedure:

\begin{equation}\label{eqn:z-scale}
    z(x) = 2 \left( x - \frac{\mathrm{min}}{\mathrm{max} - \mathrm{min}}\right) -1,
\end{equation}

where $z$ is the normalized value of a given data value $x$ for each dataset (see Appendix C, Table~\ref{tab:PPDF_info_table} for the unnormalized values used to build the priors for each method). The normalization is determined by chosen minimum and maximum limits for each method, reported in Table~\ref{tab:z-scaling parameters}. The min-max scaling is applied such that $z=0$ corresponds to the threshold between NF distinctions.

Similarly, the uncertainties ($\sigma$) for each method are normalized using the same min and max values, such that the normalized $\sigma$ is scaled from 0; i.e. $\sigma_{\mathrm{norm}}$ = $|z(\sigma)-z(0)|$. This scaling is analogous to calculating the uncertainty of the normalized dataset distribution. 

After the data have been normalized, we model the shape of each method's NF distribution for a given cloud as a 1-D Gaussian of the form:

\begin{equation}
    P(x) = A\cdot\mathrm{exp} \left( \frac{-(x-\mu)^{2}}{2\sigma^{2}}   \right)   , 
\end{equation}
with relative probability $P(x)$ which peaks at value $A$. The offset $\mu$ corresponds to the normalized NF position ($z(x)$), and the width of the distribution, $\sigma$, is set by the normalized uncertainty ($z(\sigma)$). Each method's distribution is normalized to integral unit probability. 

The choice to model the priors as Gaussian was made to flexibly represent the overall distribution for each cloud. The normalized distributions for most methods appear relatively Gaussian or log-normal. Additionally, most methods report a likely NF position with some uncertainty, which is easily represented with a Gaussian formalism. We highlight a few examples of the normalized distributions for different methods compared to Gaussian priors with various uncertainty estimations in Section~\ref{sec:prior_gauss_justification}. The Bayesian framework presented here is intended to be adjustable, such that other researchers may opt for different distribution shapes. 


Finally, we assign a weight to each method based on the relative confidence in the methods' NF position estimates. 
The weights of the available datasets for each source are totaled and then individual weights rescaled to be used in the Bayesian combination. 

Both the construction of prior distributions and the weighting prescription are flexible, allowing for any researcher to determine relative confidence in any combination of datasets. Altering the weights will impact the relative probability (i.e. the Gaussian A variable) of each prior distribution, and can influence the overall posterior distributions depending on which method is given higher confidence. Thus, the weighting of each method must be justified on an individual basis. 

We discuss the normalization parameters used to build prior Gaussian distributions, and weighting justifications for each method below. The parameters are summarized in Table~\ref{tab:z-scaling parameters}.\\

\textbf{Flux difference:} To build the Gaussian priors, we must obtain normalized uncertainties ($\sigma$) and offsets ($\mu$), which for the 8\micron flux difference correspond to the $\sigma_{\mathrm{MAD}}$ and median flux difference values for each cloud. 

The 8\micron flux difference method defines the threshold between near and far positions as the central flux value in the variable CMZ minus the constant 8\micron foreground emission ($f$), i.e. $\Delta\mathrm{F_{center}} = \frac{\mathrm{CMZ}}{2} - f \approx -108$~MJy/sr. More positive (negative) values denote more likely near-sided (far-sided) positions. To normalize the median flux difference values to a -1 (far) to +1 (near) scale, we use Equation~\ref{eqn:z-scale} with min and max extents of -150~MJy/sr and 150~MJy/sr, respectively. The asymmetric min and max bounds about $\Delta\mathrm{F_{center}}$ are a consequence of the assumed foreground, which is taken as the averaged minimum intensity of the Brick and Sgr B2. The flux difference is thus a quantitative measure of extinction relative to this value. The overall sample distribution is approximately Gaussian, offset by the value of $\Delta\mathrm{F_{center}}$. This offset NF threshold of -108~MJy/sr has a normalized value of $-0.72$, which is subtracted from the normalized $\mu$ values such that $z=0$ corresponds to the -108~MJy/sr. 

The Gaussian $\sigma$ for the flux difference is set by the normalized median absolute deviation ($\sigma_{\mathrm{MAD}}$) of the flux difference distribution for each cloud in the catalog, obtained using Equation~\ref{eqn:z-scale}.

Lastly, we assign the flux difference a relative weighting of 1, as there are significant uncertainties pertaining to assumptions for modeling constant foreground and  background emission.
\\

\textbf{Flux ratio:} The 8\micron flux ratio method from \paperIV is defined such that a value of 1 denotes an exact match to the 8\micron background, a value of 0 denotes complete obscuring of the background by a near-sided cloud, and the threshold between near and far is equal to 0.5. Because the flux ratio has clearly defined bounds for NF positions, we use the minimum and maximum of the flux ratio dataset to define the limits for Equation~\ref{eqn:z-scale}. 

The uncertainty for the flux ratio are determined from the normalized $\sigma_{\mathrm{MAD}}$ for each cloud using Equation~\ref{eqn:z-scale}. 

Similar to the flux difference method, we assign the flux ratio a relative weighting of 1, as it is subject to the same uncertainties from modeling the constant foreground and background emission.
\\

\textbf{Correlation coefficient:} By definition, the Pearson correlation coefficient, $r$, is bound between -1 (anti-correlation) and +1 (1:1 correlation) \citep{Pearson1895}. \paperIV defines masks where $r \geq 0.3$ as near-sided. For $|r| \geq 0.3$, we assign a Gaussian uncertainty equal to the standard deviation of all correlation coefficient values in the dataset.

The range between $-0.3 < r < 0.3$ is an ambiguous distinction, that requires more information to discern the NF position. Many astrophysical studies assume $|r|<0.4$ as a weak or poor correlation \citep[e.g.][]{Mokeddem2026,Aizawa2025}. We assign clouds with notably weak correlation a larger uncertainty, as results in this range cannot confidently indicate either near or far-sided positions. For clouds within these ranges, we assign a normalized uncertainty to be 3 times the standard deviation of all values in the dataset, i.e. $\sigma_{r} = 3~\cdot~ \mathrm{stdv(r)}$ 
 
The uncertainties and assumptions for the background estimates needed to compute the extinction column densities differ from those used for the 8\micron methods. Since the correlation coefficient is based off of the statistical correlation of column densities for each pixel in the cloud mask, we determine it a more confident measure of NF distinction, and assign a relative weight of 2. 
\\

\textbf{Molecular line absorption:} Line absorption fractions from \paperIII are defined such that fractional absorption values $>1$ indicate a far-sided position, and values $<1$ indicate a near-sided position. We normalize the absorption data using a minimum value of 0 and maximum value of 2. 

Uncertainties for the absorption method are obtained by calculating the error of the individual fractional absorption maps. For a given cloud mask, the fractional absorption is calculated using via: (median GBT C-Band Continuum + a cosmic microwave background estimate) / abs(H$_{2}$CO Min Intensity). For the C-band continuum, we estimate the noise as the standard deviation of the C-band map of the cloud mask ($\sigma_{\mathrm{GBT}}$). For the H$_{2}$CO uncertainty, we use the signal-to-noise ratio (SNR$_{\mathrm{H_{2}CO}}$) of the masked H$_{2}$CO data relative to the RMS of the cube. The absorption uncertainty $\sigma_{\mathrm{absorp}}$ for the cloud is then calculated via:
\begin{align}
\begin{split}
&\sigma_{\mathrm{absorp}} =\\&
\Big( \frac{\sigma_{\mathrm{GBT}} }{\mathrm{SNR}_{\mathrm{H_{2}CO}}}\Big) \sqrt{
\Big( \frac{\sigma_{\mathrm{GBT}} }{\mathrm{med ~GBT}}\Big)^2 + \Big(\frac{\mathrm{SNR}_{\mathrm{H_{2}CO}}}{\mathrm{min ~H_{2}CO}}\Big)^2 }
\end{split}
\end{align}


The molecular line absorption method is assigned a relative confidence weight of 2, based on the strong absorption features seen in the H$_{2}$CO spectra. 

We note that two cloud IDs 9 (20 km/s cloud) and 25 (SgrB2) have high fractional absorption, however their NF distinctions were manually overruled in \paperIII due to embedded star formation confusing the absorption analysis. These two clouds are also excluded from the absorption dataset for the PPDF analysis.\\

\textbf{Star count ratio:} The star count ratio is defined such that values $>1$ indicate a far-sided position, where the star count exceeds the model; and values $<1$ indicate a near-sided position, where the star count is less than the expected stellar distribution model. We normalize the star count ratio using a minimum value of 0 and maximum value of 2. 

The largest uncertainty in the star count method is due to the modeled stellar distribution. We assign a common normalized Gaussian $\sigma$ of 3 times the median $\sigma_{\mathrm{MAD}}$ of the normalized data obtained using Equation~\ref{eqn:z-scale}. Similar to the 8\micron methods, we assign a relative weight of 1 to the prior distributions, due to the large uncertainties from background modeling. 
\\

\textbf{X-ray echoes:} Unlike the other NF distinction methods, X-ray echoes provide a range of line of sight distance estimates for illuminated objects based on a given age of the illuminating event. The near-side distance constraints for the Sticks and Stone clouds from X-rays are quite stringent. Positions closer than 20~pc in front of the GC would reduce the delay between the past illumination and its reverberation in the cloud, so the event responsible for the echoes would have been detected directly by X-ray telescopes, which is not the case. The far-side limit is less constrained due to uncertainties in the age and number of events. Current estimates place the clouds up to 30~pc behind the GC \citep[e.g][]{Clavel(2013),Chuard2018,Stel2025}. It is possible for the clouds to lie anywhere within this range, but unlikely elsewhere. Thus, we characterize the X-ray distance constraints as a flat top-hat function, rather than a Gaussian curve or a piece-wise combination of a flat-top with broadened Gaussian tails.

As with the previous data sets, we convert the top-hat function to the -1 to +1 scaling. Normalizing the LOS distances onto a NF scaling requires a way to compare the distances to the relative positions of the clouds for which true distance estimates are unavailable. At this step in the procedure, there is no direct way to compare the reported distances and NF values. However, we may introduce an prior assumption for the minimum and maximum values along the LOS by projecting the the distances onto an orbital ring of an assumed radius. We use a $150$~pc galactocentric radius as a standard assumption for the CMZ radius, similar to other orbital model predictions in the literature \citep[e.g.][]{Kruijssen2015,Sofue2025}. The $\pm150$~pc estimate is used as the maximum and minimum values for Equation~\ref{eqn:z-scale}. The top-hat function is then normalized to integral unity and treated as a PDF for the Bayesian combination. 

The initial assumption for the CMZ radius has a non-negligible impact on the NF scaling, as it will place the cloud closer or further from the center of the ring (see Section~\ref{sec:limitations_PPDF} for a detailed discussion of uncertainties). 

Since the X-ray distances are well-constrained, and provide some of the more confident measures for line-of-sight (LOS) positions of clouds, we assign the method a relative weighting of 2.
\\
 
\textbf{Distances from proper motions:} The estimated distances to the 20 and 50\kms~ clouds from \citet{Nogueras-Lara2026} are converted to  the -1 to 1 NF scaling similar to the X-ray distance constraints, using $\pm150$~pc as the maximum and minimum values for Equation~\ref{eqn:z-scale}. However, we model the prior distributions as Gaussian, scaling the associated uncertainties as done when constructing the other Gaussian priors. 

Similar to the X-ray methods, converting the LOS distances to NF scaling requires an initial assumption of a CMZ radius. Altering this value will impact the normalization for this method, placing the clouds closer or further from \sgra depending on the chosen radius.

We assign a relative weight of 2 to the priors created from this method. Priors for both the stellar kinematics and X-rays are have significant systematics associated with projecting the LOS distances onto an assumed orbital radius (See Section~\ref{sec:limitations} for detailed discussion of uncertainties). The LOS distance-related metrics still provide useful limitations for the relative NF positions of these clouds, which is especially useful when interpreting the region near \sgra. We choose to include the x-ray and stellar kinematics methods in this analysis for a preliminary estimation of their positions, which may be further constrained in future iterations. \\

\begin{figure*}[t!]
    \includegraphics[width=\textwidth]{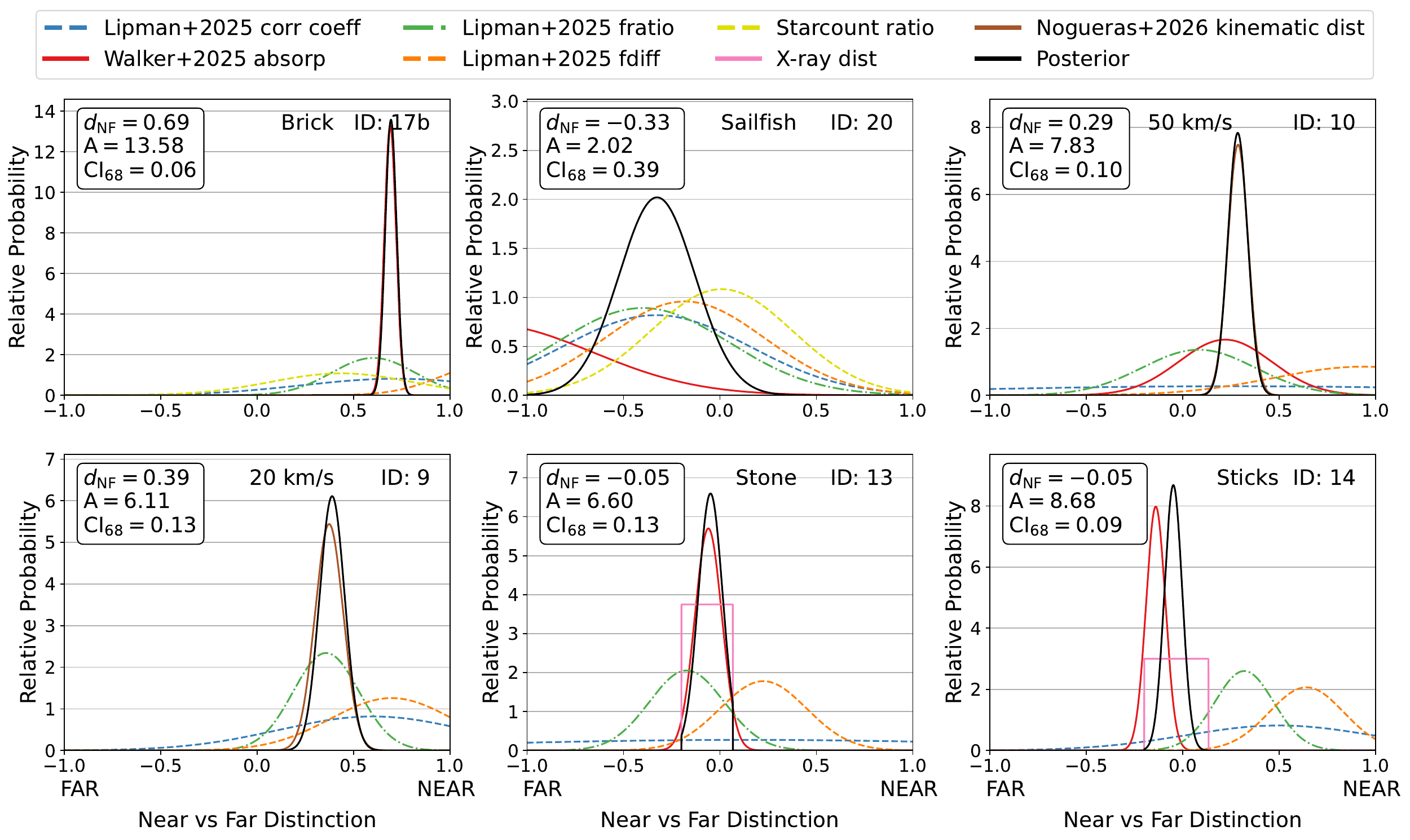}
    \caption{The methods used to infer NF positions of individual clouds in the CMZ can be combined in a Bayesian framework to create a posterior distribution of NF likely positions for a given cloud on a normalized scale of -1 (far) to +1 (near). The flux difference (orange dashed line) is used as a likelihood function, and is multiplied by distributions for the flux ratio (green), correlation coefficient (blue), absorption fraction (red), and star count ratios (yellow). A handful of clouds have prior data from X-rays (pink), or stellar kinematics (brown). 
    After taking the product of the priors, we report the location of the peak ($\mu$), the relative probability peak (A), and the 68\% confidence interval ($\mathrm{CI}_{68}$), which are used to obtain a Positional PDF posterior distribution (PPDF; black solid line). The peak of the posterior indicates the most likely NF position of the cloud. All distributions are normalized to unit integral probability.}
    \label{fig:ppdf_example}
\end{figure*}

After combining the priors, the posterior PPDF is normalized to unit integral probability. Examples of normalized prior distributions for each method, as well as the normalized posterior distributions are shown in Figures~\ref{fig:ppdf_example} and~\ref{fig:PPDFs_appendix1} as colored and black lines, respectively. The location of the posterior peak on the x-axis (d$_{NF}$) indicates the most likely, normalized NF position of the cloud. The peak value (A) gives the relative probability of the cloud being located at that NF position. We also report a 68\% confidence interval (CI68) for each posterior PPDF, which will be used as uncertainties in the orbital fitting procedure in Section~\ref{subsec:fitting_procedures}.

\subsection{Minimization procedure to find a best fitting \texorpdfstring{x$_2$}~ \text{ellipse}}\label{subsec:fitting_procedures}

Material in the CMZ is thought to lie on a closed ellipse, often treated as x$_2$ orbits which arise from the Galactic bar potential as described by numerical solutions derived in \citet{Binney1991}. It is not fully understood what sets the extents of the x$_2$ orbits, though there is evidence that the sizes are correlated to the location of Inner Linblad Resonances of the Galactic potential \citep[e.g.][]{Contopoulos_Grosbol_1989,Athanassoula(1992),Buta_Combes1996}. A simple parametric definition of an x$_2$ ellipse can be described in cartesian coordinates as follows:
\begin{align}\label{eqn:ell_eqns}
    x &= a \cos(\phi) \\
    y &= -b \sin(\phi) \\
    z &= z_0 \sin(-2\phi + \alpha)
\end{align}
with radii determined by semi-major axis $a$ and semi-minor axis $b$. The vertical height of the orbit is determined by $z_0$. The sinusoidal term of the $z$ component includes a factor of -2, which makes the ellipse appear as the twisted infinity shape in a plane-of-sky $\ell b$ projection (note this implies that the vertical oscillation frequency is a multiple of the orbital period, which is unlikely, though we use it here to test a simple geometry). The azimuthal angle $\phi$ is measured from the starting point of the ellipse, and $\alpha$ is a phase offset. This modeled ellipse assumes a constant angular momentum (i.e. $L_{z} = v_{\phi}R$ is a constant), and allows for vertical oscillations. The assumption of constant angular momentum, although not exact, is a reasonable approximation for x$_2$ orbits and the PPV shape of the observed gas distribution (see \paperIII). Our assumptions differ from the closed ellipse proposed by \citet{Molinari2011} in two ways: 1) we center the ellipse on \sgra (i.e. the minimum of the gravitational potential) while theirs is not, and 2) we assume a constant angular momentum, while \citet{Molinari2011} assumes constant orbital velocity. A constant angular momentum allows for higher (lower) velocities at peri(apo) center, and more closely resembles a true x$_2$ orbit \citep[e.g. see][]{Levy2022}. 

We examine the parameter space for the standard parametric ellipse by projecting the equations into a LOS derived geometry, such that we fit the ellipse as it is seen in $\ell bv$ space. We fit using the following parameters: $a$, $b$, and $z_0$ to describe the size of the ellipse; the initial tangential velocity ($v_0$); the angle of the ellipse measured from the Sun-Galactic center line ($\theta$); and phase angle $\alpha$. In \paperIII, the parameters of the toy model were chosen by-eye to match the observed $\ell bv$ gas distribution, and resulted in reasonable fits to NF positions determined by the molecular line analysis in \paperIII and the dust extinction analysis in \paperIV. 

To find a best fitting x$_2$ orbital model, we employ a minimization routine that incorporates the normalized posterior distributions for each source obtained in Section~\ref{subsec:bayseian_posteriors}, and the x$_2$ ellipse model defined above. 

We use the Non-Linear Least-Squares Minimization and Curve-Fitting python package \texttt{lmfit} \citep{LMFIT} as a framework for our orbital fitting procedure. The \texttt{lmfit.minimize()} function performs standard $\chi^{2}$ minimization, and provides a flexible framework to adjust parameters and fitting methods. We use a Nelder-Mead fitting method, which tends to more accurately and efficiently sample the available parameter space compared to a least-squares regression \citep{Kumar2023}. 

The target of the minimization procedure is a residual function of a given modeled ellipse to the dataset of posterior PPDFs. The procedure is as follows:
\begin{enumerate}
    \item We initialize an elliptical orbital model based the parametric equations~\ref{eqn:ell_eqns}-7. The ellipse function takes parameters of a semi-major axis (a), semi-minor axis (b), orbital height ($z_0$), initial tangential velocity ($v_{0}$), a rotation about the z-axis ($\theta$), and phase angle ($\alpha$). 
    \item In order to compare the initialized orbital model with the likely NF cloud distributions, we must first provide a NF model estimate. We follow a similar normalization procedure as described in Section~\ref{subsec:bayseian_posteriors} to translate the model's $\ell$, $b$, and $v$ position arrays onto the -1 (far) to +1 (near) scaling. We perform the scaling procedure using Equation~\ref{eqn:z-scale}, where the min and max values for normalization are based on limits of observed $\ell bv$ projections of the CMZ ring: ($\ell_{min}$, $\ell_{max}$) = (-0.6\deg, 1.7\deg); ($b_{min}$, $b_{max}$) = (-0.3\deg, 0.1\deg); and ($v_{min}$, $v_{max}$) = (-100 km/s, 100 km/s). The central $\ell bv$ positions for each catalog cloud are normalized using the same limits.
    \item Next, we find the minimum distance between each cloud in the catalog and the model ring using the \texttt{scipy.cdist} package with a Mahalanobis distance metric \citep[][also see e.g. \citet{Blaylock-Squibbs2023}]{Reprint_Mahalanobis2018}. The Mahalanobis distance measures the distance between the model and the data, normalized by the data uncertainty. 
    \item Due to the tight correlation of points in $\ell bv$ space, the Mahalanobis distance alone fails to constrain the latitude distribution of observed material. To address this, we employ a Wasserstein distance metric \citep{monge1781}, also referred to as the ``earth mover's distance" in computer science and optimal transport problems \citep[see][]{kantorovich1942,villani2021}. The Wasserstein distance is a measure of similarity between two probability distributions, and can be written as: 
    \begin{equation}
        W_p(P,Q):= \inf_{\substack{\pi \in \Gamma(P,Q)}} \Bigg(\int_{\mathbb{R}\times\mathbb{R}} ||x-y||^{p} d\pi(x,y)\Bigg)^{1/p}
    \end{equation}
    Where P and Q are probability distributions on $\mathbb{R}$ with $p$ moments. The 1D Wasserstein distance computes the distance between probability distributions, and quantifies the ``cost" of moving points from the model to the observed data (in the earth mover's context, the physical metric of work needed to convert one pile of dirt to another). The cost is calculated as the product of the probability mass to be transported and the distance the mass is moved. For our purposes, we sample the ellipse model's latitude distribution by the number of data points used in the model fitting (22), and then uniformly distribute the total catalog cloud mass along the latitude of the model. We then evaluate the cost to move the mass on the ellipse to match the data, and vice versa. The sum of the transportation costs for each cloud is returned as a scalar. We add this output scalar to each of the minimum Mahalanobis distances for a given cloud, resulting in an optimized distance which corrects for the observed latitude-distribution of the data.
    \item Using the model point with the minimum Mahalanobis distance to the cloud, we assess the model and cloud's NF agreement using a k-nearest neighbors approach with the \texttt{scitkit-learn} package's \texttt{NearestNeighbors} functionality. We use $k = \sqrt{N} = 600$, where $N$ is the number of data points in the model. The NF positions of the nearest neighbors are compiled, and the most common NF position value is used for the nearest model point, d$_{\mathrm{NF, model}}$. This step is most important near the edges of ellipse model, where the the nearest neighbors may span between the near and far sides.
    \item We then calculate a residual penalty between the NF position of the nearest $\ell b v$ point on the ellipse to the posterior likelihood distribution for the cloud. We approximate each cloud's PPDF as a Gaussian distribution which gives the relative probability that the cloud lies at a given distance, $P(\mathrm{d})$: 
    \begin{equation}
    P(\mathrm{d}) =
    A \cdot \mathrm{exp}\Big(\frac{-(\mathrm{d}-d_{\mathrm{NF}})^{2}}{2 (\mathrm{CI68})^{2}}\Big),
    \end{equation}
    where A, d$_{\mathrm{NF}}$, and CI68 are the peak probability, most likely NF position, and CI68 of a given cloud as reported in Table~\ref{tab:los_synth_table}. Using the above formalism, which utilizes the full shape of the Gaussian PPDF, we find the relative probability that a cloud exists at the $\ell b v$ location of the nearest model point ($P(\mathrm{d}_{\mathrm{NF,model}})$). We then find the difference between the model point's relative probability and the peak probability of the cloud at its most likely position from the PPDF ($P(\mathrm{d}_{\mathrm{NF}}) = $A). Thus, we compare relative probabilities of the model position to the peak PPDF value. This difference quantifies how far the closest NF point on the ellipse model is from a cloud's PPDF peak. The result is treated as a NF penalty value.
    \item Finally, each cloud's optimized distance to the model (Mahalanobis distance + Wasserstein output scalar) and NF penalty are added together to create a list of residuals for the input fitting model. The calculated reduced chi-square, \redchisq, is used to determine the best fit to the dataset. 
    
\end{enumerate}


\subsubsection{Points excluded from orbital fitting}\label{subsec:excluded_clouds}

While the cloud catalog provides a list of the densest structures in the CMZ, not all reported structures are necessarily appropriate to include in the x$_2$ orbital fitting. The full catalog has a total of 47 objects, including 17 sub-masks for clouds with multiple velocity components detected in the molecular line data. To avoid artificially weighting these multi-component sources over the single-component clouds, we only include the masks containing the most pixels for clouds with multiple velocity components. The IDs relating to the main masks used in the fitting are noted by an asterisk in Table~\ref{tab:los_synth_table}.

Additionally, we only include clouds that could be reasonably associated with the main CMZ orbit, and have sufficient data to inform the fitting. Cloud IDs 27 and greater lie off of the typically assumed 100~pc projected radius. These clouds also lack data from the absorption method, and show the most conflict between the dust extinction methods. We exclude these clouds from the fitting based on the inconsistencies between methods and lack of data in these regions. 

Lastly, there are a handful of clouds near \sgra in projection that likely do not lie on a typical x$_2$ orbit, but perhaps sit closer to \sgra or the CND. The 20 and 50 km/s clouds (IDs 9 and 10, respectively) have been theorized to be interacting with or falling in towards the CND based on their line of sight velocities \citep{Lee2008, Karlsson2003, Yan2017}. Likewise, the Stone (ID 13) and Sticks (ID 14) clouds, have been confidently constrained to be near \sgra via X-ray fluorescence, as discussed above \citep{Clavel(2013),Chuard2018,Stel2025,Marin2023}. Enough evidence points towards these four clouds not being a part of the main x$_2$ orbit that the majority of other structures seem to trace. Including clouds which do not lie on x${2}$ orbital paths into the minimization procedure can skew the fitting and greatly impact the potential best-fit CMZ sizes. The impact is especially important for these clouds, which have well-constrained PPDFs. Thus, we remove IDs 9, 10, 13, and 14 from the orbital fitting procedure and instead use the fitting results to interpret their relative positions to other clouds based on their posterior PPDFs in Section~\ref{sec:discussion}.

\section{Results}\label{sec:results}

\subsection{Synthesis of Near vs Far Positions for CMZ Catalog Clouds}\label{subsec:results_posteriors}

The posterior PPDFs described in Section~\ref{subsec:bayseian_posteriors} peak at the location of the most likely NF position for each cloud, and are measures of the relative NF positions of clouds between each other. We report the posterior PPDF parameters in Table~\ref{tab:los_synth_table}. The PPDF peak $\mathrm{A}$ represents the relative probability of the cloud being located at a given position. The d$_{\mathrm{NF}}$ value indicates the normalized NF position and CI68 is the 68\% confidence interval of the highest peak of the posterior distribution.

In many cases, the absorption has a significant impact on the position and uncertainty of the posterior due to the more constrained uncertainties (e.g. IDs 17b, 13, and 14 in Fig~\ref{fig:ppdf_example}). The posterior PPDFs are mostly Gaussian, although there are some distributions which show double peaks (e.g. ID 15) or sharp edges due to the step-function priors for X-ray data (e.g. ID 13). Overall, the Bayesian combination results in reasonable distributions, given the available constraints, and can be used to confidently determine relative positions of the cataloged clouds.

\subsection{Best fitting \texorpdfstring{x$_2$}~ \text{orbits}}\label{subsec:results_bestfit_x2}

 The fitting procedure described in Section~\ref{subsec:fitting_procedures} outputs parameters of a semi-major axis (a) in kpc, semi-minor axis (b) in kpc, orbital height (z$_0$) in kpc, initial tangential velocity ($v_{0}$) in \kms, a rotation about the z-axis ($\theta$) in degrees, and phase offset of the ellipse $\alpha$ in radians. During initial tests, the fitting algorithm produced various CMZ sizes depending on the initial conditions, particularly the input a and b radii, which are well-correlated. 

\begin{figure*}[t!]
    \centering
    \includegraphics[width=\textwidth]{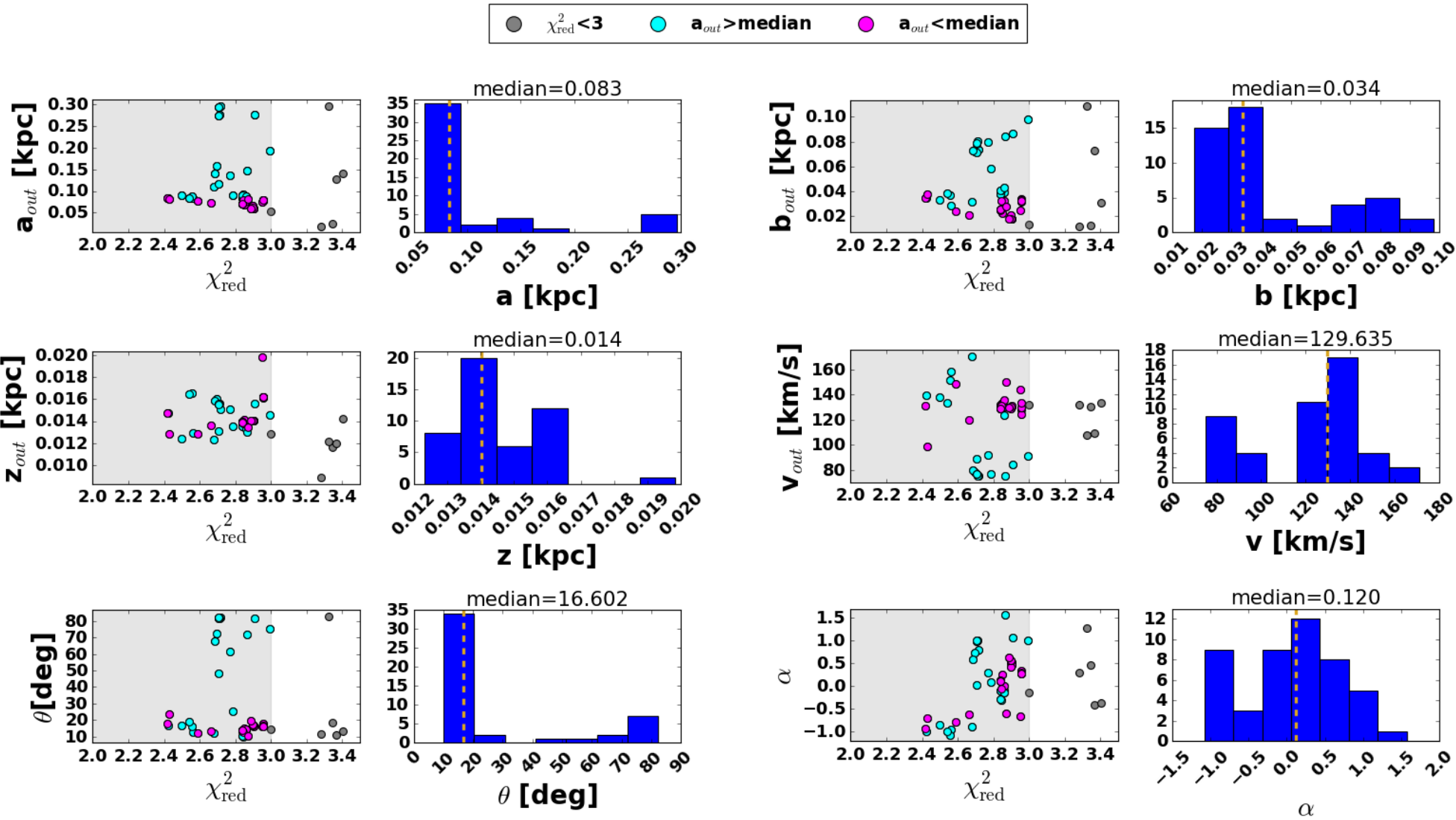}
    \caption{Left-hand panels show \redchisq values  for output parameter values from the parameter grid search. The gray shaded region covers the \redchisq range $1<$ \redchisq $< 3$ where the majority of outputs converge. Cyan (magenta) points correspond to outputs for ellipses with a major axis greater (less) than the median a$_{\mathrm{out}}$ value. Gray points note output values with \redchisq $>3$, indicating poor fits which are excluded from the histograms. Right-hand panels show histogram distributions of the output parameter values which lie within the gray shaded region from the left panel plots, along with the median output value (gold dotted line). 
    }
    \label{fig:redchiqsq_converge}
\end{figure*}

To find convergent values for each of the parameters, we run the fitting procedure for a grid of initial ellipse sizes with major axes ranging from 10 -- 300~pc in 5~pc increments. The input minor axis is set using the initial b/a ratio from the \paperIII ellipse (b/a = 0.61). Using elliptical initial values avoids cases where the algorithm may sample minor axes larger than the major axes. We do not explore the parameter space of initial conditions for z, $v_{0}$, $\theta$, or $\alpha$ as we are most interested in the possible CMZ sizes, and find these values from \paperIII to be reasonable initial guesses for the fitting procedure. Additionally, there are degeneracies in the projections from $\theta$ and $\alpha$, meaning it is more efficient to open a broad parameter space for these variables to sample in the optimization, rather than vary them in the parameter grid. After setting the input values, the minimization routine searches for ellipse parameters between set lower and upper bounds, reported in the first two columns of Table~\ref{tab:orbital_fitting_summaries}, without further constraints.
The output distribution of the parameter grid test is shown in Figure~\ref{fig:redchiqsq_converge}. The gray shaded region in the left column covers the \redchisq range $1<$ \redchisq $< 3$ where a majority of the outputs converge. Cyan (magenta) points correspond to outputs for ellipses with a major axis greater (less) than the median a$_{out}$ value. Gray points are outputs with \redchisq $>3$. The right column shows histogram distributions of the points between $1<$ \redchisq $< 3$, with a gold line denoting the median value of each output parameter.

The parameter grid results in fits that are within \redchisq ranges of $2.4 <$ \redchisq $<3.5$, with \redchisq$= 3$ used as a typical indicator for reasonable fits. The best convergence for each parameter seems to fall within the range of $2.4<$ \redchisq $< 3.0$. We find the minimum, median, and maximum value of each output parameter value based on a ranges where the \redchisq~converged for the grid. We assume the median fit value as the best-fit for the parameter search, though the bulk gas may exist on orbits anywhere between the inner and outer extents. We summarize the outputs from the grid search in Table~\ref{tab:orbital_fitting_summaries}. The smallest semi-major axis obtained within the convergent \redchisq~ range is $ a = 60$~pc and the largest is $ a = 296$~pc. 
\setlength\LTcapwidth{\textwidth}

\begin{figure*}[th!]
    \centering
\includegraphics[width=1\textwidth]{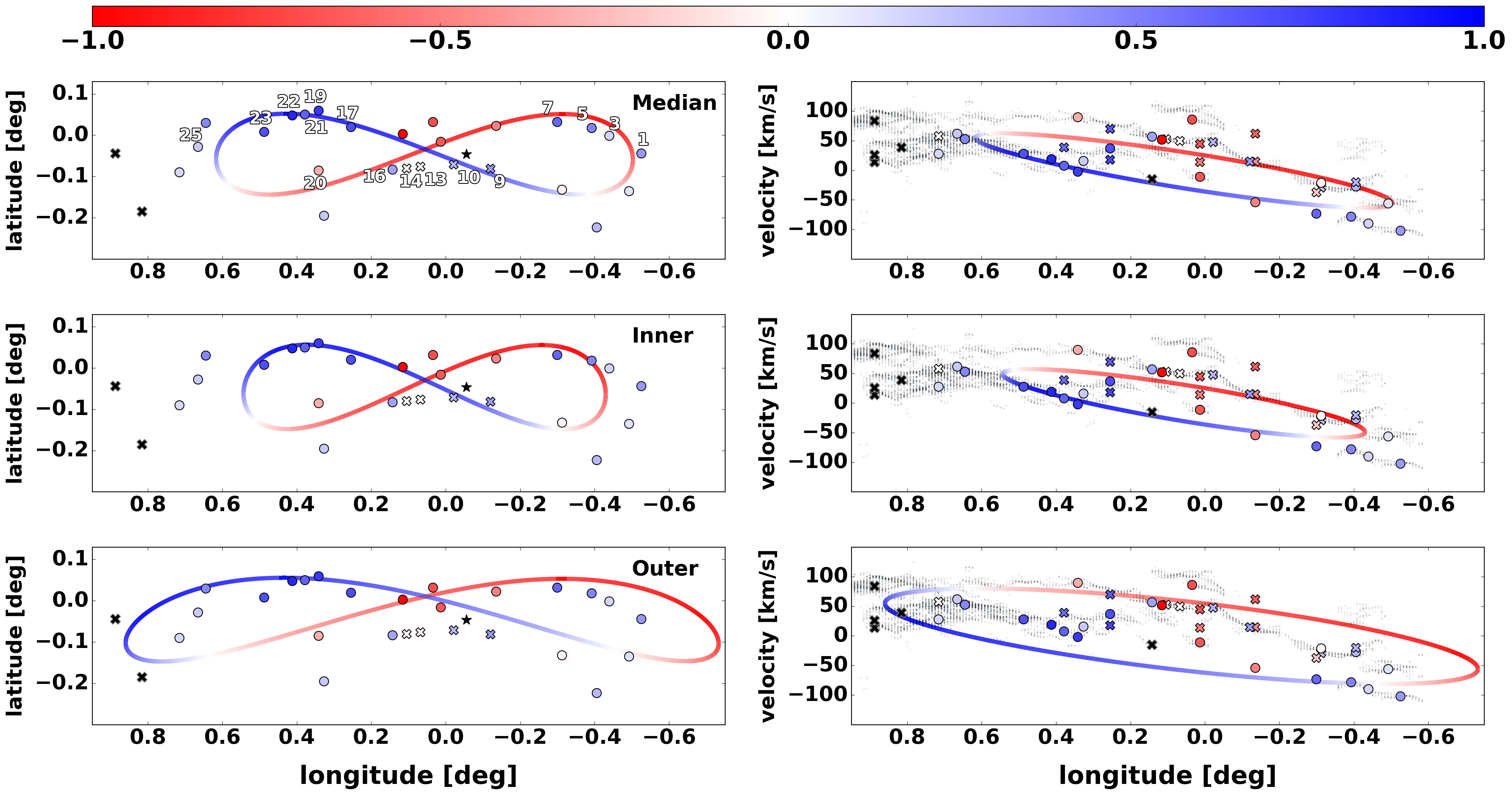}

\caption{The parameter grid search results in a range of best fitting x$_2$ elliptical orbits for the star forming ring. This figure shows ($\ell$, b) (left) and ($\ell$, v) (right) projections of the best fitting parameters summarized in the right-hand section of Table~\ref{tab:orbital_fitting_summaries}. Blue colors indicate more-likely near-sided positions, while red indicates more likely far-sided. The black star indicates the location of \sgra. Marker colors correspond to their most likely posterior PPDF positions. Colored x markers indicate clouds that have NF distinctions based on their PPDFs, but are not included in the fitting procedure. Black x markers are clouds which are not included in the fitting procedure, and for which a formal NF distinction is not shown, due to the limited data and association with the majority of CMZ data points. Gray background points correspond to the spectral decomposition of MOPRA HNCO data from \citet{Henshaw2016_GasKin_250pc}. ID numbers for well-known clouds are noted in the top left panel.}
\label{fig:6panel_lblv_fittings}
\end{figure*}

\begin{longtable*}{l|ccc|cccc|cc}
\caption{Summary of the upper and lower bounds used for the orbital ellipse minimization routine, example parameters from Paper III, and the output fitted parameters obtained from the parameter grid search. The ellipse function takes parameters of a semi-major axis (a), semi-minor axis (b), orbital height ($z_0$), initial tangential velocity ($v_{0}$), position angle of the semi-major axis with respect to the line-of-sight ($\theta$), and phase offset of the ellipse ($\alpha$). The input parameters for a and b are set by a ratio (b/a = 0.61) using parameter grid sampling inputs between $10<\mathrm{a}<300$~pc in 5~pc increments. All other inputs match those in \textit{Paper III}. We report the median parameter values from the parameter grid evaluation for results with $1 <$ \redchisq $< 3$. The median output fit values from the parameter grid search are taken as the best-fitting parameters for an average orbit.  The inner and outer extent values are used to create the top-down projection in Fig~\ref{fig:x2_topdown_fits}.}
\label{tab:orbital_fitting_summaries}\\
\hline
  & \multicolumn{3}{c|}{\textbf{Input Parameters}}& \multicolumn{4}{c|}{\textbf{Parameter Grid Search}} & \multicolumn{2}{c}{\textbf{Top-Down Extents}}  \\ 
\hline

Fitting  & Lower & Upper & Value from &Min &  Median & Max & Standard & Inner  & Outer   \\

\newline
parameter &Bound& Bound&Paper III & Fit Value   &  Fit Value & Fit Value  &  Deviation & Extent & Extent \\
\hline
a [kpc]         &   0.01  & 0.3   &   0.9        & 0.060  &  0.083  & 0.296  & 0.066  &  0.072  & 0.146    \\
b [kpc]         &   0.01  & 0.2   &   0.055      & 0.018  &  0.034  & 0.098  & 0.022  &  0.026  & 0.058    \\
$z_0$ [kpc]     &   0     & 0.05  &   0.0125     & 0.0123 &  0.0140 & 0.0198 & 0.0014 &  0.0146 & 0.0144    \\
$v_{0}$ [km/s]  &   75   & 175    &   130        & 75.0   &  129.6  & 170.8  & 25.3   &  130.9   & 109.0      \\
$\theta$ [deg]  &   -180     & 180   &   25      & 10.0    &  16.6   & 82.3   & 25.1   &  15.8    & 42.9      \\
$\alpha$ [rad]  &   $-\pi/2$ & $\pi/2$  &   0.4  & -1.08  &  0.12   & 1.57   & 0.66   & 0.04    &  0.11      \\
\hline
\end{longtable*}
Most of the parameter grid test converged on a median value of $ a = 83$~pc. We also report the average values from the parameter grids for output with major axis values above and below the median $\mathrm{a} = \mathrm{a_{median}}$, corresponding to the outer and inner extents of the ellipse, respectively. We use these as inner and outer extents of the ellipse orbits with major axis sizes of $72~\mathrm{pc}< \mathrm{a} < 146~\mathrm{pc}$. The $\ell b$ and $\ell v$ projections of the median, inner, and outer ellipse extents are shown in Figure~\ref{fig:6panel_lblv_fittings}. Colored markers represent NF positions, where bluer points indicate more near-sided positions (i.e. more positive d$_{\mathrm{NF}}$ values), and red is more far-sided values (negative d$_{\mathrm{NF}}$).

\subsection{Relative positions of clouds and comparison to best fitting \texorpdfstring{x$_2$}~ \text{orbits}}\label{sec:disc_relative_positions}

\begin{figure*}[t!]
    \includegraphics[width=\textwidth]{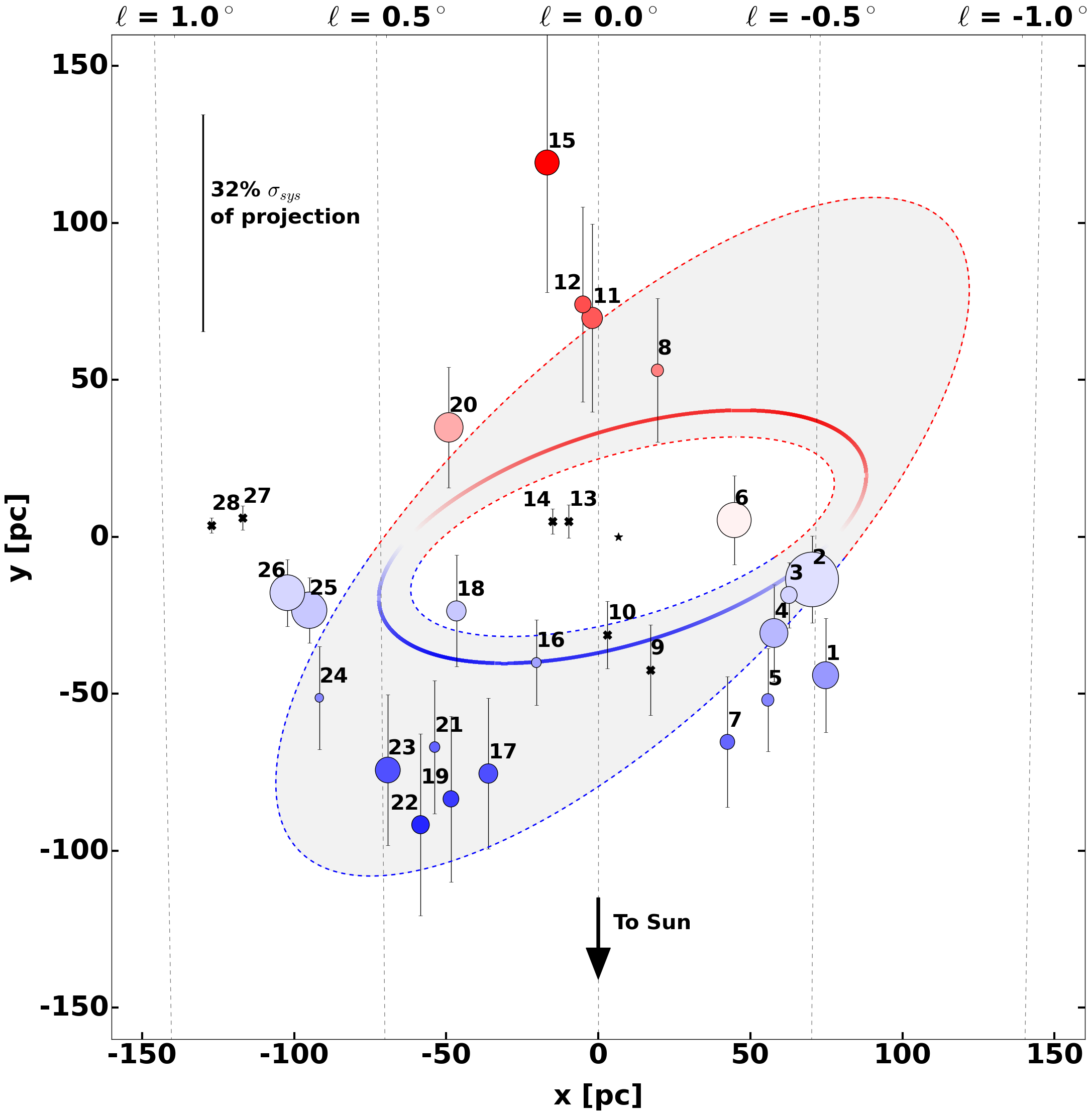}
    \caption{We present a preliminary top-down view of the best fitting CMZ orbits, as well as projected positions of all cataloged molecular clouds based on the outer ellipse model from the parameter grid search. The above figure shows a top-down projection of the orbits summarized in Table~\ref{tab:orbital_fitting_summaries}. The middle ellipse corresponds median x$_2$ orbit from the parameter grid search. The gray shaded region is bound by the best-fit inner and outer extents, and indicates a range of possible x$_2$ orbits. Blue and red colors indicate more-likely near or far side positions, respectively. Positions of CMZ clouds which were included in the x$_2$ fitting are denoted by circular markers, with colors corresponding to the location of the Gaussian peak in the posterior PPDF for each cloud, and sizes corresponding to the effective radius of each cloud as reported in \citet{Walker_2025}. Black `x' markers denote clouds which were not incorporated in the x$_2$ fitting. Error bars correspond to the uncertainties of the projected positions calculated using the PPDF CI68 values and standard deviation in the projected distances on all three ellipses for each cloud. A 32\% systematic error bar for the LOS distance projections is shown in the top left. The black star denotes the location of \sgra.\\
    \\
    }
    \label{fig:x2_topdown_fits}
\end{figure*}

The posterior PPDFs can be projected onto the best fitting orbital models to obtain LOS distances for all CMZ clouds. To explore the 3D distribution CMZ clouds, we use the median output ellipse parameters and inner/outer extents reported in Table~\ref{tab:orbital_fitting_summaries} to create a top-down view of the CMZ. To project the NF positions to LOS distances for a given orbit, we use the nearest LOS point on the ellipse ($\mathrm{R_{\mathrm{min}}}$), and the furthest point ($\mathrm{R_{\mathrm{max}}}$) based on the outer ellipse extents to project the normalized -1 (far) to +1 (near) scale into a distance from the GC in parsecs 
using a simple formula:
\begin{equation}
    y_{\mathrm{pc}} =  \frac{(y_{\mathrm{NF}} + 1)}{2} \cdot (\mathrm{R_{\mathrm{max}}} - \mathrm{R_{\mathrm{min}}}) + \mathrm{R_{\mathrm{min}}}
\end{equation}
Where $y_{\mathrm{NF}}$ is the top-down near (+1) vs far (-1) position on the same scaling as the PPDFs. For each cloud in the catalog, we then project the PPDF peak position ($\mathrm{d}_{\mathrm{NF}}$) onto the ellipse, which converts the normalized $\mathrm{d}_{\mathrm{NF}}$ value into parsecs. Similarly, we convert the CI68 to physical units by multiplying this value by the LOS width of the ellipse ( $\sigma \cdot (\mathrm{R_{\mathrm{max}}} - \mathrm{R_{\mathrm{min}}})$ ). 

We report the LOS positions obtained from projecting each cloud in Table~\ref{tab:los_synth_table} onto the outer ellipse. The LOS uncertainties are calculated as the standard deviation of the positions from projecting onto each of the three ellipse extents (the standard deviation of $\mathrm{y}_{\mathrm{min}}$, $\mathrm{y}_{\mathrm{median}}$, $\mathrm{y}_{\mathrm{max}}$). The top-down projected positions and uncertainties for each cloud are shown in Figure~\ref{fig:x2_topdown_fits}. Blue and red colors indicate more-likely near or far-sided positions, respectively. Clouds which were included in the fitting are denoted by circular markers, with colors corresponding to the posterior PPDF peak, and sizes based on the effective radii obtained in \paperIII. Black `x' markers denote clouds which were not incorporated in the orbital fitting. Error bars correspond to the LOS uncertainties of the projected positions. The conversion of PPDFs to distances introduces an approximately 32\% systematic error to the LOS estimates, based on the spread of LOS values between the three ellipse extents. A 32\% systematic error bar is shown in the upper left of Figure~\ref{fig:x2_topdown_fits}. We ask readers to regard the projected LOS distances qualitatively and refer to the relative positions of clouds as a more confident measure of positions from the fitting.

\section{Discussion}\label{sec:discussion}

\subsection{The 100~pc ring}\label{subsec:disc_100pc_ring}

The posterior PPDF results indicate the relative NF positions of CMZ clouds in the catalog on the normalized -1 to +1 scale. The nearest clouds include the contiguous dust ridge between the Brick and SgrB2 ($0.2 \lesssim \ell \lesssim 0.6$). The densest CMZ cloud, the Brick (ID 17b) has a normalized $\mathrm{d}_{\mathrm{NF}} \sim 0.69$, placing it confidently in front of the GC. ID 22 (clouds e\&f) has the nearest value in the catalog of $\mathrm{d}_{\mathrm{NF}} \sim 0.84$, with a higher relative probability ($A \sim 40$) than the Brick's peak value ($A \sim 13$), due to its very strong absorption, flux ratio, and correlation coefficient measurements. 

In addition to the dust ridge clouds, there is a stream of near-sided structures between $-0.5 \lesssim \ell \lesssim -0.1$ and $b \lesssim -0.05$, colloquially coined ``the wiggles" due to the area's corrugated velocity pattern, potentially related to gravitational instabilities \citep{Henshaw2020,Henshaw2016_Seeding}. The wiggles region covers catalog IDs 1, 3, 5, and 7, all of which have likely near-side PPDF positions. However, they overlap with a far-sided part of the best fitting ellipse. When viewed top-down in Figure~\ref{fig:x2_topdown_fits}, these clouds appear near the edge of the orbit, possibly transitioning between the near and far sides. The orbital model from \citet{Kruijssen2015} also finds the NF distinction of the orbit in this area ambiguous, as their model's streams 4 and 2 are indistinguishable in this region. 

Only 10 masks in the catalog have far side distinctions, mostly located near the \sgra region. The furthest cloud is ID 15, located at $(\ell, b) = (0.014, -0.016)$ with likely position $\mathrm{d}_{\mathrm{NF}} = -1.107$ which is heavily influenced by strong absorption measurements. Determining likely positions for objects in the Sgr A region, between $-0.1 \lesssim \ell \lesssim 0.15$, is particularly difficult since many of the NF methods, including the flux difference, flux ratio, and absorption measures are quite sensitive to uncertainties and variation in the bright background emission. The clouds in this region, as well as the Sailfish (ID 20) seem to align well with the far-side stream of the ellipse, with fairly high relative probabilities.

\subsection{Comparisons with current LOS distance estimates to CMZ clouds}\label{subsec:inflow}

The LOS positions for a handful of clouds have been widely discussed in the literature. Using the top-down projection as described in Section~\ref{subsec:disc_100pc_ring}, we compare their relative positions to the x$_2$ orbits and the GC. 

\textbf{The Brick:} Our top-down projection places the Brick at a distance of $75\pm 26~\mathrm{pc}$ on the near side of the CMZ. The position of the brick has been debated in the literature, with confident nearside placement based on its stark absorption features \citep{Nogueras2021,Walker_2025,Lipman_2025}, though some constraints place it on the nearest edge just outside of the CMZ \citep{Zoccali2021}. We find the Brick is not the most near-sided cloud, as the PPDF for Cloud e\&f peaks at an even more near-sided position. Our results corroborate estimates placing the brick on the near side on the GC, but still within the CMZ.

\textbf{20 and 50\kms Clouds:} The 20 (ID 9) and 50\kms clouds (ID 10) were excluded from the orbital fitting, but their posterior PPDFs allow us to interpret their relative positions in the CMZ. Our top-down projection places them at $43\pm 19~\mathrm{pc}$ and $31\pm 14~\mathrm{pc}$ in front of \sgra, respectively. The clouds are also connected by a ridge of material \citep[see][]{HerrnsteinHo2005,Nogueras-Lara2026}. Although their location is controversial, many models place the clouds quite close to \sgra or curving slightly behind. The close proximity to \sgra is typically justified by the non-illumination from X-rays \citep{Chuard2018,Stel2025}, evidence of interactions with the CND \citep[e.g.][]{Tsuboi2018}, and interactions between the 50\kms Cloud and the SgrA East supernova remnant \citep{Genzel1990, HerrnsteinHo2005, Lee2008} which has been proposed to lie close to the CND and behind \sgra \citep{Yusef-Zadeh1987,Pedlar1989}.

However, dynamical models and absorption measurements strongly favor near-side interpretations for both clouds. \citet{Sofue2025} projects their distances onto inner spiral arms of the CMZ, placing the 20\kms~cloud 42~pc in front and the 50\kms~cloud between $7-42$~pc in front, depending on its association with Arms III or V. Additionally, the \citet{Kruijssen2015} open stream orbital model places them $\sim 50$~pc in front of \sgra. Mills et al. (accepted) justify the near-side placement by proposing an alternative interpretation of radio constraints, with Sgr A East being seen in absorption in front the GC synchrotron background, placing it in front of Sgr A West. Our results agree with these near-side interpretations, and support the interpretation that the clouds could be actively infalling off of an inner x$_2$ orbit towards the CND, making them kinematically distinct from the main elliptical structure. 

\textbf{The Three Little Pigs:} The Stone (ID 13) and Sticks (ID 14) clouds are within 15~pc of \sgra in the ($\ell$,b) projection, while the Straw cloud (ID 16) lies at about 40~pc. X-ray observations confidently constrain the LOS positions between 30~pc behind to 20~pc (10~pc) in front of the GC for the stone (sticks) cloud \citep[e.g][]{Clavel(2013),Chuard2018,Stel2025}. The clouds are unlikely to be more than 20~pc in front of the GC, as they would not have been observed as illuminated by a flaring event otherwise. Our results place both clouds within $\sim$10~pc behind \sgra.

The estimated distances imply the clouds do not lie on an x$_2$ orbit, but are instead part of an inner spatial region which may be kinematically distinct from the star forming ring. Our projected model places the Straw cloud even further in front, meaning it may eventually be illuminated by the same X-ray event as the stone and sticks clouds in the future.


\textbf{Sgr B2 and Sgr C:} The X-ray data for SgrB2 (ID 25) is well-documented and suggests a lower limit for the LOS distance of $\sim 50~\mathrm{pc}$ in front of the GC \citep{Chuard2018, Stel2025}, with various systematic uncertainties. Our results place the cloud at  $23\pm15~\mathrm{pc}$ in front of the GC. However, the systematic uncertainty of this estimate is quite high, due to the cloud being on the edge of the orbit where there is less deviation in its projected distance between the ellipses. We also note high uncertainties in this area due to the complexity of the Sgr B2 region and variable ISM conditions. 
X-ray constraints for Sgr C (ID 2) place strict lower limits for parts of the region of at least $\sim 30$~pc in front of the GC \citep{Chuard2018, Stel2025} with some estimates placing it on a nearer edge of the CMZ around $\sim 70 - 100$~pc \citep{Ryu2013}. Our projected model places Sgr C at $14\pm18$~pc on the near side, which is within range of uncertainties for lower X-ray limits. Sgr C also appears to lie at the edge of our best fitting orbital model, potentially wrapping around near to far sides of the outer orbital extent. \\

Overall, our results agree with current literature constraints, within our reported errors. The agreement supports our method's ability to utilize LOS distance constraints in combination with NF placements, and produce LOS distance estimates that are reasonably in line with preliminary results from various datasets.

\subsection{CMZ size and morphology}\label{sec:morphology}
\subsubsection{Is the CMZ a simple ring?}\label{sec: discussion_family_of_orbits}

\begin{figure*}[th!]
\includegraphics[width=1\textwidth]{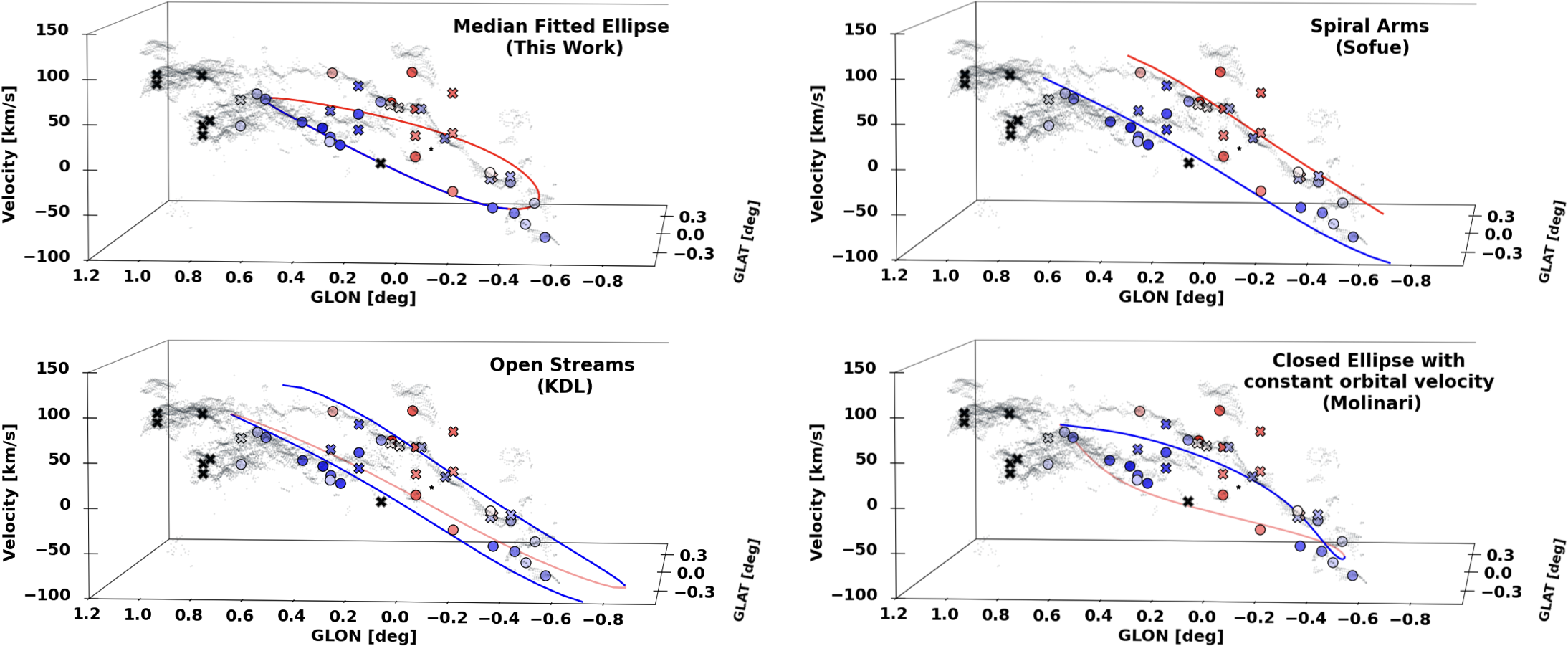}

\caption{We present the NF positions of our cataloged clouds against four different 3D orbital models in PPV space plotted over spectral decomposition of MOPRA HNCO data from \citet{Henshaw2016_GasKin_250pc} (gray background points). The models include the median fitted ellipse from this work (top left); two spiral arms similar to those proposed by \citet{Sofue1995} (top right); open streams based off of an open x$_2$ orbit from \citet{Kruijssen2015} (bottom left); and a closed x$_2$ orbit which assumes a constant orbital velocity, proposed by \citet{Molinari2011} (bottom right). Markers are the same as presented in Figure~\ref{fig:6panel_lblv_fittings}. Blue and red colors indicate likely near or far positions, respectively. The black star denotes the position of \sgra.}
\label{fig:model_comps}
\end{figure*}

While it is known that the size of the CMZ is controlled by the Galaxy's gravitational potential and the Galactic bar \citep[e.g.][]{Sormani(2015)_gasflow_barred_potl}, the exact size and shape is still debated. Most outputs from the parameter search result in major axes between $60 < \mathrm{R}<296$~pc, with our reported inner and outer extents between $\sim 72 < \mathrm{R}<146$~pc. This implies a thickness of about two times the inner radial extent, and appears visually comparable to extragalactic CMZs with thick nuclear ring such as NGC 1300 or NGC 1433. 

Observational studies of the CMZ report various estimates for the outer extents. Many studies tend to focus on the 100~pc stream. However, there is evidence to support that both the contentious 1.3\deg complex as well as seemingly bright, contiguous material out to 1.5\deg should be considered part of the CMZ \citep[see][]{Battersby_2025_3DCMZI, gramze2023}. Theoretical estimates of CMZ sizes are also quite variable, with the largest possible x$_2$ orbits lying anywhere from $R=200$~pc to $R=1$~kpc \citep{Athanassoula(1992), Sormani(2015)_gasflow_barred_potl}. Additionally, while the idealized x$_2$ orbits are exactly perpendicular to the Galactic bar, the observed orientation of the bar's major axis and the Sun-GC line is approximately $20 < \theta < 30$\deg. Our median orbit falls in the observed range within a 1$\sigma$ uncertainty. Many effects can change the orientation of x$_2$ orbits, such as gas self-gravity which produces angles similar to those seen in simulations from \citet{Tress2020}. The x$_2$ ellipse fitting returns reasonable sizes and a variety of orbits for gas on the main star forming ring, rather than a single best fit. We do not explicitly test for a consistent family of orbits in this study. However, the various extents from the parameter search could also be interpreted as corresponding to families of x$_2$ orbits. 

We also compare the NF positions of our cataloged clouds in PPV space against four different 3D orbital geometries in Figure~\ref{fig:model_comps}. The models include our median fitted ellipse (top left); spiral arms or spirals similar to those proposed by \citet{Sofue1995,Sofue2022,Sofue2025} \citep[see also][]{Sawada2004,Ridley2017} (Sofue; top right); open streams based off of an open x$_2$ orbit from \citet{Kruijssen2015} (KDL; bottom left); and a closed elliptical x$_2$ orbit which assumes a constant orbital velocity, proposed by \citet{Molinari2011} (Molinari; bottom right). Note the projection used for the Sofue model is an approximation based on the KDL streams \citep[see][]{Henshaw2016_GasKin_250pc}. 
 
The morphology of the CMZ is complex and unlikely to lie perfectly on a closed ellipse \citep{Kruijssen2015,Tress2020}. Some of these orbits may be populated by gas, while others may be cleared out due to variations in the environment.
However, a family of elliptical x$_2$ orbits, similar to those presented in \citet{Binney1991}, can fit the average observed gas distribution quite well, as seen by the KDL and our closed ellipse models. These geometries have similar distance and kinematic interpretations for the dust ridge and wiggles region, but all struggle to consistently explain the motion of material further inside the ring that could be infalling towards the CND and \sgra. Alternatively, the Sofue spiral arms model allows for separate streams to address conflicting kinematics. MHD simulations of the CMZ seem to show clear features for both average x$_2$ elliptical orbits and inspiralling features towards the center \citep[e.g.][]{Tress2024}, meaning that modeling the inner region as a spiral extending off of the x$_2$ orbit could result in a reasonable geometry. 

Both our model fittings and the 3D geometries in the literature seem to suggest that the 3D structure of the CMZ is kinematically and spatially complex, and cannot be simplified to a single contiguous orbital stream.

\subsubsection{Asymmetries along the line-of-sight}\label{subsec:asymmetries}

\begin{figure*}[th!]
    \centering
\includegraphics[width=1\textwidth]{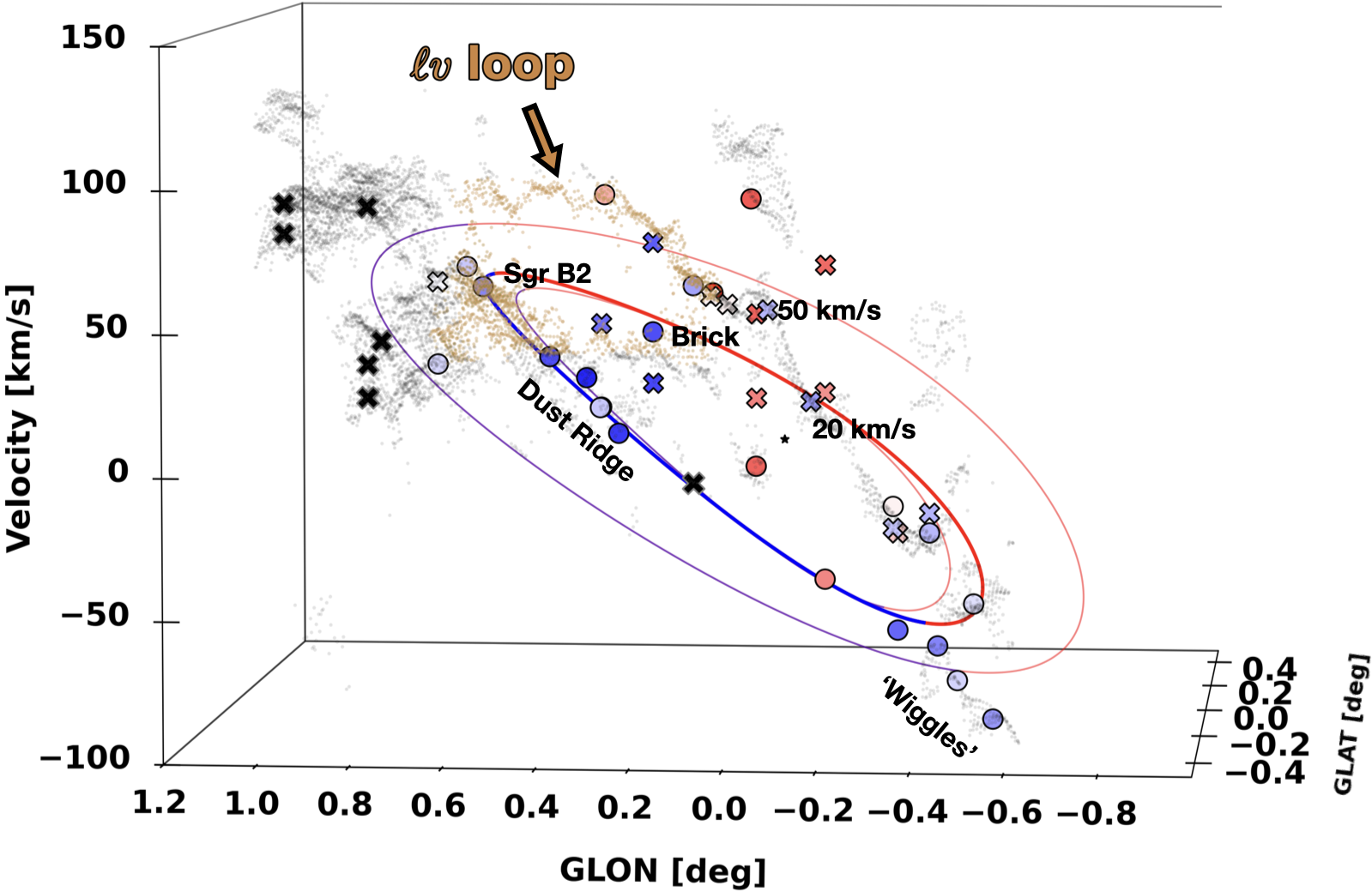}

\caption{The output of the orbital modeling is determined by fitting the clouds in PPV space, as well as their NF posteriors from the PPDFs. This figure shows the best fitting orbits and clouds on top of the spectral decomposition of MOPRA HNCO data from \citet{Henshaw2016_GasKin_250pc} (gray background points). The median ellipse is shown as a thicker line compared to the inner and outer extents. We highlight a closed loop of gas, colored with gold points, referred to here as the `$\ell v$ loop', which may be indicative of shearing or disruption events causing gas to deviate from the orbital streams. Markers are the same as presented in Figure~\ref{fig:6panel_lblv_fittings}. Blue and red colors indicate likely near or far positions, respectively. The black star denotes the position of \sgra.}
\label{fig:PPV_loop_3ellipse}
\end{figure*}

Previous studies of the CMZ have noted that gas is preferentially distributed at positive longitudes  \citep{Bally2010,Longmore2013,Battersby_2025_3DCMZI, Walker_2025}. Our results highlight another asymmetry: most of the clouds in the catalog seem to be placed on the near side of the GC, which is consistent with other elliptical models \citep{Kruijssen2015,Walker_2025}. The near-side features include the dust ridge, coinciding with the dense areas of the CMZ with more active star formation compared to the negative longitudes. However, we also see that most of the clouds in the western region, which are not as prolifically star forming, are also on the near side, though relatively further back than the dust ridge. The far-side clouds are those in the Sgr~A region, an area with high systematic uncertainties due to proximity to \sgra. 

Although our analysis places the majority of CMZ clouds on the near side, we argue this is not the result of observational bias. Both Herschel column density maps \citep{Battersby_2025_3DCMZI} and molecular gas \citep{Jones2012,Ginsburg2015,Ginsburg2016} show more diffuse emission which is coincident with the far side of our model. 
Despite the systematic uncertainties, it is reasonable to assume that our detection of structures is not biased or incomplete at these wavelengths. Additionally, the individual NF metrics independently report NF asymmetries, as highlighted in \paperIII and \paperIV. Both analyses find most clouds on the main CMZ orbit lie on the near side, with only a handful of clouds having confident far-side distinctions. The star count ratio finds a similar distribution as the 8\micron methods. Extragalactic CMZs also show asymmetric gas distributions, such as the central rings of NGC 3351 \citep{Sun2024}, NGC 1300 and NGC 1672 \citep{Gleis2026} which show closed rings with asymmetric CO emission; or M83 for which about two-thirds of gas lies on one side of the orbit \citep{Callanan2021}. While systematic uncertainties could present biases for individual NF methods, the overall LOS asymmetries are likely real and not necessarily unexpected.

Strong LOS asymmetries are supported by theory, as gas on x$_2$ orbits is unlikely to be uniformly distributed. From a computational standpoint, MHD simulations show that the CMZ may go through epochs where gas is skewed to one side \citep{Tress2020,Tress2024}, meaning large anisotropies may point towards collisions or merging of material from the bar lanes disrupting large amounts of gas off of the average orbit. 

The non-uniform distribution of gas in the CMZ is also seen in velocity space. In particular, the spectral decomposition of MOPRA HNCO data from \citet{Henshaw2016_GasKin_250pc} shows distinct streams which match our best-fit model extents reasonably well, as shown in the PPV diagram of Figure~\ref{fig:PPV_loop_3ellipse}. We also note a closed loop connecting gas between longitudes of $0.1^{\circ} < \ell < 1^{\circ}$ which stretches between the ellipses. We refer to this structure as the `$\ell v$ loop', highlighted as golden points in Figure~\ref{fig:PPV_loop_3ellipse}. The physical interpretation and origin of this loop are unclear. Similar features have been seen in simulations such as \citet{Tress2020,Tress2024} where gas is sheared from collisions with the dust lanes, causing large disruptions in the orbital motions of bulk gas. The $\ell v$ loop could also be indicative of gas falling onto inner x$_2$ orbits or stretching between local minima. These features make fitting a precise ellipse to the data difficult, as the gas becomes spread out between multiple orbits.

\section{Uncertainties}\label{sec:limitations}

\subsection{Uncertainties and Limitations of the PPDF method}\label{sec:limitations_PPDF}

The Bayesian framework presented in Section~\ref{subsec:bayseian_posteriors} flexibly combines datasets with position or distance-based constraints for NF estimates in the CMZ. While the framework's main advantages are its flexible parameters and weighting scheme, the creation of prior distributions is largely based on user interpretation, which requires justifications and consideration of systematic bias. 

The quality of PPDF results are dependent on the precision of the input data. More constrained data will have a stronger impact on the posterior fits, as evidence by the confident absorption data which dominate the posterior fittings. While the higher weighting of this method also affects the resulting combinations, the spread and amplitude of the priors have much larger impacts. The most uncertain results are typically in areas where data is omitted due to systematic uncertainties (e.g. the Sgr A region) or is unavailable (e.g. no absorption estimates for IDs 27 -- 31). Some methods may systematically place clouds on either the near or far side, and the lack of confident metrics in these areas could lead to systematic NF preferences. The inclusion of more data across the entire catalog, especially LOS distance estimates from NIR dust extinction methods, proper motions, or parallax measurements, would greatly constrain the PPDF posteriors and mitigate the systematics from less certain methods. 

The second limitation of the PPDF method is the normalization procedure. All data are normalized to the (-1,1) scaling before the posterior combination, which estimates positional constraints well for the majority of clouds. However, at present, there does not exist a clear method to compare LOS distances with NF positions of cloud for which distance estimates are unavailable. Physical distance priors must be projected onto the (-1,1) scaling, which requires a prior assumed CMZ radius. In this case, we use the typical assumption from the literature of a 150~pc CMZ radius. Using different initial radii will impact the LOS metrics, as larger (smaller) initial guesses will place clouds further from (closer to) the center. The effects of an initial guess must therefore be well-justified if using LOS distances as priors. Future iterations of the method may be improved by incorporating future distance constraints for a large sample of CMZ clouds, making it possible develop LOS likelihood functions for the PPDFs.



\subsection{Uncertainties of the ellipse fitting method}\label{sec:limitations_orbitalfit}

The x$_2$ ellipse fitting generally produces reasonable fits to the observed data, and does not converge on a single orbit. Overall, the fitting procedure is quite sensitive to both the initial parameters and the PPDF priors. For example, incorporating 50-100\% systematic errors for all methods results in the posterior PPDFs being pulled closer 0 with large uncertainties. Those broad PPDFs can greatly impact the parameter grid outputs, resulting in incredibly compact CMZs ($20 < \mathrm{R}<80$~pc). While the PPDFs are good indicators for relative positions of clouds, we present the fitting procedure as an exploratory method to compile multiwavelength measures of CMZ cloud positions. The reported CMZ model extents are best used as starting points for future improvements.

The spread in the reported CMZ extents could be affected by the minimization method. In this study, we use the Nelder-Mead minimization method which performs with higher accuracy and broader sampling of parameter space compared to the typical least-squares method. However, the Nelder-Mead simplex can be quite sensitive to initial guesses, and can lead to any parameter falling into local minima regardless of the input.

The inclusion new of structures could also impact the fitting, and potentially result in different ellipse sizes. For example, Cloud IDs 27 -- 31 were not included in the fitting procedure due to the lack of a variety of NF metrics (the clouds only have results from dust extinction and star count metrics, but no available absorption data) and their separation from the majority of clouds near \sgra. Including these points could impact the size and orientation of the ellipse. Inclusion of foreground structures could also impact the extents and orientation of the ellipse, particularly if their PPDFs are tightly constrained. For example, it is possible IDs 4a and 4b could be foreground structures based on their latitude positions. However, their PPDFs are quite consistent with the surrounding clouds in that region, so their individual impact on the fitting is not high. A more impactful limitation of the current method is the simplification of each cloud to a single point for the orbital fitting, which neglects extended features. Future work is planned to modify the fitting procedure for a pixel-by-pixel approach, and incorporate extended filamentary features \citep[e.g.][Battersby et al. subm]{} to help constrain the extents of the gas streams.


We also note large systematic uncertainties associated with the top-down projection presented in Figure~\ref{fig:x2_topdown_fits}. While we confidently constrain the relative NF positions of clouds, there are many ways one could project the PPDFs onto the orbital model to obtain LOS distances. The distance estimates shown in the figure and reported in Table~\ref{tab:los_synth_table} are based on projection of the PPDF positions on to the outer-most fitted ellipse extent. Thus, the distance estimates are dependent on the inner and outer extents, which set the possible LOS extents for the dataset. While the relative NF positions are robust for relative comparison of cloud locations, the distance estimates and LOS uncertainties are projection-dependent.

\section{Conclusions and Summary}\label{sec_conclusions}

We introduce a flexible Bayesian framework to combine different datasets of various near versus far position and line of sight distance estimates for CMZ clouds onto a single quantitative scale. We use estimates from dust extinction, molecular line absorption, star count ratios, X-ray echoes, and stellar kinematics to find the posterior Position Probability Density Functions (PPDFs) for each structure in the CMZ cloud catalog presented in this paper series. We then create an optimized fitting procedure with the \texttt{lmfit} python library to find a best fitting x$_2$ orbit for CMZ clouds by minimizing the residuals between the orbit and posterior PPDFs.

The PPDF and x$_2$ fitting methods are intended to be continually updated and modified with new data. The relevant code for this paper series are publicly available via: \url{https://centralmolecularzone.github.io/3D_CMZ/}. The maps and data products can be found on Harvard Dataverse: \url{https://dataverse.harvard.edu/dataverse/3D_CMZ}

Our main findings from the combined PPDF and orbital fitting methods include the following:

\begin{itemize}
    \item By exploring a range of input major axes between 10--300~pc, we find that the fitting procedure converges on a population of ellipses spanning major axis radii of $60 < a < 296$~pc. We find an inner and outer average semi-major extents of the CMZ to be 72~pc and 146~pc, and a median extent of 83~pc. These results support the interpretation that gas orbits in the CMZ are more complex than a single 100~pc ellipse. 
    
    \item The median best fitting orbit and estimated LOS distances are consistent with other elliptical orbital models \citep[e.g.][]{Kruijssen2015,Molinari2011}. However, elliptical models struggle to kinematically explain material which may lie closer to \sgra along the LOS. Many CMZ clouds may exist on a family of x$_2$ orbital ellipses between the inner and outer extents identified in the fitting procedure. An x$_2$ orbital model may require additional complexity inside of the average orbit, such as an inspiralling component, to explain infall.  

    \item We present a top-down view of our CMZ model and estimate LOS positions to clouds by projecting the PPDF peak positions onto the outer ellipse extents. 
    
    \item Our projected LOS distances agree with many current distance constraints from the literature. In particular, our results support evidence that the three pigs may lie much closer to \sgra, within 10~pc. Additionally, our results place the 50 and 20\kms~clouds at $31\pm14$~pc and $42\pm19$~pc in front of \sgra, which disagrees with models placing the clouds near the CND or behind \sgra. These structures may be actively falling inwards from an inner x$_2$ orbit, making them kinematically separate from the average gas flows. 

    \item We highlight prominent asymmetries along the line of sight, with most clouds located on the near side of the CMZ. An overall near-side preference for the gas is found independently in all NF methods, and mirrors similar distributions seen in extragalactic CMZs \citep[e.g.][]{Callanan2021,Gleis2026,Sun2024} and MHD simulations \citep[e.g.][]{Tress2020} when large parcels of gas from the bar lanes collide with orbiting material.

    \item We note the existence of an $\ell v$ loop feature seen in molecular data from MOPRA, which may be the result of shearing events disrupting gas on x$_2$ orbits, and possibly inducing infall between orbits. 
    
\end{itemize}

Our analysis highlights the complexity of the CMZ gas distributions, which may not be simply fit by a single elliptical orbit. We note the need for further study of the family of possible orbits, and suggest that combining both elliptical x$_2$ and spiral features may provide a reasonable geometry to explain both average CMZ mass flows and infalling material. More detailed analysis of individual clouds, including LOS distances will help strengthen the modeling, and could help determine which structures belong to various orbits.

\section*{Acknowledgments}

D.\ Lipman and C.\ Battersby gratefully acknowledge funding from the National Science Foundation under Award Nos. 1816715, 2108938, 2206510, and CAREER 2145689. D.\ Lipman also acknowledges funding from the NASA Connecticut Space Grant Consortium under PTE Federal Award No: 80NSSC20M0129.
C.\ Battersby also  acknowledges  funding  from  National Aeronautics and Space Administration through the Astrophysics Data Analysis Program under Award ``3-D MC: Mapping Circumnuclear Molecular Clouds from X-ray to Radio,” Grant No. 80NSSC22K1125.

D.\ Walker  acknowledges support from the UK ALMA Regional Centre (ARC) Node, which is supported by the Science and Technology Facilities Council [grant numbers ST/Y004108/1 and ST/T001488/1].

F.\ Nogueras-Lara gratefully acknowledges financial support from grant PID2024-162148NA-I00, funded by MCIN/AEI/10.13039/501100011033 and the European Regional Development Fund (ERDF) “A way of making Europe”, from the Ramón y Cajal programme (RYC2023-044924-I) funded by MCIN/AEI/10.13039/501100011033 and FSE+, and from the Severo Ochoa grant CEX2021-001131-S, funded by MCIN/AEI/10.13039/501100011033.

R.\ Klessen acknowledges financial support from the ERC via Synergy Grant ``ECOGAL'' (project ID 855130),  from the German Excellence Strategy via the Heidelberg Cluster ``STRUCTURES'' (EXC 2181 - 390900948), from the German Science Foundation under grant KL 1358/22-1, and from the German Ministry for Economic Affairs and Climate Action in project ``MAINN'' (funding ID 50OO2206). 

A.\ Ginsburg acknowledges support from the NSF via AAG 2206511 and CAREER 2142300.

M.\ Clavel acknowledges financial support from the Centre National d’Etudes Spatiales (CNES).

M.~C.\ Sormani acknowledges financial support from the European Research Council under the ERC Starting Grant ``GalFlow'' (grant 101116226) and from Fondazione Cariplo under the grant ERC attrattivit\`{a} n. 2023-3014.

This work made use of Astropy:\footnote{http://www.astropy.org} a community-developed core Python package and an ecosystem of tools and resources for astronomy \citep{astropy:2013, astropy:2018, astropy:2022}.


\appendix 
\counterwithin{figure}{section}
\vspace{-0.5cm}
\section{Justification for modeling PPDF priors as Gaussian}\label{sec:prior_gauss_justification}

The Bayesian synthesis of observational data presented in Section~\ref{subsec:bayseian_posteriors} is a flexible framework that allows the user to assume a distribution shape of their choice for any given prior. For this study, we chose to model most of the prior distributions as Gaussian, given the methods report a single median value for a cloud, rather than a continuous distribution.

\begin{figure*}[t!]
    \includegraphics[width=\textwidth]{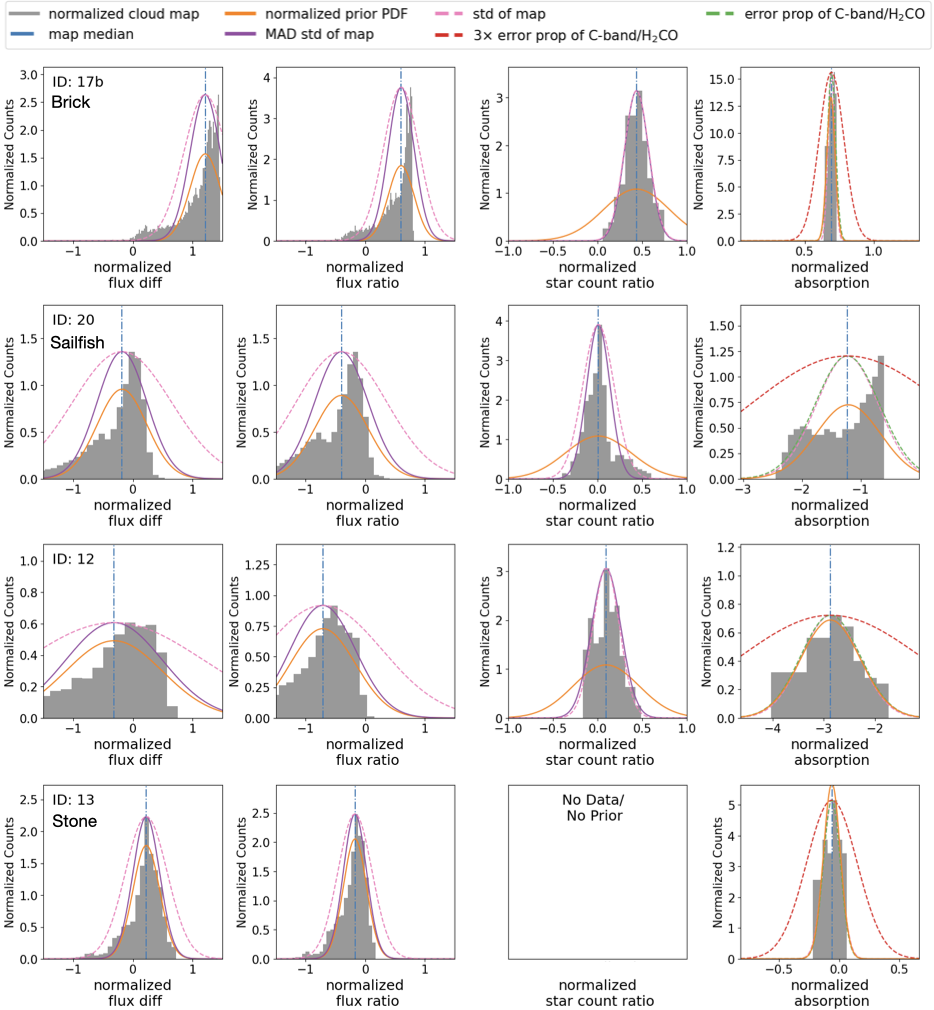}
    \caption{Most normalized NF distributions can be approximated by a single Gaussian. We highlight four examples of clouds showing the distribution of normalized counts for the flux difference (first column), flux ratio (second column), star count ratio (third column), and absorption methods (fourth column). Each plot shows a histogram of the cloud maps with NF values normalized using Eqn~\ref{eqn:z-scale} (gray), the median of the normalized map (blue dash-dotted line), and the normalized PPDF prior ultimately used for each method (orange line).}
    \label{fig:appendix_NF_prior_gaussian}
\end{figure*}

We explored different uncertainty estimations for the flux difference, flux ratio, star count ratio, and absorption methods, summarized in Figure~\ref{fig:appendix_NF_prior_gaussian}. For each method, we normalize a given cloud map using Eqn~\ref{eqn:z-scale}, and show the normalized counts (gray histogram) along with the median of the distribution (blue dash-dotted line) corresponding to the NF position for the prior ($\mu$). The normalized prior used for the Bayesian synthesis is shown by the solid orange line.
Most distributions are approximately Gaussian or log-normal. Each method has undetermined systematic uncertainties, so we focus less on perfectly representing the shape of the normalized maps, and more on creating a prior which represents the spread of the data overall. For the 8\micron flux methods and star count ratio, we show unnormalized Gaussian distributions created using the median of the normalized map, and uncertainties determined by the standard deviation (pink dotted) or $\sigma_{\mathrm{MAD}}$ (purple). For these methods, the $\sigma_{\mathrm{MAD}}$ provides a better overall fit to the spread in the data, and is thus used for building the priors.

\paperIII reported the absorption fraction value over the entire cloud masks. Here, we create absorption fraction maps to obtain statistical uncertainty estimates. The pixel-wise absorption fraction maps are created via: (GBT C-Band Continuum map + a cosmic microwave background estimate) / abs(H$_{2}$CO Min Intensity over the masked cube). The uncertainty estimate for the absorption method was originally set at $3\sigma$ based on the standard deviation of the un-normalized dataset. However, upon review, we investigated the error propagation of the normalized maps, as presented in Section~\ref{subsec:bayseian_posteriors}. We show the un-normalized distributions with uncertainties from the error propagation (green dotted line) and $3\times$ the error (red dotted). The 1$\sigma$ estimation almost exactly matches the standard deviation of the maps. We choose to model the absorption fraction Gaussian priors with the 1$\sigma$ uncertainty from the error propagation, for the most robust estimation of the statistical errors.

The 70\micron correlation coefficient is a single value summarizing the overall statistical correlation of the calculated cloud extinction and observed emission maps and is best reflected as a Gaussian with large uncertainties. As discussed in \paperIV, the method could be improved with more confident 70\micron emission models to generate a unique, continuous NF position probability distribution. However, building such a distribution requires carefully constructed radiative transfer models for 70\micron emission, which is outside the scope of this series of papers.

\section{PPDF Posterior Plots}\label{sec:PPDF_appendix}

The PPDFs for all CMZ clouds are presented in Figure~\ref{fig:PPDFs_appendix1} below. The methods used to infer NF positions of individual clouds in the CMZ can be combined in a Bayesian framework to create a posterior distribution of NF likely positions for a given cloud on a normalized scale of -1 (far) to +1 (near). Here we show example prior and posterior PPDFs for six clouds in the catalog. The flux difference (orange dashed line) is used as a likelihood function, and is multiplied by distributions for the flux ratio (green), correlation coefficient (blue), absorption fraction (red), and star count ratios (yellow). 

IDs 13 and 14 have prior data from X-rays (pink), and IDs 9 and 10 have prior data from stellar densities (brown). ID 16 is an empty submask, and thus has no prior or posterior information plotted.

After taking the product of the priors, we report the location of the peak ($\mu$, vertical gray dash-dotted line), the relative probability peak (A), and the 68\% confidence interval ($\mathrm{CI}_{68}$, horizontal solid gray line), which are used to obtain a Positional PDF posterior distribution (PPDF; black solid line). The peak of the PPDF indicates the most likely NF position of the cloud. All distributions are normalized to unit integral probability. 

\begin{figure*}[!t]
    \includegraphics[width=\textwidth]{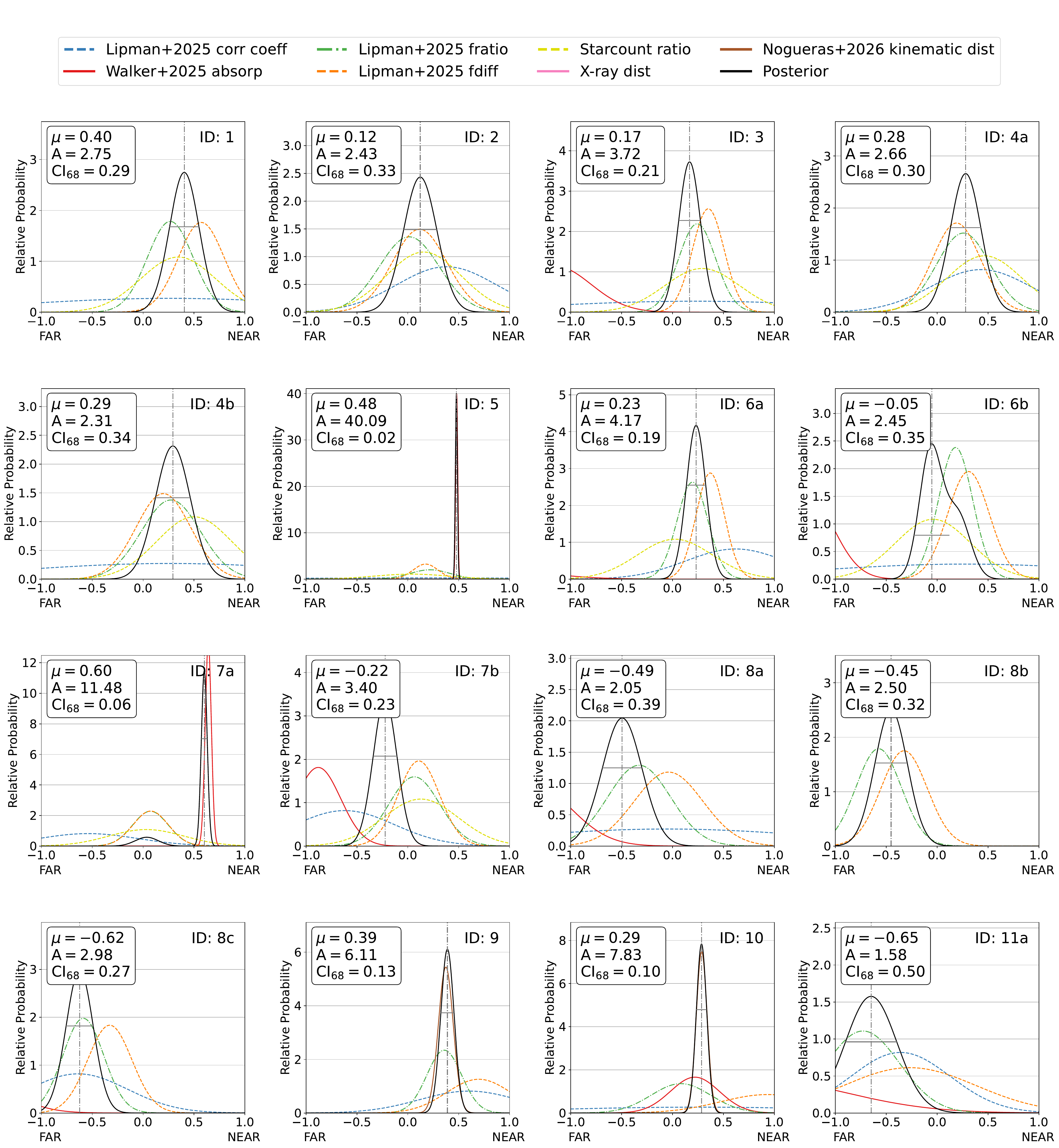}
    \caption{Prior and posterior PPDFs for all cloud masks in the catalog. Each panel shows priors for different methods as colored lines, and the posterior is indicated by a black solid line. We report the location of the peak ($\mu$, vertical gray dash-dotted line), the relative probability peak (A), and the 68\% confidence interval ($\mathrm{CI}_{68}$, horizontal solid gray line), which are used to obtain a posterior PPDF distribution used for the orbital fitting procedure in Section~\ref{subsec:fitting_procedures}. }
    \label{fig:PPDFs_appendix1}
\end{figure*}
\makeatletter 
\renewcommand{\thefigure}{B.1}
\makeatother
\begin{figure*}[!t]
    \includegraphics[width=\textwidth]{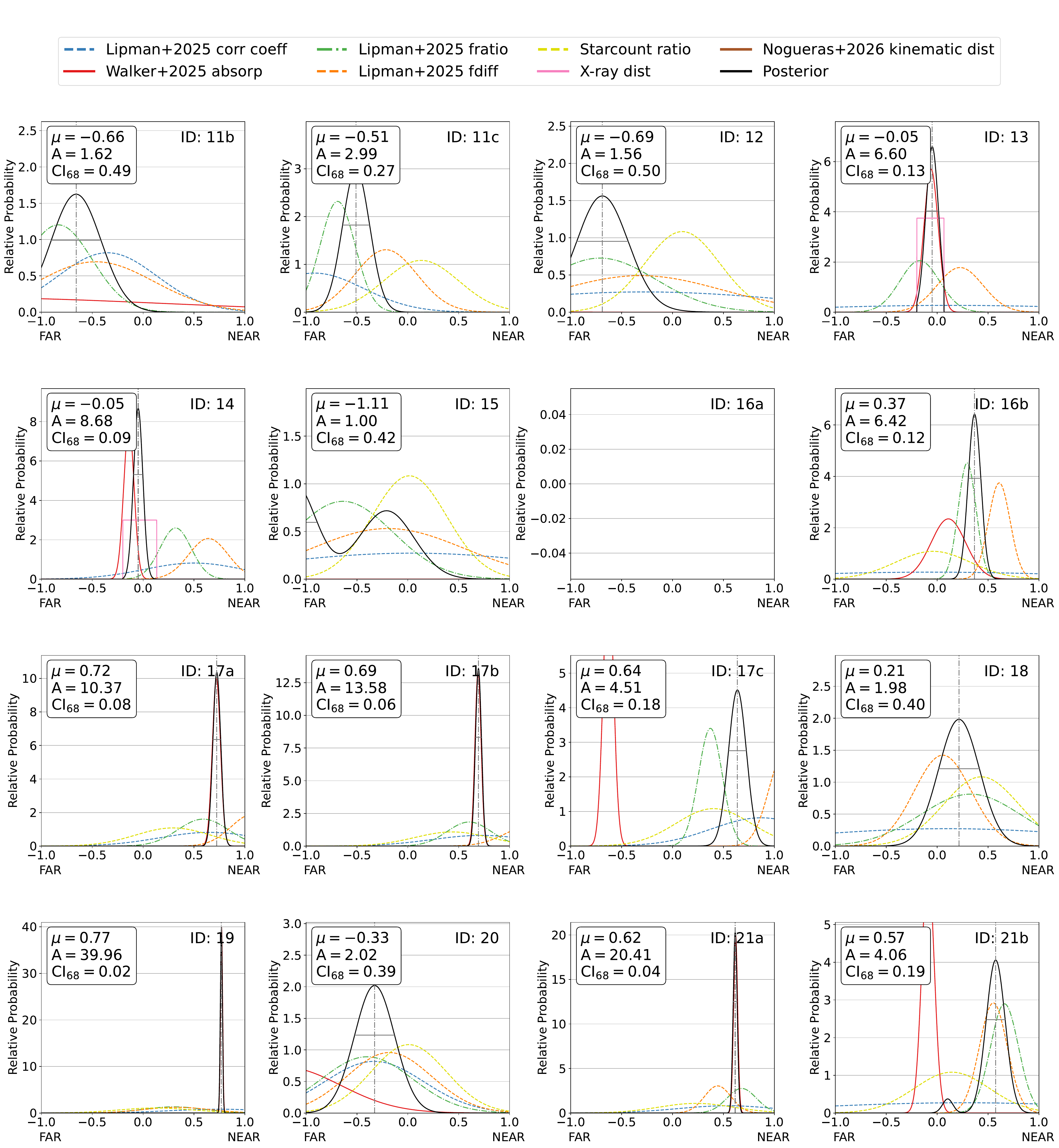}
    \caption{(cont.)}
    \label{fig:PPDFs_appendix2}
\end{figure*}
\makeatletter 
\renewcommand{\thefigure}{B.1}
\makeatother
\begin{figure*}[!t]
    \includegraphics[width=\textwidth]{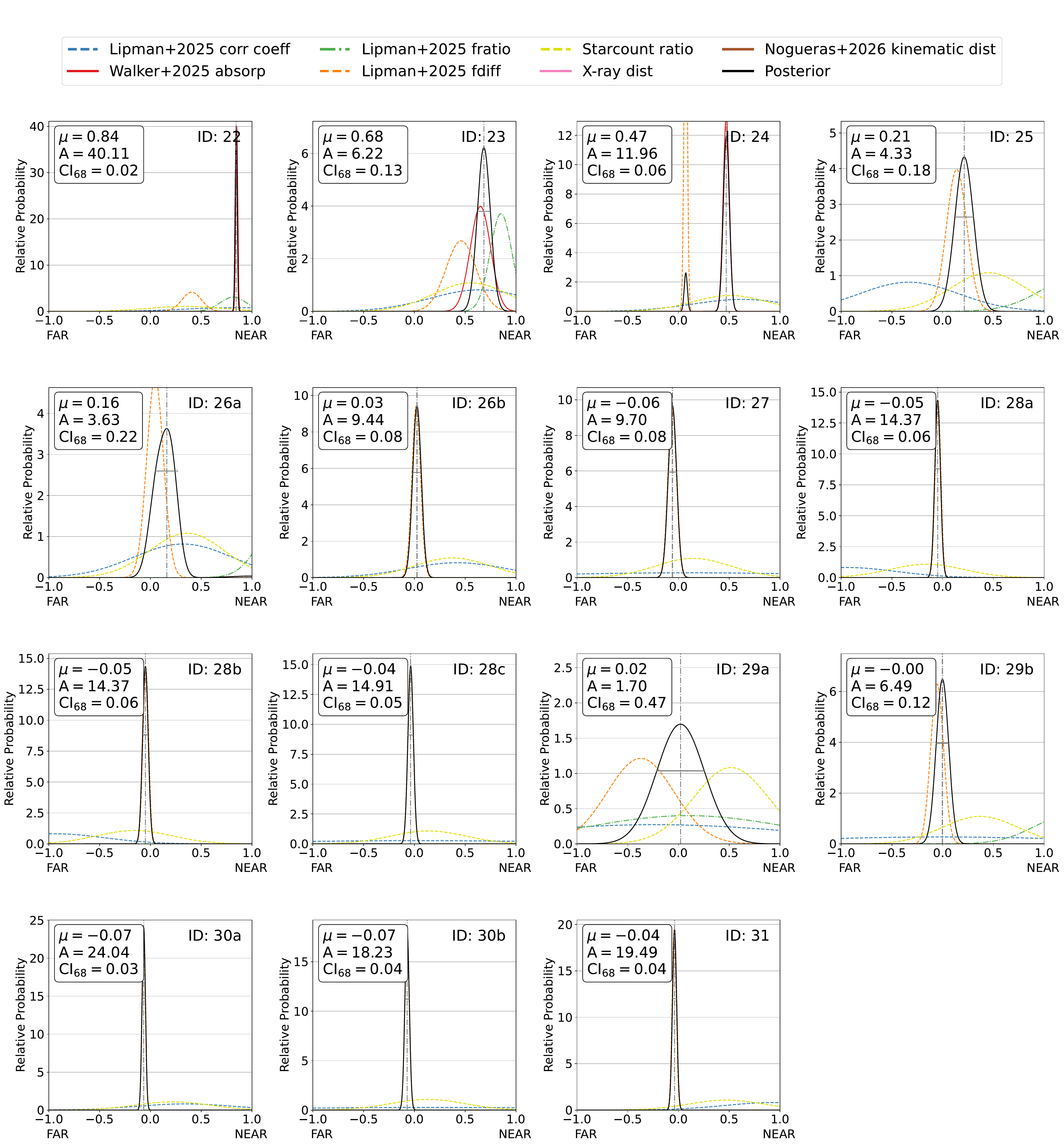}
    \caption{(cont.)}
    \label{fig:PPDFs_appendix3}
\end{figure*}

\section{Data Table for Building PPDF Priors}

All information needed to build the prior PPDFs for near/far distinction methods is reported in Table~\ref{tab:PPDF_info_table}. We provide this information for any method which distinguishes NF positions for clouds: the flux difference, flux ratio, correlation coefficient, absorption fraction, and star count method. These priors are all created by normalizing the values in the table using Equations~\ref{eqn:z-scale}, and constructing normalized Gaussian profiles, as discussed in Section~\ref{subsec:bayseian_posteriors}. The priors from distance X-ray and stellar kinematic distance estimates are built in a different way, involving projection onto an assumed initial guess for an orbital model to normalized them to the same scaling as the NF distinction datasets. This is only done for five clouds, and refer the reader to Section~\ref{subsec:bayseian_posteriors} for details of how to construct these priors.

The first five columns of Table~\ref{tab:PPDF_info_table} are taken from Table 1 of \citet{Walker_2025}, including the leaf IDs of the cataloged clouds, the associated cloud names, central coordinates in degrees ($l, b$), and central velocity from HNCO (v). The data metrics available to construct the PPDF priors for NF methods include: the median flux difference value, flux difference standard deviation, median flux ratio, flux ratio standard deviation, correlation coefficient, absorption fraction, absorption fraction sigma value, star count ratio, and star count ratio standard deviation. If a given method was not available or reported for a given cloud, we report the value as a ``nan'' in the table column. Otherwise, all numerical values for NF metrics are the reported values used to construct the NF method's prior for that cloud.

\clearpage 

\begin{longrotatetable}
\begin{deluxetable}{cccccccccccccc}
\label{tab:PPDF_info_table}

\tablecaption{Information needed to build the prior PPDFs from the near/far distinction methods.}

\tablewidth{0pt}

\tablehead{
\colhead{leaf id} & \colhead{cloud name} & \colhead{l} & \colhead{b} & \colhead{v} & \colhead{flux diff} & \colhead{flux diff}& \colhead{flux ratio} & \colhead{flux ratio} & \colhead{corr coeff} & \colhead{absorption} & \colhead{absorption} & \colhead{star count} & \colhead{star count} \\
\colhead{} & \colhead{} & \colhead{} & \colhead{} & \colhead{} & \colhead{} & \colhead{stdv} & \colhead{} & \colhead{stdv} & \colhead{} & \colhead{fraction} & \colhead{fraction $\sigma$} & \colhead{ratio} & \colhead{ratio stdv}\\
\hline
\colhead{} & \colhead{} & \colhead{deg} & \colhead{deg} & \colhead{km/s} & \colhead{MJy/sr} & \colhead{MJy/sr}& \colhead{} & \colhead{} & \colhead{} & \colhead{} & \colhead{} & \colhead{} & \colhead{} 
}
\startdata
1 & G359.475-0.044 & -0.525 & -0.044 & -102.0 & -20.97 & 33.92 & 0.37 & 0.112 & 0.26 & nan & nan & 0.656 & 0.196 \\
2 & G359.508-0.135 & -0.492 & -0.135 & -56.0 & -73.25 & 40.054 & 0.46 & 0.147 & 0.38 & nan & nan & 0.844 & 0.139 \\
3 & G359.561-0.001 & -0.439 & -0.001 & -90.0 & -59.31 & 23.36 & 0.38 & 0.091 & 0.22 & 2.15 & 0.35 & 0.708 & 0.157 \\
4a & G359.595-0.223 & -0.405 & -0.223 & -27.0 & -79.52 & 34.939 & 0.37 & 0.131 & 0.43 & nan & nan & 0.536 & 0.249 \\
4b & G359.595-0.223 & -0.405 & -0.223 & -20.0 & -78.19 & 40.247 & 0.36 & 0.145 & 0.26 & nan & nan & 0.498 & 0.277 \\
5 & G359.608+0.018 & -0.392 & 0.018 & -78.0 & -79.45 & 18.421 & 0.38 & 0.099 & 0.18 & 0.52 & 0.01 & 0.962 & 0.119 \\
6a & G359.688-0.132 & -0.312 & -0.132 & -29.0 & -52.41 & 20.782 & 0.4 & 0.076 & 0.62 & 3.3 & 0.66 & 0.977 & 0.154 \\
6b & G359.688-0.132 & -0.312 & -0.132 & -21.0 & -61.84 & 30.72 & 0.41 & 0.084 & 0.28 & 2.28 & 0.26 & 1.041 & 0.097 \\
7a & G359.701+0.032 & -0.299 & 0.032 & -73.0 & -96.86 & 26.109 & 0.46 & 0.087 & -0.54 & 0.36 & 0.03 & 0.969 & 0.095 \\
7b & G359.701+0.032 & -0.299 & 0.032 & -37.0 & -91.8 & 30.474 & 0.47 & 0.125 & -0.62 & 1.88 & 0.22 & 0.872 & 0.112 \\
8a & G359.865+0.023 & -0.135 & 0.023 & -54.0 & -113.79 & 50.639 & 0.66 & 0.154 & -0.05 & 2.4 & 0.39 & nan & nan \\
8b & G359.865+0.023 & -0.135 & 0.023 & 15.0 & -156.73 & 34.166 & 0.79 & 0.112 & nan & 34.87 & 601.01 & nan & nan \\
8c & G359.865+0.023 & -0.135 & 0.023 & 62.0 & -157.19 & 32.632 & 0.79 & 0.101 & -0.64 & 2.81 & 0.41 & nan & nan \\
9 & G359.88-0.081 & -0.12 & -0.081 & 15.0 & -5.06 & 47.509 & 0.31 & 0.085 & 0.6 & 3.45 & 3.55 & nan & nan \\
10 & G359.979-0.071 & -0.021 & -0.071 & 48.0 & 38.37 & 69.886 & 0.46 & 0.147 & 0.26 & 0.78 & 0.24 & nan & nan \\
11a & G0.014-0.016 & 0.014 & -0.016 & -11.0 & -148.38 & 97.611 & 0.86 & 0.18 & -0.36 & 2.78 & 0.87 & nan & nan \\
11b & G0.014-0.016 & 0.014 & -0.016 & 45.0 & -177.87 & 86.135 & 0.92 & 0.166 & -0.35 & 2.8 & 1.98 & nan & nan \\
11c & G0.014-0.016 & 0.014 & -0.016 & 14.0 & -140.24 & 45.786 & 0.84 & 0.086 & -0.92 & 2.62 & 0.05 & 0.869 & 0.068 \\
12 & G0.035+0.032 & 0.035 & 0.032 & 86.0 & -156.53 & 121.741 & 0.86 & 0.274 & -0.28 & 3.87 & 0.58 & 0.904 & 0.164 \\
13 & G0.068-0.076 & 0.068 & -0.076 & 50.0 & -65.96 & 33.614 & 0.57 & 0.097 & 0.14 & 1.06 & 0.07 & nan & nan \\
14 & G0.105-0.08 & 0.105 & -0.08 & 53.0 & -14.0 & 28.949 & 0.34 & 0.077 & 0.5 & 1.14 & 0.05 & nan & nan \\
15 & G0.116+0.003 & 0.116 & 0.003 & 52.0 & -144.13 & 112.682 & 0.83 & 0.244 & 0.04 & 3.38 & 0.32 & 0.984 & 0.163 \\
16a & G0.143-0.083 & 0.143 & -0.083 & -15.0 & nan & nan & nan & nan & nan & nan & nan & nan & nan \\
16b & G0.143-0.083 & 0.143 & -0.083 & 57.0 & -16.52 & 15.977 & 0.35 & 0.044 & -0.06 & 0.89 & 0.17 & 1.041 & 0.103 \\
17a & G0.255+0.02 & 0.255 & 0.02 & 18.0 & 53.12 & 31.617 & 0.2 & 0.124 & 0.65 & 0.28 & 0.04 & 0.707 & 0.177 \\
17b & G0.255+0.02 & 0.255 & 0.02 & 37.0 & 74.09 & 38.177 & 0.2 & 0.108 & 0.72 & 0.31 & 0.03 & 0.564 & 0.134 \\
17c & G0.255+0.02 & 0.255 & 0.02 & 70.0 & 56.12 & 22.608 & 0.31 & 0.059 & 0.87 & 1.63 & 0.05 & 0.597 & 0.036 \\
18 & G0.327-0.195 & 0.327 & -0.195 & 16.0 & -98.34 & 42.062 & 0.34 & 0.245 & 0.1 & nan & nan & 0.567 & 0.203 \\
19 & G0.342+0.06 & 0.342 & 0.06 & -2.0 & -57.98 & 48.867 & 0.34 & 0.15 & 0.75 & 0.23 & 0.01 & 0.878 & 0.111 \\
20 & G0.342-0.085 & 0.342 & -0.085 & 90.0 & -126.21 & 62.454 & 0.68 & 0.224 & -0.33 & 2.21 & 0.55 & 0.991 & 0.13 \\
21a & G0.379+0.05 & 0.379 & 0.05 & 8.0 & -41.39 & 19.633 & 0.16 & 0.072 & 0.54 & 0.38 & 0.02 & 0.769 & 0.04 \\
21b & G0.379+0.05 & 0.379 & 0.05 & 39.0 & -25.27 & 20.49 & 0.17 & 0.069 & 0.26 & 1.09 & 0.06 & 0.856 & 0.077 \\
22 & G0.413+0.048 & 0.413 & 0.048 & 19.0 & -49.32 & 14.361 & 0.09 & 0.064 & 0.84 & 0.15 & 0.01 & 0.656 & 0.127 \\
23 & G0.488+0.008 & 0.488 & 0.008 & 28.0 & -39.11 & 22.333 & 0.08 & 0.054 & 0.64 & 0.35 & 0.1 & 0.446 & 0.086 \\
24 & G0.645+0.03 & 0.645 & 0.03 & 53.0 & -97.56 & 2.348 & -0.25 & 0.019 & 0.62 & 0.53 & 0.03 & 0.491 & 0.1 \\
25 & G0.666-0.028 & 0.666 & -0.028 & 62.0 & -78.28 & 15.046 & -0.18 & 0.193 & -0.33 & 2.12 & 0.84 & 0.549 & 0.117 \\
26a & G0.716-0.09 & 0.716 & -0.09 & 28.0 & -101.03 & 12.295 & -0.17 & 0.117 & 0.32 & nan & nan & 0.639 & 0.084 \\
26b & G0.716-0.09 & 0.716 & -0.09 & 58.0 & -105.37 & 6.36 & -0.24 & 0.067 & 0.42 & nan & nan & 0.627 & 0.074 \\
27 & G0.816-0.185 & 0.816 & -0.185 & 39.0 & -117.03 & 6.211 & -0.39 & 0.089 & 0.1 & nan & nan & 0.856 & 0.096 \\
28a & G0.888-0.044 & 0.888 & -0.044 & 14.0 & -114.84 & 4.182 & -0.23 & 0.043 & -0.94 & nan & nan & 1.156 & 0.072 \\
28b & G0.888-0.044 & 0.888 & -0.044 & 26.0 & -114.84 & 4.182 & -0.23 & 0.043 & -0.94 & nan & nan & 1.156 & 0.072 \\
28c & G0.888-0.044 & 0.888 & -0.044 & 84.0 & -113.75 & 4.022 & -0.28 & 0.065 & 0.04 & nan & nan & 0.863 & 0.181 \\
29a & G1.075-0.049 & 1.075 & -0.049 & 74.0 & -163.86 & 49.3 & 0.45 & 0.497 & -0.23 & nan & nan & 0.482 & 0.123 \\
29b & G1.075-0.049 & 1.075 & -0.049 & 85.0 & -116.07 & 9.506 & -0.13 & 0.175 & -0.01 & nan & nan & 0.629 & 0.28 \\
30a & G1.601+0.012 & 1.601 & 0.012 & 48.0 & -117.92 & 2.494 & -0.76 & 0.047 & 0.34 & nan & nan & 0.781 & 0.23 \\
30b & G1.601+0.012 & 1.601 & 0.012 & 58.0 & -118.97 & 3.29 & -0.74 & 0.059 & 0.15 & nan & nan & 0.872 & 0.261 \\
31 & G1.652-0.052 & 1.652 & -0.052 & 50.0 & -114.32 & 3.079 & -0.8 & 0.053 & 0.94 & nan & nan & 0.531 & 0.15 \\
\enddata
\end{deluxetable}
\end{longrotatetable}

\clearpage

\bibliography{references.bib}{}

@ARTICLE{Bally2010,
       author = {{Bally}, John and {Aguirre}, James and {Battersby}, Cara and {Bradley}, Eric Todd and {Cyganowski}, Claudia and {Dowell}, Darren and {Drosback}, Meredith and {Dunham}, Miranda K. and {Evans}, II, Neal J. and {Ginsburg}, Adam and {Glenn}, Jason and {Harvey}, Paul and {Mills}, Elisabeth and {Merello}, Manuel and {Rosolowsky}, Erik and {Schlingman}, Wayne and {Shirley}, Yancy L. and {Stringfellow}, Guy S. and {Walawender}, Josh and {Williams}, Jonathan},
        title = "{The Bolocam Galactic Plane Survey: {\ensuremath{\lambda}} = 1.1 and 0.35 mm Dust Continuum Emission in the Galactic Center Region}",
      journal = {\apj},
         year = 2010,
        month = sep,
       volume = {721},
       number = {1},
        pages = {137-163},
          doi = {10.1088/0004-637X/721/1/137},
archivePrefix = {arXiv},
       eprint = {1011.0932},
 primaryClass = {astro-ph.GA},
       adsurl = {https://ui.adsabs.harvard.edu/abs/2010ApJ...721..137B}
}

@INPROCEEDINGS{Levy2022,
       author = {{Levy}, Rebecca},
        title = "{The Morpho-Kinematic Architecture of Super Star Clusters in the Center of NGC253}",
    booktitle = {American Astronomical Society Meeting \#240},
         year = 2022,
       series = {American Astronomical Society Meeting Abstracts},
       volume = {240},
        month = jun,
          eid = {130.07},
        pages = {130.07},
       adsurl = {https://ui.adsabs.harvard.edu/abs/2022AAS...24013007L}
}

@INPROCEEDINGS{Nagayama2003,
       author = {{Nagayama}, Takahiro and {Nagashima}, Chie and {Nakajima}, Yasushi and {Nagata}, Tetsuya and {Sato}, Shuji and {Nakaya}, Hidehiko and {Yamamuro}, Tomoyasu and {Sugitani}, Koji and {Tamura}, Motohide},
        title = "{SIRIUS: a near infrared simultaneous three-band camera}",
    booktitle = {Instrument Design and Performance for Optical/Infrared Ground-based Telescopes},
         year = 2003,
       editor = {{Iye}, Masanori and {Moorwood}, Alan F.~M.},
       series = {Society of Photo-Optical Instrumentation Engineers (SPIE) Conference Series},
       volume = {4841},
        month = mar,
        pages = {459-464},
          doi = {10.1117/12.460770},
       adsurl = {https://ui.adsabs.harvard.edu/abs/2003SPIE.4841..459N}
}

@ARTICLE{Launhardt2002,
       author = {{Launhardt}, R. and {Zylka}, R. and {Mezger}, P.~G.},
        title = "{The nuclear bulge of the Galaxy. III. Large-scale physical characteristics of stars and interstellar matter}",
      journal = {\aap},
         year = 2002,
        month = mar,
       volume = {384},
        pages = {112-139},
          doi = {10.1051/0004-6361:20020017},
archivePrefix = {arXiv},
       eprint = {astro-ph/0201294},
 primaryClass = {astro-ph},
       adsurl = {https://ui.adsabs.harvard.edu/abs/2002A&A...384..112L}
}

@ARTICLE{Schodel2014,
       author = {{Sch{\"o}del}, R. and {Feldmeier}, A. and {Kunneriath}, D. and {Stolovy}, S. and {Neumayer}, N. and {Amaro-Seoane}, P. and {Nishiyama}, S.},
        title = "{Surface brightness profile of the Milky Way's nuclear star cluster}",
      journal = {\aap},
         year = 2014,
        month = jun,
       volume = {566},
          eid = {A47},
        pages = {A47},
          doi = {10.1051/0004-6361/201423481},
archivePrefix = {arXiv},
       eprint = {1403.6657},
 primaryClass = {astro-ph.GA},
       adsurl = {https://ui.adsabs.harvard.edu/abs/2014A&A...566A..47S}
}

@ARTICLE{Genzel2010,
       author = {{Genzel}, Reinhard and {Eisenhauer}, Frank and {Gillessen}, Stefan},
        title = "{The Galactic Center massive black hole and nuclear star cluster}",
      journal = {Reviews of Modern Physics},
         year = 2010,
        month = oct,
       volume = {82},
       number = {4},
        pages = {3121-3195},
          doi = {10.1103/RevModPhys.82.3121},
archivePrefix = {arXiv},
       eprint = {1006.0064},
 primaryClass = {astro-ph.GA},
       adsurl = {https://ui.adsabs.harvard.edu/abs/2010RvMP...82.3121G}
}

@ARTICLE{Hatano2013,
       author = {{Hatano}, Hirofumi and {Nishiyama}, Shogo and {Kurita}, Mikio and {Kanai}, Saori and {Nakajima}, Yasushi and {Nagata}, Tetsuya and {Tamura}, Motohide and {Kandori}, Ryo and {Kato}, Daisuke and {Sato}, Yaeko and {Yoshikawa}, Tatsuhito and {Suenaga}, Takuya and {Sato}, Shuji},
        title = "{The Efficiency and Wavelength Dependence of Near-infrared Interstellar Polarization toward the Galactic Center}",
      journal = {\aj},
         year = 2013,
        month = apr,
       volume = {145},
       number = {4},
          eid = {105},
        pages = {105},
          doi = {10.1088/0004-6256/145/4/105},
archivePrefix = {arXiv},
       eprint = {1303.0456},
 primaryClass = {astro-ph.GA},
       adsurl = {https://ui.adsabs.harvard.edu/abs/2013AJ....145..105H}
}

@ARTICLE{Yasui2015,
       author = {{Yasui}, Kazuki and {Nishiyama}, Shogo and {Yoshikawa}, Tatsuhito and {Nagatomo}, Schun and {Uchiyama}, Hideki and {Tsuru}, Takeshi Go and {Koyama}, Katsuji and {Tamura}, Motohide and {Kwon}, Jungmi and {Sugitani}, Koji and {Sch{\"o}del}, Rainer and {Nagata}, Tetsuya},
        title = "{Number density distribution of near-infrared sources on a sub-degree scale in the Galactic center: Comparison with the Fe XXV K{\ensuremath{\alpha}} line at 6.7 keV}",
      journal = {\pasj},
         year = 2015,
        month = dec,
       volume = {67},
       number = {6},
          eid = {123},
        pages = {123},
          doi = {10.1093/pasj/psv100},
archivePrefix = {arXiv},
       eprint = {1510.06832},
 primaryClass = {astro-ph.GA},
       adsurl = {https://ui.adsabs.harvard.edu/abs/2015PASJ...67..123Y}
}

@ARTICLE{Nishiyama_Schodel_2013,
       author = {{Nishiyama}, S. and {Sch{\"o}del}, R.},
        title = "{Young, massive star candidates detected throughout the nuclear star cluster of the Milky Way}",
      journal = {\aap},
         year = 2013,
        month = jan,
       volume = {549},
          eid = {A57},
        pages = {A57},
          doi = {10.1051/0004-6361/201219773},
archivePrefix = {arXiv},
       eprint = {1210.6125},
 primaryClass = {astro-ph.GA},
       adsurl = {https://ui.adsabs.harvard.edu/abs/2013A&A...549A..57N}
}

@ARTICLE{Nishiyama2013,
       author = {{Nishiyama}, Shogo and {Yasui}, Kazuki and {Nagata}, Tetsuya and {Yoshikawa}, Tatsuhito and {Uchiyama}, Hideki and {Sch{\"o}del}, Rainer and {Hatano}, Hirofumi and {Sato}, Shuji and {Sugitani}, Koji and {Suenaga}, Takuya and {Kwon}, Jungmi and {Tamura}, Motohide},
        title = "{Magnetically Confined Interstellar Hot Plasma in the Nuclear Bulge of Our Galaxy}",
      journal = {\apjl},
         year = 2013,
        month = jun,
       volume = {769},
       number = {2},
          eid = {L28},
        pages = {L28},
          doi = {10.1088/2041-8205/769/2/L28},
archivePrefix = {arXiv},
       eprint = {1305.0347},
 primaryClass = {astro-ph.GA},
       adsurl = {https://ui.adsabs.harvard.edu/abs/2013ApJ...769L..28N}
}

@ARTICLE{Binney1991,
       author = {{Binney}, James and {Gerhard}, Ortwin E. and {Stark}, Antony A. and {Bally}, John and {Uchida}, Keven I.},
        title = "{Understanding the kinematics of Galactic Centre gas.}",
      journal = {\mnras},
         year = 1991,
        month = sep,
       volume = {252},
        pages = {210},
          doi = {10.1093/mnras/252.2.210},
       adsurl = {https://ui.adsabs.harvard.edu/abs/1991MNRAS.252..210B}
}

@ARTICLE{Nogueras2021,
       author = {{Nogueras-Lara}, F. and {Sch{\"o}del}, R. and {Neumayer}, N. and {Schultheis}, M.},
        title = "{Distance to three molecular clouds in the central molecular zone}",
      journal = {\aap},
         year = 2021,
        month = mar,
       volume = {647},
          eid = {L6},
        pages = {L6},
          doi = {10.1051/0004-6361/202140554},
archivePrefix = {arXiv},
       eprint = {2103.02513},
 primaryClass = {astro-ph.GA},
       adsurl = {https://ui.adsabs.harvard.edu/abs/2021A&A...647L...6N}
}

@ARTICLE{Henshaw2016_Seeding,
       author = {{Henshaw}, J.~D. and {Longmore}, S.~N. and {Kruijssen}, J.~M.~D.},
        title = "{Seeding the Galactic Centre gas stream: gravitational instabilities set the initial conditions for the formation of protocluster clouds}",
      journal = {\mnras},
         year = 2016,
        month = nov,
       volume = {463},
       number = {1},
        pages = {L122-L126},
          doi = {10.1093/mnrasl/slw168},
archivePrefix = {arXiv},
       eprint = {1609.01721},
 primaryClass = {astro-ph.GA},
       adsurl = {https://ui.adsabs.harvard.edu/abs/2016MNRAS.463L.122H}
}

@ARTICLE{Pedlar1989,
       author = {{Pedlar}, A. and {Anantharamaiah}, K.~R. and {Ekers}, R.~D. and {Goss}, W.~M. and {van Gorkom}, J.~H. and {Schwarz}, U.~J. and {Zhao}, Jun-Hui},
        title = "{Radio Studies of the Galactic Center. I. The Sagittarius A Complex}",
      journal = {\apj},
         year = 1989,
        month = jul,
       volume = {342},
        pages = {769},
          doi = {10.1086/167635},
       adsurl = {https://ui.adsabs.harvard.edu/abs/1989ApJ...342..769P}
}

@ARTICLE{Yusef-Zadeh1987,
       author = {{Yusef-Zadeh}, F. and {Morris}, Mark},
        title = "{Structural Details of the Sagittarius A Complex: Evidence for a Large-Scale Poloidal Magnetic Field in the Galactic Center Region}",
      journal = {\apj},
         year = 1987,
        month = sep,
       volume = {320},
        pages = {545},
          doi = {10.1086/165572},
       adsurl = {https://ui.adsabs.harvard.edu/abs/1987ApJ...320..545Y}
}

@ARTICLE{Ridley2017,
       author = {{Ridley}, Matthew G.~L. and {Sormani}, Mattia C. and {Tre{\ss}}, Robin G. and {Magorrian}, John and {Klessen}, Ralf S.},
        title = "{Nuclear spirals in the inner Milky Way}",
      journal = {\mnras},
         year = 2017,
        month = aug,
       volume = {469},
       number = {2},
        pages = {2251-2262},
          doi = {10.1093/mnras/stx944},
archivePrefix = {arXiv},
       eprint = {1704.03665},
 primaryClass = {astro-ph.GA},
       adsurl = {https://ui.adsabs.harvard.edu/abs/2017MNRAS.469.2251R}
}

@ARTICLE{Sawada2004,
       author = {{Sawada}, Tsuyoshi and {Hasegawa}, Tetsuo and {Handa}, Toshihiro and {Cohen}, R.~J.},
        title = "{A molecular face-on view of the Galactic Centre region}",
      journal = {\mnras},
         year = 2004,
        month = apr,
       volume = {349},
       number = {4},
        pages = {1167-1178},
          doi = {10.1111/j.1365-2966.2004.07603.x},
archivePrefix = {arXiv},
       eprint = {astro-ph/0401286},
 primaryClass = {astro-ph},
       adsurl = {https://ui.adsabs.harvard.edu/abs/2004MNRAS.349.1167S}
}

@ARTICLE{Henshaw2020,
       author = {{Henshaw}, Jonathan D. and {Kruijssen}, J.~M. Diederik and {Longmore}, Steven N. and {Riener}, Manuel and {Leroy}, Adam K. and {Rosolowsky}, Erik and {Ginsburg}, Adam and {Battersby}, Cara and {Chevance}, M{\'e}lanie and {Meidt}, Sharon E. and {Glover}, Simon C.~O. and {Hughes}, Annie and {Kainulainen}, Jouni and {Klessen}, Ralf S. and {Schinnerer}, Eva and {Schruba}, Andreas and {Beuther}, Henrik and {Bigiel}, Frank and {Blanc}, Guillermo A. and {Emsellem}, Eric and {Henning}, Thomas and {Herrera}, Cynthia N. and {Koch}, Eric W. and {Pety}, J{\'e}r{\^o}me and {Ragan}, Sarah E. and {Sun}, Jiayi},
        title = "{Ubiquitous velocity fluctuations throughout the molecular interstellar medium}",
      journal = {Nature Astronomy},
         year = 2020,
        month = jul,
       volume = {4},
        pages = {1064-1071},
          doi = {10.1038/s41550-020-1126-z},
archivePrefix = {arXiv},
       eprint = {2007.01877},
 primaryClass = {astro-ph.GA},
       adsurl = {https://ui.adsabs.harvard.edu/abs/2020NatAs...4.1064H}
}

@ARTICLE{Reid2009,
       author = {{Reid}, M.~J. and {Menten}, K.~M. and {Zheng}, X.~W. and {Brunthaler}, A. and {Xu}, Y.},
        title = "{A Trigonometric Parallax of Sgr B2}",
      journal = {\apj},
         year = 2009,
        month = nov,
       volume = {705},
       number = {2},
        pages = {1548-1553},
          doi = {10.1088/0004-637X/705/2/1548},
archivePrefix = {arXiv},
       eprint = {0908.3637},
 primaryClass = {astro-ph.GA},
       adsurl = {https://ui.adsabs.harvard.edu/abs/2009ApJ...705.1548R}
}

@article{Pearson1895,
author = {Pearson, Karl  and Galton, Francis },
title = {VII. Note on regression and inheritance in the case of two parents},
journal = {Proceedings of the Royal Society of London},
volume = {58},
number = {347-352},
pages = {240-242},
year = {1895},
doi = {10.1098/rspl.1895.0041},
URL = {https://royalsocietypublishing.org/doi/abs/10.1098/rspl.1895.0041},
eprint = {https://royalsocietypublishing.org/doi/pdf/10.1098/rspl.1895.0041}
,}

@software{astrodendro,
       author = {{Robitaille}, Thomas and {Rice}, Tom and {Beaumont}, Chris and {Ginsburg}, Adam and {MacDonald}, Braden and {Rosolowsky}, Erik},
        title = "{astrodendro: Astronomical data dendrogram creator}",
 howpublished = {Astrophysics Source Code Library, record ascl:1907.016},
         year = 2019,
        month = jul,
          eid = {ascl:1907.016},
archivePrefix = {ascl},
       eprint = {1907.016},
       adsurl = {https://ui.adsabs.harvard.edu/abs/2019ascl.soft07016R}
}

@ARTICLE{Nogueras-Lara2026,
       author = {{Nogueras-Lara}, Francisco and {Barnes}, Ashley T. and {Henshaw}, Jonathan D. and {Fiteni}, Karl and {Sofue}, Yoshiaki and {Sch{\"o}del}, Rainer and {Mart{\'\i}nez-Arranz}, {\'A}lvaro and {Sormani}, Mattia C. and {Armijos-Abenda{\~n}o}, Jairo and {Colzi}, Laura and {Jim{\'e}nez-Serra}, Izaskun and {Rivilla}, V{\'\i}ctor M. and {Garc{\'\i}a}, Pablo and {Ginsburg}, Adam and {Hu}, Yue and {Klessen}, Ralf S. and {Kruijssen}, J.~M. Diederik and {Tolls}, Volker and {Lazarian}, Alex and {Lipman}, Dani R. and {Longmore}, Steven N. and {Lu}, Xing and {Mart{\'\i}n}, Sergio and {Riquelme-V{\'a}squez}, Denise and {Pineda}, Jaime E. and {S{\'a}nchez-Monge}, {\'A}lvaro and {Vasini}, Arianna and {Mills}, Elisabeth A.~C.},
        title = "{Unveiling the 3D structure of the central molecular zone from stellar kinematics and photometry: The 50 and 20 km/s clouds}",
      journal = {\aap},
         year = 2026,
        month = jan,
       volume = {706},
          eid = {A18},
        pages = {A18},
          doi = {10.1051/0004-6361/202556047},
archivePrefix = {arXiv},
       eprint = {2601.05252},
 primaryClass = {astro-ph.GA},
       adsurl = {https://ui.adsabs.harvard.edu/abs/2026A&A...706A..18N}
}

@ARTICLE{Gleis2026,
       author = {{Gleis}, Damian R. and {Stuber}, Sophia K. and {Schinnerer}, Eva and {Neumann}, Justus and {Meidt}, Sharon E. and {Querejeta}, Miguel and {Emsellem}, Eric and {Leroy}, Adam K. and {Barnes}, Ashley T. and {Bigiel}, Frank and {Burton}, Charlie and {Chevance}, M{\'e}lanie and {Dale}, Daniel A. and {Grasha}, Kathryn and {Klessen}, Ralf S. and {Levy}, Rebecca C. and {Neumann}, Lukas and {Pan}, Hsi-An and {Ruiz-Garc{\'\i}a}, Marina and {Sormani}, Mattia C. and {Sun}, Jiayi and {Teng}, Yu-Hsuan and {Williams}, Thomas G.},
        title = "{Molecular gas and star formation in central rings across nearby galaxies}",
      journal = {arXiv e-prints},
         year = 2026,
        month = jan,
          eid = {arXiv:2601.11127},
        pages = {arXiv:2601.11127},
          doi = {10.48550/arXiv.2601.11127},
archivePrefix = {arXiv},
       eprint = {2601.11127},
 primaryClass = {astro-ph.GA},
       adsurl = {https://ui.adsabs.harvard.edu/abs/2026arXiv260111127G}
}

@ARTICLE{Callanan2021,
       author = {{Callanan}, Daniel and {Longmore}, Steven N. and {Kruijssen}, J.~M. Diederik and {Schruba}, Andreas and {Ginsburg}, Adam and {Krumholz}, Mark R. and {Bastian}, Nate and {Alves}, Jo{\~a}o and {Henshaw}, Jonathan D. and {Knapen}, Johan H. and {Chevance}, M{\'e}lanie},
        title = "{The centres of M83 and the Milky Way: opposite extremes of a common star formation cycle}",
      journal = {\mnras},
         year = 2021,
        month = aug,
       volume = {505},
       number = {3},
        pages = {4310-4337},
          doi = {10.1093/mnras/stab1527},
archivePrefix = {arXiv},
       eprint = {2105.09761},
 primaryClass = {astro-ph.GA},
       adsurl = {https://ui.adsabs.harvard.edu/abs/2021MNRAS.505.4310C}
}

@ARTICLE{Aizawa2025,
       author = {{Aizawa}, Masataka and {Muto}, Takayuki and {Momose}, Munetake},
        title = "{Axisymmetric Modeling of DSHARP Dusty Disks: Asymmetric Structures and Inner-Disk Dispersal}",
      journal = {arXiv e-prints},
         year = 2025,
        month = dec,
          eid = {arXiv:2512.16091},
        pages = {arXiv:2512.16091},
          doi = {10.48550/arXiv.2512.16091},
archivePrefix = {arXiv},
       eprint = {2512.16091},
 primaryClass = {astro-ph.EP},
       adsurl = {https://ui.adsabs.harvard.edu/abs/2025arXiv251216091A}
}

@ARTICLE{Mokeddem2026,
       author = {{Mokeddem}, Rahima and {Lopes}, Maria and {Avila}, Felipe and {Bernui}, Armando and {Hip{\'o}lito-Ricaldi}, Wiliam S.},
        title = "{Probing cosmic isotropy: Hubble constant and matter density large-angle variations with the Pantheon+SH0ES data}",
      journal = {Physics of the Dark Universe},
         year = 2026,
        month = feb,
       volume = {51},
          eid = {102185},
        pages = {102185},
          doi = {10.1016/j.dark.2025.102185},
archivePrefix = {arXiv},
       eprint = {2504.00903},
 primaryClass = {astro-ph.CO},
       adsurl = {https://ui.adsabs.harvard.edu/abs/2026PDU....5102185M}
}

@article{monge1781,
  title={M{\'e}moire sur la th{\'e}orie des d{\'e}blais et des remblais},
  author={Monge, Gaspard},
  journal={Mem. Math. Phys. Acad. Royale Sci.},
  pages={666--704},
  year={1781}
}

@article{kantorovich1942,
  title={oOn the translocation of masses, pDokl},
  author={Kantorovich, L},
  journal={Akad. Nauk SSSR},
  volume={37},
  year={1942}
}

@book{villani2021,
  title={Topics in optimal transportation},
  author={Villani, C{\'e}dric},
  volume={58},
  year={2021},
  publisher={American Mathematical Soc.}
}

@article{Reprint_Mahalanobis2018,
  title={Reprint of: Mahalanobis, P.C. (1936) "On the Generalised Distance in Statistics."},
  author={Mahalanobis, P.C.},
  journal={Sankhya A},
  year={1936},
  volume={80},
  pages={1 - 7},
  url={https://api.semanticscholar.org/CorpusID:239595337}
}

@ARTICLE{Blaylock-Squibbs2023,
       author = {{Blaylock-Squibbs}, George A. and {Parker}, Richard J.},
        title = "{The evolution of phase space densities in star-forming regions}",
      journal = {\mnras},
         year = 2023,
        month = mar,
       volume = {519},
       number = {3},
        pages = {3643-3655},
          doi = {10.1093/mnras/stac3683},
archivePrefix = {arXiv},
       eprint = {2301.03472},
 primaryClass = {astro-ph.GA},
       adsurl = {https://ui.adsabs.harvard.edu/abs/2023MNRAS.519.3643B}
}

@INPROCEEDINGS{Henshaw2023a,
       author = {{Henshaw}, J.~D. and {Barnes}, A.~T. and {Battersby}, C. and {Ginsburg}, A. and {Sormani}, M.~C. and {Walker}, D.~L.},
        title = "{Star Formation in the Central Molecular Zone of the Milky Way}",
    booktitle = {Protostars and Planets VII},
         year = 2023,
       editor = {{Inutsuka}, S. and {Aikawa}, Y. and {Muto}, T. and {Tomida}, K. and {Tamura}, M.},
       series = {Astronomical Society of the Pacific Conference Series},
       volume = {534},
        month = jul,
        pages = {83},
          doi = {10.48550/arXiv.2203.11223},
archivePrefix = {arXiv},
       eprint = {2203.11223},
 primaryClass = {astro-ph.GA},
       adsurl = {https://ui.adsabs.harvard.edu/abs/2023ASPC..534...83H}
}

@article{Pillai2015,
	Adsurl = {http://adsabs.harvard.edu/abs/2015ApJ...799...74P},
	Archiveprefix = {arXiv},
	Author = {{Pillai}, T. and {Kauffmann}, J. and {Tan}, J.~C. and {Goldsmith}, P.~F. and {Carey}, S.~J. and {Menten}, K.~M.},
	Doi = {10.1088/0004-637X/799/1/74},
	Eid = {74},
	Eprint = {1410.7390},
	Journal = {\apj},
	Month = jan,
	Pages = {74},
	Title = {{Magnetic Fields in High-mass Infrared Dark Clouds}},
	Volume = 799,
	Year = 2015}

@ARTICLE{Butterfield2024,
       author = {{Butterfield}, Natalie O. and {Chuss}, David T. and {Guerra}, Jordan A. and {Morris}, Mark R. and {Par{\'e}}, Dylan and {Wollack}, Edward J. and {Dowell}, C. Darren and {Hankins}, Matthew J. and {Karpovich}, Kaitlyn and {Siah}, Javad and {Staguhn}, Johannes and {Zweibel}, Ellen},
        title = "{SOFIA/HAWC+ Far-Infrared Polarimetric Large Area CMZ Exploration Survey. I. General Results from the Pilot Program}",
      journal = {\apj},
         year = 2024,
        month = mar,
       volume = {963},
       number = {2},
          eid = {130},
        pages = {130},
          doi = {10.3847/1538-4357/ad12b9},
archivePrefix = {arXiv},
       eprint = {2306.01681},
 primaryClass = {astro-ph.GA},
       adsurl = {https://ui.adsabs.harvard.edu/abs/2024ApJ...963..130B}
}

@ARTICLE{Gravity2021,
       author = {{GRAVITY Collaboration} and {Abuter}, R. and {Amorim}, A. and {Baub{\"o}ck}, M. and {Berger}, J.~P. and {Bonnet}, H. and {Brandner}, W. and {Cl{\'e}net}, Y. and {Davies}, R. and {de Zeeuw}, P.~T. and {Dexter}, J. and {Dallilar}, Y. and {Drescher}, A. and {Eckart}, A. and {Eisenhauer}, F. and {F{\"o}rster Schreiber}, N.~M. and {Garcia}, P. and {Gao}, F. and {Gendron}, E. and {Genzel}, R. and {Gillessen}, S. and {Habibi}, M. and {Haubois}, X. and {Hei{\ss}el}, G. and {Henning}, T. and {Hippler}, S. and {Horrobin}, M. and {Jim{\'e}nez-Rosales}, A. and {Jochum}, L. and {Jocou}, L. and {Kaufer}, A. and {Kervella}, P. and {Lacour}, S. and {Lapeyr{\`e}re}, V. and {Le Bouquin}, J. -B. and {L{\'e}na}, P. and {Lutz}, D. and {Nowak}, M. and {Ott}, T. and {Paumard}, T. and {Perraut}, K. and {Perrin}, G. and {Pfuhl}, O. and {Rabien}, S. and {Rodr{\'\i}guez-Coira}, G. and {Shangguan}, J. and {Shimizu}, T. and {Scheithauer}, S. and {Stadler}, J. and {Straub}, O. and {Straubmeier}, C. and {Sturm}, E. and {Tacconi}, L.~J. and {Vincent}, F. and {von Fellenberg}, S. and {Waisberg}, I. and {Widmann}, F. and {Wieprecht}, E. and {Wiezorrek}, E. and {Woillez}, J. and {Yazici}, S. and {Young}, A. and {Zins}, G.},
        title = "{Improved GRAVITY astrometric accuracy from modeling optical aberrations}",
      journal = {\aap},
         year = 2021,
        month = mar,
       volume = {647},
          eid = {A59},
        pages = {A59},
          doi = {10.1051/0004-6361/202040208},
archivePrefix = {arXiv},
       eprint = {2101.12098},
 primaryClass = {astro-ph.GA},
       adsurl = {https://ui.adsabs.harvard.edu/abs/2021A&A...647A..59G}
}

@ARTICLE{Krieger2017,
       author = {{Krieger}, Nico and {Ott}, J{\"u}rgen and {Beuther}, Henrik and {Walter}, Fabian and {Kruijssen}, J.~M. Diederik and {Meier}, David S. and {Mills}, Elisabeth A.~C. and {Contreras}, Yanett and {Edwards}, Phil and {Ginsburg}, Adam and {Henkel}, Christian and {Henshaw}, Jonathan and {Jackson}, James and {Kauffmann}, Jens and {Longmore}, Steven and {Mart{\'\i}n}, Sergio and {Morris}, Mark R. and {Pillai}, Thushara and {Rickert}, Matthew and {Rosolowsky}, Erik and {Shinnaga}, Hiroko and {Walsh}, Andrew and {Yusef-Zadeh}, Farhad and {Zhang}, Qizhou},
        title = "{The Survey of Water and Ammonia in the Galactic Center (SWAG): Molecular Cloud Evolution in the Central Molecular Zone}",
      journal = {\apj},
         year = 2017,
        month = nov,
       volume = {850},
       number = {1},
          eid = {77},
        pages = {77},
          doi = {10.3847/1538-4357/aa951c},
archivePrefix = {arXiv},
       eprint = {1710.06902},
 primaryClass = {astro-ph.GA},
       adsurl = {https://ui.adsabs.harvard.edu/abs/2017ApJ...850...77K}
}

@ARTICLE{Mills2018a,
       author = {{Mills}, E.~A.~C. and {Ginsburg}, A. and {Immer}, K. and {Barnes}, J.~M. and {Wiesenfeld}, L. and {Faure}, A. and {Morris}, M.~R. and {Requena-Torres}, M.~A.},
        title = "{The Dense Gas Fraction in Galactic Center Clouds}",
      journal = {\apj},
         year = 2018,
        month = nov,
       volume = {868},
       number = {1},
          eid = {7},
        pages = {7},
          doi = {10.3847/1538-4357/aae581},
archivePrefix = {arXiv},
       eprint = {1810.00266},
 primaryClass = {astro-ph.GA},
       adsurl = {https://ui.adsabs.harvard.edu/abs/2018ApJ...868....7M}
}

@ARTICLE{Molinari2016,
       author = {{Molinari}, S. and {Schisano}, E. and {Elia}, D. and {Pestalozzi}, M. and {Traficante}, A. and {Pezzuto}, S. and {Swinyard}, B.~M. and {Noriega-Crespo}, A. and {Bally}, J. and {Moore}, T.~J.~T. and {Plume}, R. and {Zavagno}, A. and {di Giorgio}, A.~M. and {Liu}, S.~J. and {Pilbratt}, G.~L. and {Mottram}, J.~C. and {Russeil}, D. and {Piazzo}, L. and {Veneziani}, M. and {Benedettini}, M. and {Calzoletti}, L. and {Faustini}, F. and {Natoli}, P. and {Piacentini}, F. and {Merello}, M. and {Palmese}, A. and {Del Grande}, R. and {Polychroni}, D. and {Rygl}, K.~L.~J. and {Polenta}, G. and {Barlow}, M.~J. and {Bernard}, J. -P. and {Martin}, P.~G. and {Testi}, L. and {Ali}, B. and {Andr{\'e}}, P. and {Beltr{\'a}n}, M.~T. and {Billot}, N. and {Carey}, S. and {Cesaroni}, R. and {Compi{\`e}gne}, M. and {Eden}, D. and {Fukui}, Y. and {Garcia-Lario}, P. and {Hoare}, M.~G. and {Huang}, M. and {Joncas}, G. and {Lim}, T.~L. and {Lord}, S.~D. and {Martinavarro-Armengol}, S. and {Motte}, F. and {Paladini}, R. and {Paradis}, D. and {Peretto}, N. and {Robitaille}, T. and {Schilke}, P. and {Schneider}, N. and {Schulz}, B. and {Sibthorpe}, B. and {Strafella}, F. and {Thompson}, M.~A. and {Umana}, G. and {Ward-Thompson}, D. and {Wyrowski}, F.},
        title = "{Hi-GAL, the Herschel infrared Galactic Plane Survey: photometric maps and compact source catalogues. First data release for the inner Milky Way: +68{\textdegree} {\ensuremath{\geq}} l {\ensuremath{\geq}} -70{\textdegree}}",
      journal = {\aap},
         year = 2016,
        month = jul,
       volume = {591},
          eid = {A149},
        pages = {A149},
          doi = {10.1051/0004-6361/201526380},
archivePrefix = {arXiv},
       eprint = {1604.05911},
 primaryClass = {astro-ph.GA},
       adsurl = {https://ui.adsabs.harvard.edu/abs/2016A&A...591A.149M}
}

@ARTICLE{Molinari2010,
       author = {{Molinari}, S. and {Swinyard}, B. and {Bally}, J. and {Barlow}, M. and {Bernard}, J. -P. and {Martin}, P. and {Moore}, T. and {Noriega-Crespo}, A. and {Plume}, R. and {Testi}, L. and {Zavagno}, A. and {Abergel}, A. and {Ali}, B. and {Anderson}, L. and {Andr{\'e}}, P. and {Baluteau}, J. -P. and {Battersby}, C. and {Beltr{\'a}n}, M.~T. and {Benedettini}, M. and {Billot}, N. and {Blommaert}, J. and {Bontemps}, S. and {Boulanger}, F. and {Brand}, J. and {Brunt}, C. and {Burton}, M. and {Calzoletti}, L. and {Carey}, S. and {Caselli}, P. and {Cesaroni}, R. and {Cernicharo}, J. and {Chakrabarti}, S. and {Chrysostomou}, A. and {Cohen}, M. and {Compiegne}, M. and {de Bernardis}, P. and {de Gasperis}, G. and {di Giorgio}, A.~M. and {Elia}, D. and {Faustini}, F. and {Flagey}, N. and {Fukui}, Y. and {Fuller}, G.~A. and {Ganga}, K. and {Garcia-Lario}, P. and {Glenn}, J. and {Goldsmith}, P.~F. and {Griffin}, M. and {Hoare}, M. and {Huang}, M. and {Ikhenaode}, D. and {Joblin}, C. and {Joncas}, G. and {Juvela}, M. and {Kirk}, J.~M. and {Lagache}, G. and {Li}, J.~Z. and {Lim}, T.~L. and {Lord}, S.~D. and {Marengo}, M. and {Marshall}, D.~J. and {Masi}, S. and {Massi}, F. and {Matsuura}, M. and {Minier}, V. and {Miville-Desch{\^e}nes}, M. -A. and {Montier}, L.~A. and {Morgan}, L. and {Motte}, F. and {Mottram}, J.~C. and {M{\"u}ller}, T.~G. and {Natoli}, P. and {Neves}, J. and {Olmi}, L. and {Paladini}, R. and {Paradis}, D. and {Parsons}, H. and {Peretto}, N. and {Pestalozzi}, M. and {Pezzuto}, S. and {Piacentini}, F. and {Piazzo}, L. and {Polychroni}, D. and {Pomar{\`e}s}, M. and {Popescu}, C.~C. and {Reach}, W.~T. and {Ristorcelli}, I. and {Robitaille}, J. -F. and {Robitaille}, T. and {Rod{\'o}n}, J.~A. and {Roy}, A. and {Royer}, P. and {Russeil}, D. and {Saraceno}, P. and {Sauvage}, M. and {Schilke}, P. and {Schisano}, E. and {Schneider}, N. and {Schuller}, F. and {Schulz}, B. and {Sibthorpe}, B. and {Smith}, H.~A. and {Smith}, M.~D. and {Spinoglio}, L. and {Stamatellos}, D. and {Strafella}, F. and {Stringfellow}, G.~S. and {Sturm}, E. and {Taylor}, R. and {Thompson}, M.~A. and {Traficante}, A. and {Tuffs}, R.~J. and {Umana}, G. and {Valenziano}, L. and {Vavrek}, R. and {Veneziani}, M. and {Viti}, S. and {Waelkens}, C. and {Ward-Thompson}, D. and {White}, G. and {Wilcock}, L.~A. and {Wyrowski}, F. and {Yorke}, H.~W. and {Zhang}, Q.},
        title = "{Clouds, filaments, and protostars: The Herschel Hi-GAL Milky Way}",
      journal = {\aap},
         year = 2010,
        month = jul,
       volume = {518},
          eid = {L100},
        pages = {L100},
          doi = {10.1051/0004-6361/201014659},
archivePrefix = {arXiv},
       eprint = {1005.3317},
 primaryClass = {astro-ph.GA},
       adsurl = {https://ui.adsabs.harvard.edu/abs/2010A&A...518L.100M}
}

@ARTICLE{Sofue2022,
       author = {{Sofue}, Yoshiaki},
        title = "{Three-dimensional structure of the central molecular zone}",
      journal = {\mnras},
         year = 2022,
        month = oct,
       volume = {516},
       number = {1},
        pages = {907-923},
          doi = {10.1093/mnras/stac2243},
archivePrefix = {arXiv},
       eprint = {2208.02451},
 primaryClass = {astro-ph.GA},
       adsurl = {https://ui.adsabs.harvard.edu/abs/2022MNRAS.516..907S}
}

@ARTICLE{Ellsworth-Bowers2013,
       author = {{Ellsworth-Bowers}, Timothy P. and {Glenn}, Jason and {Rosolowsky}, Erik and {Mairs}, Steven and {Evans}, Neal J., II and {Battersby}, Cara and {Ginsburg}, Adam and {Shirley}, Yancy L. and {Bally}, John},
        title = "{The Bolocam Galactic Plane Survey. VIII. A Mid-infrared Kinematic Distance Discrimination Method}",
      journal = {\apj},
         year = 2013,
        month = jun,
       volume = {770},
       number = {1},
          eid = {39},
        pages = {39},
          doi = {10.1088/0004-637X/770/1/39},
archivePrefix = {arXiv},
       eprint = {1304.4597},
 primaryClass = {astro-ph.GA},
       adsurl = {https://ui.adsabs.harvard.edu/abs/2013ApJ...770...39E}
}

@INPROCEEDINGS{Ponti2010,
       author = {{Ponti}, Gabriele and {Terrier}, R. and {Goldwurm}, A. and {Belanger}, G. and {Trap}, G.},
        title = "{XMM-Newton Study of the Fe K Emission from Molecular Clouds in the Galactic Centre}",
    booktitle = {AAS/High Energy Astrophysics Division \#11},
         year = 2010,
       series = {AAS/High Energy Astrophysics Division},
       volume = {11},
        month = mar,
          eid = {11.12},
        pages = {11.12},
       adsurl = {https://ui.adsabs.harvard.edu/abs/2010HEAD...11.1112P}
}

@ARTICLE{KimElmegreen2017,
       author = {{Kim}, Woong-Tae and {Elmegreen}, Bruce G.},
        title = "{Nuclear Spiral Shocks and Induced Gas Inflows in Weak Oval Potentials}",
      journal = {\apjl},
         year = 2017,
        month = may,
       volume = {841},
       number = {1},
          eid = {L4},
        pages = {L4},
          doi = {10.3847/2041-8213/aa70a1},
archivePrefix = {arXiv},
       eprint = {1705.00863},
 primaryClass = {astro-ph.GA},
       adsurl = {https://ui.adsabs.harvard.edu/abs/2017ApJ...841L...4K}
}

@ARTICLE{Shlosman1989,
       author = {{Shlosman}, Isaac and {Frank}, Juhan and {Begelman}, Mitchell C.},
        title = "{Bars within bars: a mechanism for fuelling active galactic nuclei}",
      journal = {\nat},
         year = 1989,
        month = mar,
       volume = {338},
       number = {6210},
        pages = {45-47},
          doi = {10.1038/338045a0},
       adsurl = {https://ui.adsabs.harvard.edu/abs/1989Natur.338...45S}
}

@ARTICLE{BalbusHawley1998,
       author = {{Balbus}, Steven A. and {Hawley}, John F.},
        title = "{Instability, turbulence, and enhanced transport in accretion disks}",
      journal = {Reviews of Modern Physics},
         year = 1998,
        month = jan,
       volume = {70},
       number = {1},
        pages = {1-53},
          doi = {10.1103/RevModPhys.70.1},
       adsurl = {https://ui.adsabs.harvard.edu/abs/1998RvMP...70....1B}
}

@ARTICLE{Davies2007,
       author = {{Davies}, R.~I. and {M{\"u}ller S{\'a}nchez}, F. and {Genzel}, R. and {Tacconi}, L.~J. and {Hicks}, E.~K.~S. and {Friedrich}, S. and {Sternberg}, A.},
        title = "{A Close Look at Star Formation around Active Galactic Nuclei}",
      journal = {\apj},
         year = 2007,
        month = dec,
       volume = {671},
       number = {2},
        pages = {1388-1412},
          doi = {10.1086/523032},
archivePrefix = {arXiv},
       eprint = {0704.1374},
 primaryClass = {astro-ph},
       adsurl = {https://ui.adsabs.harvard.edu/abs/2007ApJ...671.1388D}
}

@ARTICLE{Henshaw2016_GasKin_250pc,
       author = {{Henshaw}, J.~D. and {Longmore}, S.~N. and {Kruijssen}, J.~M.~D. and {Davies}, B. and {Bally}, J. and {Barnes}, A. and {Battersby}, C. and {Burton}, M. and {Cunningham}, M.~R. and {Dale}, J.~E. and {Ginsburg}, A. and {Immer}, K. and {Jones}, P.~A. and {Kendrew}, S. and {Mills}, E.~A.~C. and {Molinari}, S. and {Moore}, T.~J.~T. and {Ott}, J. and {Pillai}, T. and {Rathborne}, J. and {Schilke}, P. and {Schmiedeke}, A. and {Testi}, L. and {Walker}, D. and {Walsh}, A. and {Zhang}, Q.},
        title = "{Molecular gas kinematics within the central 250 pc of the Milky Way}",
      journal = {\mnras},
         year = 2016,
        month = apr,
       volume = {457},
       number = {3},
        pages = {2675-2702},
          doi = {10.1093/mnras/stw121},
archivePrefix = {arXiv},
       eprint = {1601.03732},
 primaryClass = {astro-ph.GA},
       adsurl = {https://ui.adsabs.harvard.edu/abs/2016MNRAS.457.2675H}
}

@ARTICLE{Kruijssen2015,
       author = {{Kruijssen}, J.~M. Diederik and {Dale}, James E. and {Longmore}, Steven N.},
        title = "{The dynamical evolution of molecular clouds near the Galactic Centre - I. Orbital structure and evolutionary timeline}",
      journal = {\mnras},
         year = 2015,
        month = feb,
       volume = {447},
       number = {2},
        pages = {1059-1079},
          doi = {10.1093/mnras/stu2526},
archivePrefix = {arXiv},
       eprint = {1412.0664},
 primaryClass = {astro-ph.GA},
       adsurl = {https://ui.adsabs.harvard.edu/abs/2015MNRAS.447.1059K}
}

@ARTICLE{Tress2020,
       author = {{Tress}, Robin G. and {Sormani}, Mattia C. and {Glover}, Simon C.~O. and {Klessen}, Ralf S. and {Battersby}, Cara D. and {Clark}, Paul C. and {Hatchfield}, H. Perry and {Smith}, Rowan J.},
        title = "{Simulations of the Milky Way's central molecular zone - I. Gas dynamics}",
      journal = {\mnras},
         year = 2020,
        month = dec,
       volume = {499},
       number = {3},
        pages = {4455-4478},
          doi = {10.1093/mnras/staa3120},
archivePrefix = {arXiv},
       eprint = {2004.06724},
 primaryClass = {astro-ph.GA},
       adsurl = {https://ui.adsabs.harvard.edu/abs/2020MNRAS.499.4455T}
}

@ARTICLE{Edenhofer2024,
       author = {{Edenhofer}, Gordian and {Zucker}, Catherine and {Frank}, Philipp and {Saydjari}, Andrew K. and {Speagle}, Joshua S. and {Finkbeiner}, Douglas and {En{\ss}lin}, Torsten A.},
        title = "{A parsec-scale Galactic 3D dust map out to 1.25 kpc from the Sun}",
      journal = {\aap},
         year = 2024,
        month = may,
       volume = {685},
          eid = {A82},
        pages = {A82},
          doi = {10.1051/0004-6361/202347628},
archivePrefix = {arXiv},
       eprint = {2308.01295},
 primaryClass = {astro-ph.GA},
       adsurl = {https://ui.adsabs.harvard.edu/abs/2024A&A...685A..82E}
}

@ARTICLE{Zhang2025,
       author = {{Zhang}, M. and {Kainulainen}, J. and {Zhao}, H. and {Su}, Y. and {Fang}, M. and {Ma}, Y. and {Chen}, Z. and {Jiang}, Z.},
        title = "{Dust extinction map of the Galactic plane based on the UKIDSS survey data}",
      journal = {\mnras},
         year = 2025,
        month = nov,
       volume = {543},
       number = {4},
        pages = {3830-3848},
          doi = {10.1093/mnras/staf1704},
archivePrefix = {arXiv},
       eprint = {2510.14380},
 primaryClass = {astro-ph.GA},
       adsurl = {https://ui.adsabs.harvard.edu/abs/2025MNRAS.543.3830Z}
}

@ARTICLE{Sormani2022_NSD,
       author = {{Sormani}, Mattia C. and {Sanders}, Jason L. and {Fritz}, Tobias K. and {Smith}, Leigh C. and {Gerhard}, Ortwin and {Sch{\"o}del}, Rainer and {Magorrian}, John and {Neumayer}, Nadine and {Nogueras-Lara}, Francisco and {Feldmeier-Krause}, Anja and {Mastrobuono-Battisti}, Alessandra and {Schultheis}, Mathias and {Shahzamanian}, Banafsheh and {Vasiliev}, Eugene and {Klessen}, Ralf S. and {Lucas}, Philip and {Minniti}, Dante},
        title = "{Self-consistent modelling of the Milky Way's nuclear stellar disc}",
      journal = {\mnras},
         year = 2022,
        month = may,
       volume = {512},
       number = {2},
        pages = {1857-1884},
          doi = {10.1093/mnras/stac639},
archivePrefix = {arXiv},
       eprint = {2111.12713},
 primaryClass = {astro-ph.GA},
       adsurl = {https://ui.adsabs.harvard.edu/abs/2022MNRAS.512.1857S}
}

@ARTICLE{Sormani2020_NSD,
       author = {{Sormani}, Mattia C. and {Magorrian}, John and {Nogueras-Lara}, Francisco and {Neumayer}, Nadine and {Sch{\"o}nrich}, Ralph and {Klessen}, Ralf S. and {Mastrobuono-Battisti}, Alessandra},
        title = "{Jeans modelling of the Milky Way's nuclear stellar disc}",
      journal = {\mnras},
         year = 2020,
        month = nov,
       volume = {499},
       number = {1},
        pages = {7-24},
          doi = {10.1093/mnras/staa2785},
archivePrefix = {arXiv},
       eprint = {2007.06577},
 primaryClass = {astro-ph.GA},
       adsurl = {https://ui.adsabs.harvard.edu/abs/2020MNRAS.499....7S}
}

@ARTICLE{Sormani2020,
       author = {{Sormani}, Mattia C. and {Tress}, Robin G. and {Glover}, Simon C.~O. and {Klessen}, Ralf S. and {Battersby}, Cara D. and {Clark}, Paul C. and {Hatchfield}, H. Perry and {Smith}, Rowan J.},
        title = "{Simulations of the Milky Way's Central Molecular Zone - II. Star formation}",
      journal = {\mnras},
         year = 2020,
        month = oct,
       volume = {497},
       number = {4},
        pages = {5024-5040},
          doi = {10.1093/mnras/staa1999},
archivePrefix = {arXiv},
       eprint = {2004.06731},
 primaryClass = {astro-ph.GA},
       adsurl = {https://ui.adsabs.harvard.edu/abs/2020MNRAS.497.5024S}
}

@ARTICLE{Churchwell2009,
       author = {{Churchwell}, Ed and {Babler}, Brian L. and {Meade}, Marilyn R. and {Whitney}, Barbara A. and {Benjamin}, Robert and {Indebetouw}, Remy and {Cyganowski}, Claudia and {Robitaille}, Thomas P. and {Povich}, Matthew and {Watson}, Christer and {Bracker}, Steve},
        title = "{The Spitzer/GLIMPSE Surveys: A New View of the Milky Way}",
      journal = {\pasp},
         year = 2009,
        month = mar,
       volume = {121},
       number = {877},
        pages = {213},
          doi = {10.1086/597811},
       adsurl = {https://ui.adsabs.harvard.edu/abs/2009PASP..121..213C}
}

@ARTICLE{Benjamin2003,
       author = {{Benjamin}, Robert A. and {Churchwell}, E. and {Babler}, Brian L. and {Bania}, T.~M. and {Clemens}, Dan P. and {Cohen}, Martin and {Dickey}, John M. and {Indebetouw}, R{\'e}my and {Jackson}, James M. and {Kobulnicky}, Henry A. and {Lazarian}, Alex and {Marston}, A.~P. and {Mathis}, John S. and {Meade}, Marilyn R. and {Seager}, Sara and {Stolovy}, S.~R. and {Watson}, C. and {Whitney}, Barbara A. and {Wolff}, Michael J. and {Wolfire}, Mark G.},
        title = "{GLIMPSE. I. An SIRTF Legacy Project to Map the Inner Galaxy}",
      journal = {\pasp},
         year = 2003,
        month = aug,
       volume = {115},
       number = {810},
        pages = {953-964},
          doi = {10.1086/376696},
archivePrefix = {arXiv},
       eprint = {astro-ph/0306274},
 primaryClass = {astro-ph},
       adsurl = {https://ui.adsabs.harvard.edu/abs/2003PASP..115..953B}
}

@ARTICLE{Ellsworth-Bowers2015A,
       author = {{Ellsworth-Bowers}, Timothy P. and {Glenn}, Jason and {Riley}, Allyssa and {Rosolowsky}, Erik and {Ginsburg}, Adam and {Evans}, Neal J., II and {Bally}, John and {Battersby}, Cara and {Shirley}, Yancy L. and {Merello}, Manuel},
        title = "{The Bolocam Galactic Plane Survey. XIII. Physical Properties and Mass Functions of Dense Molecular Cloud Structures}",
      journal = {\apj},
         year = 2015,
        month = jun,
       volume = {805},
       number = {2},
          eid = {157},
        pages = {157},
          doi = {10.1088/0004-637X/805/2/157},
archivePrefix = {arXiv},
       eprint = {1504.01388},
 primaryClass = {astro-ph.GA},
       adsurl = {https://ui.adsabs.harvard.edu/abs/2015ApJ...805..157E}
}

@ARTICLE{Chuard2018,
       author = {{Chuard}, D. and {Terrier}, R. and {Goldwurm}, A. and {Clavel}, M. and {Soldi}, S. and {Morris}, M.~R. and {Ponti}, G. and {Walls}, M. and {Chernyakova}, M.},
        title = "{Glimpses of the past activity of Sgr A$^{★}$ inferred from X-ray echoes in Sgr C}",
      journal = {\aap},
         year = 2018,
        month = feb,
       volume = {610},
          eid = {A34},
        pages = {A34},
          doi = {10.1051/0004-6361/201731864},
archivePrefix = {arXiv},
       eprint = {1712.02678},
 primaryClass = {astro-ph.HE},
       adsurl = {https://ui.adsabs.harvard.edu/abs/2018A&A...610A..34C}
}

@ARTICLE{Gallego-Cano2020,
       author = {{Gallego-Cano}, E. and {Sch{\"o}del}, R. and {Nogueras-Lara}, F. and {Dong}, H. and {Shahzamanian}, B. and {Fritz}, T.~K. and {Gallego-Calvente}, A.~T. and {Neumayer}, N.},
        title = "{New constraints on the structure of the nuclear stellar cluster of the Milky Way from star counts and MIR imaging}",
      journal = {\aap},
         year = 2020,
        month = feb,
       volume = {634},
          eid = {A71},
        pages = {A71},
          doi = {10.1051/0004-6361/201935303},
archivePrefix = {arXiv},
       eprint = {2001.08182},
 primaryClass = {astro-ph.GA},
       adsurl = {https://ui.adsabs.harvard.edu/abs/2020A&A...634A..71G}
}

@ARTICLE{Sun2024,
       author = {{Sun}, Jiayi and {He}, Hao and {Batschkun}, Kyle and {Levy}, Rebecca C. and {Emig}, Kimberly and {Rodr{\'\i}guez}, M. Jimena and {Hassani}, Hamid and {Leroy}, Adam K. and {Schinnerer}, Eva and {Ostriker}, Eve C. and {Wilson}, Christine D. and {Bolatto}, Alberto D. and {Mills}, Elisabeth A.~C. and {Rosolowsky}, Erik and {Lee}, Janice C. and {Dale}, Daniel A. and {Larson}, Kirsten L. and {Thilker}, David A. and {Ubeda}, Leonardo and {Whitmore}, Bradley C. and {Williams}, Thomas G. and {Barnes}, Ashley T. and {Bigiel}, Frank and {Chevance}, M{\'e}lanie and {Glover}, Simon C.~O. and {Grasha}, Kathryn and {Groves}, Brent and {Henshaw}, Jonathan D. and {Indebetouw}, R{\'e}my and {Jim{\'e}nez-Donaire}, Mar{\'\i}a J. and {Klessen}, Ralf S. and {Koch}, Eric W. and {Liu}, Daizhong and {Mathur}, Smita and {Meidt}, Sharon and {Menon}, Shyam H. and {Neumann}, Justus and {Pinna}, Francesca and {Querejeta}, Miguel and {Sormani}, Mattia C. and {Tress}, Robin G.},
        title = "{Hidden Gems on a Ring: Infant Massive Clusters and Their Formation Timeline Unveiled by ALMA, HST, and JWST in NGC 3351}",
      journal = {\apj},
         year = 2024,
        month = jun,
       volume = {967},
       number = {2},
          eid = {133},
        pages = {133},
          doi = {10.3847/1538-4357/ad3de6},
archivePrefix = {arXiv},
       eprint = {2401.14453},
 primaryClass = {astro-ph.GA},
       adsurl = {https://ui.adsabs.harvard.edu/abs/2024ApJ...967..133S}
}

@ARTICLE{Sun2020,
       author = {{Sun}, Jiayi and {Leroy}, Adam K. and {Schinnerer}, Eva and {Hughes}, Annie and {Rosolowsky}, Erik and {Querejeta}, Miguel and {Schruba}, Andreas and {Liu}, Daizhong and {Saito}, Toshiki and {Herrera}, Cinthya N. and {Faesi}, Christopher and {Usero}, Antonio and {Pety}, J{\'e}r{\^o}me and {Kruijssen}, J.~M. Diederik and {Ostriker}, Eve C. and {Bigiel}, Frank and {Blanc}, Guillermo A. and {Bolatto}, Alberto D. and {Boquien}, M{\'e}d{\'e}ric and {Chevance}, M{\'e}lanie and {Dale}, Daniel A. and {Deger}, Sinan and {Emsellem}, Eric and {Glover}, Simon C.~O. and {Grasha}, Kathryn and {Groves}, Brent and {Henshaw}, Jonathan and {Jimenez-Donaire}, Maria J. and {Kim}, Jenny J. and {Klessen}, Ralf S. and {Kreckel}, Kathryn and {Lee}, Janice C. and {Meidt}, Sharon and {Sandstrom}, Karin and {Sardone}, Amy E. and {Utomo}, Dyas and {Williams}, Thomas G.},
        title = "{Molecular Gas Properties on Cloud Scales across the Local Star-forming Galaxy Population}",
      journal = {\apjl},
         year = 2020,
        month = sep,
       volume = {901},
       number = {1},
          eid = {L8},
        pages = {L8},
          doi = {10.3847/2041-8213/abb3be},
archivePrefix = {arXiv},
       eprint = {2009.01842},
 primaryClass = {astro-ph.GA},
       adsurl = {https://ui.adsabs.harvard.edu/abs/2020ApJ...901L...8S}
}

@ARTICLE{Stuber2023,
       author = {{Stuber}, Sophia K. and {Schinnerer}, Eva and {Williams}, Thomas G. and {Querejeta}, Miguel and {Meidt}, Sharon and {Emsellem}, {\'E}ric and {Barnes}, Ashley and {Klessen}, Ralf S. and {Leroy}, Adam K. and {Neumann}, Justus and {Sormani}, Mattia C. and {Bigiel}, Frank and {Chevance}, M{\'e}lanie and {Dale}, Danny and {Faesi}, Christopher and {Glover}, Simon C.~O. and {Grasha}, Kathryn and {Kruijssen}, J.~M. Diederik and {Liu}, Daizhong and {Pan}, Hsi-an and {Pety}, J{\'e}r{\^o}me and {Pinna}, Francesca and {Saito}, Toshiki and {Usero}, Antonio and {Watkins}, Elizabeth J.},
        title = "{The gas morphology of nearby star-forming galaxies}",
      journal = {\aap},
         year = 2023,
        month = aug,
       volume = {676},
          eid = {A113},
        pages = {A113},
          doi = {10.1051/0004-6361/202346318},
archivePrefix = {arXiv},
       eprint = {2305.17172},
 primaryClass = {astro-ph.GA},
       adsurl = {https://ui.adsabs.harvard.edu/abs/2023A&A...676A.113S}
}

@ARTICLE{Kruijssen2013,
       author = {{Kruijssen}, J.~M. Diederik and {Longmore}, Steven N.},
        title = "{Comparing molecular gas across cosmic time-scales: the Milky Way as both a typical spiral galaxy and a high-redshift galaxy analogue}",
      journal = {\mnras},
         year = 2013,
        month = nov,
       volume = {435},
       number = {3},
        pages = {2598-2603},
          doi = {10.1093/mnras/stt1634},
archivePrefix = {arXiv},
       eprint = {1309.0505},
 primaryClass = {astro-ph.CO},
       adsurl = {https://ui.adsabs.harvard.edu/abs/2013MNRAS.435.2598K}
}

@ARTICLE{Longmore2013,
       author = {{Longmore}, S.~N. and {Bally}, J. and {Testi}, L. and {Purcell}, C.~R. and {Walsh}, A.~J. and {Bressert}, E. and {Pestalozzi}, M. and {Molinari}, S. and {Ott}, J. and {Cortese}, L. and {Battersby}, C. and {Murray}, N. and {Lee}, E. and {Kruijssen}, J.~M.~D. and {Schisano}, E. and {Elia}, D.},
        title = "{Variations in the Galactic star formation rate and density thresholds for star formation}",
      journal = {\mnras},
         year = 2013,
        month = feb,
       volume = {429},
       number = {2},
        pages = {987-1000},
          doi = {10.1093/mnras/sts376},
archivePrefix = {arXiv},
       eprint = {1208.4256},
 primaryClass = {astro-ph.GA},
       adsurl = {https://ui.adsabs.harvard.edu/abs/2013MNRAS.429..987L}
}

@ARTICLE{Molinari2011,
       author = {{Molinari}, S. and {Bally}, J. and {Noriega-Crespo}, A. and {Compi{\`e}gne}, M. and {Bernard}, J.~P. and {Paradis}, D. and {Martin}, P. and {Testi}, L. and {Barlow}, M. and {Moore}, T. and {Plume}, R. and {Swinyard}, B. and {Zavagno}, A. and {Calzoletti}, L. and {Di Giorgio}, A.~M. and {Elia}, D. and {Faustini}, F. and {Natoli}, P. and {Pestalozzi}, M. and {Pezzuto}, S. and {Piacentini}, F. and {Polenta}, G. and {Polychroni}, D. and {Schisano}, E. and {Traficante}, A. and {Veneziani}, M. and {Battersby}, C. and {Burton}, M. and {Carey}, S. and {Fukui}, Y. and {Li}, J.~Z. and {Lord}, S.~D. and {Morgan}, L. and {Motte}, F. and {Schuller}, F. and {Stringfellow}, G.~S. and {Tan}, J.~C. and {Thompson}, M.~A. and {Ward-Thompson}, D. and {White}, G. and {Umana}, G.},
        title = "{A 100 pc Elliptical and Twisted Ring of Cold and Dense Molecular Clouds Revealed by Herschel Around the Galactic Center}",
      journal = {\apjl},
         year = 2011,
        month = jul,
       volume = {735},
       number = {2},
          eid = {L33},
        pages = {L33},
          doi = {10.1088/2041-8205/735/2/L33},
archivePrefix = {arXiv},
       eprint = {1105.5486},
 primaryClass = {astro-ph.GA},
       adsurl = {https://ui.adsabs.harvard.edu/abs/2011ApJ...735L..33M}
}

@ARTICLE{Sofue1995,
       author = {{Sofue}, Yoshiaki},
        title = "{Galactic-Center Molecular Arms, Ring, and Expanding Shell. I. Kinematical Structures in Longitude--Velocity Diagrams}",
      journal = {\pasj},
         year = 1995,
        month = oct,
       volume = {47},
        pages = {527-549},
archivePrefix = {arXiv},
       eprint = {astro-ph/9508110},
 primaryClass = {astro-ph},
       adsurl = {https://ui.adsabs.harvard.edu/abs/1995PASJ...47..527S}
}

@ARTICLE{gramze2023,
       author = {{Gramze}, Savannah R. and {Ginsburg}, Adam and {Meier}, David S. and {Ott}, Juergen and {Shirley}, Yancy and {Sormani}, Mattia C. and {Svoboda}, Brian E.},
        title = "{Evidence of a Cloud-Cloud Collision from Overshooting Gas in the Galactic Center}",
      journal = {\apj},
         year = 2023,
        month = dec,
       volume = {959},
       number = {2},
          eid = {93},
        pages = {93},
          doi = {10.3847/1538-4357/ad01be},
archivePrefix = {arXiv},
       eprint = {2309.16403},
 primaryClass = {astro-ph.GA},
       adsurl = {https://ui.adsabs.harvard.edu/abs/2023ApJ...959...93G}
}

@article{astropy:2013,
Adsurl = {http://adsabs.harvard.edu/abs/2013A%26A...558A..33A},
Archiveprefix = {arXiv},
Author = {{Astropy Collaboration} and {Robitaille}, T.~P. and {Tollerud}, E.~J. and {Greenfield}, P. and {Droettboom}, M. and {Bray}, E. and {Aldcroft}, T. and {Davis}, M. and {Ginsburg}, A. and {Price-Whelan}, A.~M. and {Kerzendorf}, W.~E. and {Conley}, A. and {Crighton}, N. and {Barbary}, K. and {Muna}, D. and {Ferguson}, H. and {Grollier}, F. and {Parikh}, M.~M. and {Nair}, P.~H. and {Unther}, H.~M. and {Deil}, C. and {Woillez}, J. and {Conseil}, S. and {Kramer}, R. and {Turner}, J.~E.~H. and {Singer}, L. and {Fox}, R. and {Weaver}, B.~A. and {Zabalza}, V. and {Edwards}, Z.~I. and {Azalee Bostroem}, K. and {Burke}, D.~J. and {Casey}, A.~R. and {Crawford}, S.~M. and {Dencheva}, N. and {Ely}, J. and {Jenness}, T. and {Labrie}, K. and {Lim}, P.~L. and {Pierfederici}, F. and {Pontzen}, A. and {Ptak}, A. and {Refsdal}, B. and {Servillat}, M. and {Streicher}, O.},
Doi = {10.1051/0004-6361/201322068},
Eid = {A33},
Eprint = {1307.6212},
Journal = {\aap},
Month = oct,
Pages = {A33},
Primaryclass = {astro-ph.IM},
Title = {{Astropy: A community Python package for astronomy}},
Volume = 558,
Year = 2013}

@ARTICLE{astropy:2018,
       author = {{Astropy Collaboration} and {Price-Whelan}, A.~M. and
         {Sip{\H{o}}cz}, B.~M. and {G{\"u}nther}, H.~M. and {Lim}, P.~L. and
         {Crawford}, S.~M. and {Conseil}, S. and {Shupe}, D.~L. and
         {Craig}, M.~W. and {Dencheva}, N. and {Ginsburg}, A. and {Vand
        erPlas}, J.~T. and {Bradley}, L.~D. and {P{\'e}rez-Su{\'a}rez}, D. and
         {de Val-Borro}, M. and {Aldcroft}, T.~L. and {Cruz}, K.~L. and
         {Robitaille}, T.~P. and {Tollerud}, E.~J. and {Ardelean}, C. and
         {Babej}, T. and {Bach}, Y.~P. and {Bachetti}, M. and {Bakanov}, A.~V. and
         {Bamford}, S.~P. and {Barentsen}, G. and {Barmby}, P. and
         {Baumbach}, A. and {Berry}, K.~L. and {Biscani}, F. and {Boquien}, M. and
         {Bostroem}, K.~A. and {Bouma}, L.~G. and {Brammer}, G.~B. and
         {Bray}, E.~M. and {Breytenbach}, H. and {Buddelmeijer}, H. and
         {Burke}, D.~J. and {Calderone}, G. and {Cano Rodr{\'\i}guez}, J.~L. and
         {Cara}, M. and {Cardoso}, J.~V.~M. and {Cheedella}, S. and {Copin}, Y. and
         {Corrales}, L. and {Crichton}, D. and {D'Avella}, D. and {Deil}, C. and
         {Depagne}, {\'E}. and {Dietrich}, J.~P. and {Donath}, A. and
         {Droettboom}, M. and {Earl}, N. and {Erben}, T. and {Fabbro}, S. and
         {Ferreira}, L.~A. and {Finethy}, T. and {Fox}, R.~T. and
         {Garrison}, L.~H. and {Gibbons}, S.~L.~J. and {Goldstein}, D.~A. and
         {Gommers}, R. and {Greco}, J.~P. and {Greenfield}, P. and
         {Groener}, A.~M. and {Grollier}, F. and {Hagen}, A. and {Hirst}, P. and
         {Homeier}, D. and {Horton}, A.~J. and {Hosseinzadeh}, G. and {Hu}, L. and
         {Hunkeler}, J.~S. and {Ivezi{\'c}}, {\v{Z}}. and {Jain}, A. and
         {Jenness}, T. and {Kanarek}, G. and {Kendrew}, S. and {Kern}, N.~S. and
         {Kerzendorf}, W.~E. and {Khvalko}, A. and {King}, J. and {Kirkby}, D. and
         {Kulkarni}, A.~M. and {Kumar}, A. and {Lee}, A. and {Lenz}, D. and
         {Littlefair}, S.~P. and {Ma}, Z. and {Macleod}, D.~M. and
         {Mastropietro}, M. and {McCully}, C. and {Montagnac}, S. and
         {Morris}, B.~M. and {Mueller}, M. and {Mumford}, S.~J. and {Muna}, D. and
         {Murphy}, N.~A. and {Nelson}, S. and {Nguyen}, G.~H. and
         {Ninan}, J.~P. and {N{\"o}the}, M. and {Ogaz}, S. and {Oh}, S. and
         {Parejko}, J.~K. and {Parley}, N. and {Pascual}, S. and {Patil}, R. and
         {Patil}, A.~A. and {Plunkett}, A.~L. and {Prochaska}, J.~X. and
         {Rastogi}, T. and {Reddy Janga}, V. and {Sabater}, J. and
         {Sakurikar}, P. and {Seifert}, M. and {Sherbert}, L.~E. and
         {Sherwood-Taylor}, H. and {Shih}, A.~Y. and {Sick}, J. and
         {Silbiger}, M.~T. and {Singanamalla}, S. and {Singer}, L.~P. and
         {Sladen}, P.~H. and {Sooley}, K.~A. and {Sornarajah}, S. and
         {Streicher}, O. and {Teuben}, P. and {Thomas}, S.~W. and
         {Tremblay}, G.~R. and {Turner}, J.~E.~H. and {Terr{\'o}n}, V. and
         {van Kerkwijk}, M.~H. and {de la Vega}, A. and {Watkins}, L.~L. and
         {Weaver}, B.~A. and {Whitmore}, J.~B. and {Woillez}, J. and
         {Zabalza}, V. and {Astropy Contributors}},
        title = "{The Astropy Project: Building an Open-science Project and Status of the v2.0 Core Package}",
      journal = {\aj},
         year = 2018,
        month = sep,
       volume = {156},
       number = {3},
          eid = {123},
        pages = {123},
          doi = {10.3847/1538-3881/aabc4f},
archivePrefix = {arXiv},
       eprint = {1801.02634},
 primaryClass = {astro-ph.IM},
       adsurl = {https://ui.adsabs.harvard.edu/abs/2018AJ....156..123A}
}

@ARTICLE{astropy:2022,
       author = {{Astropy Collaboration} and {Price-Whelan}, Adrian M. and {Lim}, Pey Lian and {Earl}, Nicholas and {Starkman}, Nathaniel and {Bradley}, Larry and {Shupe}, David L. and {Patil}, Aarya A. and {Corrales}, Lia and {Brasseur}, C.~E. and {N{"o}the}, Maximilian and {Donath}, Axel and {Tollerud}, Erik and {Morris}, Brett M. and {Ginsburg}, Adam and {Vaher}, Eero and {Weaver}, Benjamin A. and {Tocknell}, James and {Jamieson}, William and {van Kerkwijk}, Marten H. and {Robitaille}, Thomas P. and {Merry}, Bruce and {Bachetti}, Matteo and {G{"u}nther}, H. Moritz and {Aldcroft}, Thomas L. and {Alvarado-Montes}, Jaime A. and {Archibald}, Anne M. and {B{'o}di}, Attila and {Bapat}, Shreyas and {Barentsen}, Geert and {Baz{'a}n}, Juanjo and {Biswas}, Manish and {Boquien}, M{'e}d{'e}ric and {Burke}, D.~J. and {Cara}, Daria and {Cara}, Mihai and {Conroy}, Kyle E. and {Conseil}, Simon and {Craig}, Matthew W. and {Cross}, Robert M. and {Cruz}, Kelle L. and {D'Eugenio}, Francesco and {Dencheva}, Nadia and {Devillepoix}, Hadrien A.~R. and {Dietrich}, J{"o}rg P. and {Eigenbrot}, Arthur Davis and {Erben}, Thomas and {Ferreira}, Leonardo and {Foreman-Mackey}, Daniel and {Fox}, Ryan and {Freij}, Nabil and {Garg}, Suyog and {Geda}, Robel and {Glattly}, Lauren and {Gondhalekar}, Yash and {Gordon}, Karl D. and {Grant}, David and {Greenfield}, Perry and {Groener}, Austen M. and {Guest}, Steve and {Gurovich}, Sebastian and {Handberg}, Rasmus and {Hart}, Akeem and {Hatfield-Dodds}, Zac and {Homeier}, Derek and {Hosseinzadeh}, Griffin and {Jenness}, Tim and {Jones}, Craig K. and {Joseph}, Prajwel and {Kalmbach}, J. Bryce and {Karamehmetoglu}, Emir and {Ka{l}uszy{'n}ski}, Miko{l}aj and {Kelley}, Michael S.~P. and {Kern}, Nicholas and {Kerzendorf}, Wolfgang E. and {Koch}, Eric W. and {Kulumani}, Shankar and {Lee}, Antony and {Ly}, Chun and {Ma}, Zhiyuan and {MacBride}, Conor and {Maljaars}, Jakob M. and {Muna}, Demitri and {Murphy}, N.~A. and {Norman}, Henrik and {O'Steen}, Richard and {Oman}, Kyle A. and {Pacifici}, Camilla and {Pascual}, Sergio and {Pascual-Granado}, J. and {Patil}, Rohit R. and {Perren}, Gabriel I. and {Pickering}, Timothy E. and {Rastogi}, Tanuj and {Roulston}, Benjamin R. and {Ryan}, Daniel F. and {Rykoff}, Eli S. and {Sabater}, Jose and {Sakurikar}, Parikshit and {Salgado}, Jes{'u}s and {Sanghi}, Aniket and {Saunders}, Nicholas and {Savchenko}, Volodymyr and {Schwardt}, Ludwig and {Seifert-Eckert}, Michael and {Shih}, Albert Y. and {Jain}, Anany Shrey and {Shukla}, Gyanendra and {Sick}, Jonathan and {Simpson}, Chris and {Singanamalla}, Sudheesh and {Singer}, Leo P. and {Singhal}, Jaladh and {Sinha}, Manodeep and {Sip{H{o}}cz}, Brigitta M. and {Spitler}, Lee R. and {Stansby}, David and {Streicher}, Ole and {{{S}}umak}, Jani and {Swinbank}, John D. and {Taranu}, Dan S. and {Tewary}, Nikita and {Tremblay}, Grant R. and {Val-Borro}, Miguel de and {Van Kooten}, Samuel J. and {Vasovi{'c}}, Zlatan and {Verma}, Shresth and {de Miranda Cardoso}, Jos{'e} Vin{'i}cius and {Williams}, Peter K.~G. and {Wilson}, Tom J. and {Winkel}, Benjamin and {Wood-Vasey}, W.~M. and {Xue}, Rui and {Yoachim}, Peter and {Zhang}, Chen and {Zonca}, Andrea and {Astropy Project Contributors}},
        title = "{The Astropy Project: Sustaining and Growing a Community-oriented Open-source Project and the Latest Major Release (v5.0) of the Core Package}",
      journal = {apj},
         year = 2022,
        month = aug,
       volume = {935},
       number = {2},
          eid = {167},
        pages = {167},
          doi = {10.3847/1538-4357/ac7c74},
archivePrefix = {arXiv},
       eprint = {2206.14220},
 primaryClass = {astro-ph.IM},
       adsurl = {https://ui.adsabs.harvard.edu/abs/2022ApJ...935..167A}
}

@ARTICLE{Jones2012,
       author = {{Jones}, P.~A. and {Burton}, M.~G. and {Cunningham}, M.~R. and {Requena-Torres}, M.~A. and {Menten}, K.~M. and {Schilke}, P. and {Belloche}, A. and {Leurini}, S. and {Mart{\'\i}n-Pintado}, J. and {Ott}, J. and {Walsh}, A.~J.},
        title = "{Spectral imaging of the Central Molecular Zone in multiple 3-mm molecular lines}",
      journal = {\mnras},
         year = 2012,
        month = feb,
       volume = {419},
       number = {4},
        pages = {2961-2986},
          doi = {10.1111/j.1365-2966.2011.19941.x},
archivePrefix = {arXiv},
       eprint = {1110.1421},
 primaryClass = {astro-ph.GA},
       adsurl = {https://ui.adsabs.harvard.edu/abs/2012MNRAS.419.2961J}
}

@ARTICLE{Nogueras2022,
       author = {{Nogueras-Lara}, F.},
        title = "{A first glimpse at the line-of-sight structure of the Milky Way's nuclear stellar disc}",
      journal = {\aap},
         year = 2022,
        month = dec,
       volume = {668},
          eid = {L8},
        pages = {L8},
          doi = {10.1051/0004-6361/202244934},
archivePrefix = {arXiv},
       eprint = {2212.00047},
 primaryClass = {astro-ph.GA},
       adsurl = {https://ui.adsabs.harvard.edu/abs/2022A&A...668L...8N}
}

@ARTICLE{Contopoulos_Grosbol_1989,
       author = {{Contopoulos}, G. and {Grosbol}, P.},
        title = "{Orbits in barred galaxies}",
      journal = {\aapr},
         year = 1989,
        month = nov,
       volume = {1},
       number = {3-4},
        pages = {261-289},
          doi = {10.1007/BF00873080},
       adsurl = {https://ui.adsabs.harvard.edu/abs/1989A&ARv...1..261C}
}

@ARTICLE{Buta_Combes1996,
       author = {{Buta}, R. and {Combes}, F.},
        title = "{Galactic Rings}",
      journal = {\fcp},
         year = 1996,
        month = jan,
       volume = {17},
        pages = {95-281},
       adsurl = {https://ui.adsabs.harvard.edu/abs/1996FCPh...17...95B}
}

@ARTICLE{Martinez-Arranz2022,
       author = {{Mart{\'\i}nez-Arranz}, {\'A}. and {Sch{\"o}del}, R. and {Nogueras-Lara}, F. and {Shahzamanian}, B.},
        title = "{Distance to the Brick cloud using stellar kinematics}",
      journal = {\aap},
         year = 2022,
        month = apr,
       volume = {660},
          eid = {L3},
        pages = {L3},
          doi = {10.1051/0004-6361/202243263},
archivePrefix = {arXiv},
       eprint = {2204.02109},
 primaryClass = {astro-ph.GA},
       adsurl = {https://ui.adsabs.harvard.edu/abs/2022A&A...660L...3M}
}

@ARTICLE{Zoccali2021,
       author = {{Zoccali}, M. and {Valenti}, E. and {Surot}, F. and {Gonzalez}, O.~A. and {Renzini}, A. and {Valenzuela Navarro}, A.},
        title = "{A new distance to the Brick, the dense molecular cloud G0.253+0.016}",
      journal = {\mnras},
         year = 2021,
        month = mar,
       volume = {502},
       number = {1},
        pages = {1246-1252},
          doi = {10.1093/mnras/stab089},
archivePrefix = {arXiv},
       eprint = {2101.04022},
 primaryClass = {astro-ph.GA},
       adsurl = {https://ui.adsabs.harvard.edu/abs/2021MNRAS.502.1246Z}
}

@ARTICLE{Ginsburg2016,
       author = {{Ginsburg}, Adam and {Henkel}, Christian and {Ao}, Yiping and {Riquelme}, Denise and {Kauffmann}, Jens and {Pillai}, Thushara and {Mills}, Elisabeth A.~C. and {Requena-Torres}, Miguel A. and {Immer}, Katharina and {Testi}, Leonardo and {Ott}, Juergen and {Bally}, John and {Battersby}, Cara and {Darling}, Jeremy and {Aalto}, Susanne and {Stanke}, Thomas and {Kendrew}, Sarah and {Kruijssen}, J.~M. Diederik and {Longmore}, Steven and {Dale}, James and {Guesten}, Rolf and {Menten}, Karl M.},
        title = "{Dense gas in the Galactic central molecular zone is warm and heated by turbulence}",
      journal = {\aap},
         year = 2016,
        month = feb,
       volume = {586},
          eid = {A50},
        pages = {A50},
          doi = {10.1051/0004-6361/201526100},
archivePrefix = {arXiv},
       eprint = {1509.01583},
 primaryClass = {astro-ph.GA},
       adsurl = {https://ui.adsabs.harvard.edu/abs/2016A&A...586A..50G}
}

@ARTICLE{Sofue2025,
       author = {{Sofue}, Y. and {Oka}, Tomo. and {Longmore}, S.~N. and {Walker}, D. and {Ginsburg}, A. and {Henshaw}, J.~D. and {Bally}, J. and {Barnes}, A.~T. and {Battersby}, C. and {Colzi}, L. and {Ho}, P. and {Jimenez-Serra}, I. and {Kruijssen}, J.~M.~D. and {Mills}, E. and {Petkova}, M.~A. and {Sormani}, M.~C. and {Wallace}, J. and {Armijos-Abendano}, J. and {Dutkowska}, K.~M. and {Enokiya}, R. and {Fukui}, Y. and {Garcia}, P. and {Guzman}, A. and {Henkel}, C. and {Hsieh}, P. -Y. and {Hu}, Y. and {Immer}, K. and {Jeff}, D. and {Klessen}, R.~S. and {Kohno}, K. and {Krumholz}, M.~R. and {Lipman}, D. and {Martin}, S. and {Morris}, M.~R. and {Nogueras-Lara}, F. and {Nonhebel}, M. and {Ott}, J. and {Pineda}, J.~E. and {Requena-Torres}, M.~A. and {Rivilla}, V.~M. and {Riquelme-Vasquez}, D. and {Sanchez-Monge}, A. and {Santa-Maria}, M.~G. and {Smith}, H.~A. and {Tanvir}, T.~S. and {Tolls}, V. and {Wang}, Q.~D.},
        title = "{The Galactic-Centre Arms inferred from ACES (ALMA CMZ Exploration Survey)}",
      journal = {arXiv e-prints},
         year = 2025,
        month = apr,
          eid = {arXiv:2504.03331},
        pages = {arXiv:2504.03331},
          doi = {10.48550/arXiv.2504.03331},
archivePrefix = {arXiv},
       eprint = {2504.03331},
 primaryClass = {astro-ph.GA},
       adsurl = {https://ui.adsabs.harvard.edu/abs/2025arXiv250403331S}
}

@ARTICLE{Tress2024,
       author = {{Tress}, R.~G. and {Sormani}, M.~C. and {Girichidis}, P. and {Glover}, S.~C.~O. and {Klessen}, R.~S. and {Smith}, R.~J. and {Sobacchi}, E. and {Armillotta}, L. and {Barnes}, A.~T. and {Battersby}, C. and {Bogue}, K.~R.~J. and {Brucy}, N. and {Colzi}, L. and {Federrath}, C. and {Garc{\'\i}a}, P. and {Ginsburg}, A. and {G{\"o}ller}, J. and {Hatchfield}, H.~P. and {Henkel}, C. and {Hennebelle}, P. and {Henshaw}, J.~D. and {Hirschmann}, M. and {Hu}, Y. and {Kauffmann}, J. and {Kruijssen}, J.~M.~D. and {Lazarian}, A. and {Lipman}, D. and {Longmore}, S.~N. and {Morris}, M.~R. and {Nogueras-Lara}, F. and {Petkova}, M.~A. and {Pillai}, T.~G.~S. and {Rivilla}, V.~M. and {S{\'a}nchez-Monge}, {\'A}. and {Soler}, J.~D. and {Whitworth}, D. and {Zhang}, Q.},
        title = "{Magnetic field morphology and evolution in the Central Molecular Zone and its effect on gas dynamics}",
      journal = {\aap},
         year = 2024,
        month = nov,
       volume = {691},
          eid = {A303},
        pages = {A303},
          doi = {10.1051/0004-6361/202450035},
archivePrefix = {arXiv},
       eprint = {2403.13048},
 primaryClass = {astro-ph.GA},
       adsurl = {https://ui.adsabs.harvard.edu/abs/2024A&A...691A.303T}
}

@article{Walker_2025,
doi = {10.3847/1538-4357/adb5ef},
url = {https://dx.doi.org/10.3847/1538-4357/adb5ef},
year = {2025},
month = {may},
publisher = {The American Astronomical Society},
volume = {984},
number = {2},
pages = {158},
author = {Walker, Daniel L. and Battersby, Cara and Lipman, Dani and Sormani, Mattia C. and Ginsburg, Adam and Glover, Simon C. O. and Henshaw, Jonathan D. and Longmore, Steven N. and Klessen, Ralf S. and Immer, Katharina and Alboslani, Danya and Bally, John and Barnes, Ashley and Hatchfield, H Perry and Mills, Elisabeth A. C. and Smith, Rowan and Tress, Robin G. and Zhang, Qizhou},
title = {3D CMZ. III. Constraining the 3D Structure of the Central Molecular Zone via Molecular Line Emission and Absorption},
journal = {The Astrophysical Journal}
}

@article{Battersby_2025_3DCMZII,
doi = {10.3847/1538-4357/adb844},
url = {https://dx.doi.org/10.3847/1538-4357/adb844},
year = {2025},
month = {may},
publisher = {The American Astronomical Society},
volume = {984},
number = {2},
pages = {157},
author = {Battersby, Cara and Walker, Daniel L. and Barnes, Ashley and Ginsburg, Adam and Lipman, Dani and Alboslani, Danya and Hatchfield, H Perry and Bally, John and Glover, Simon C. O. and Henshaw, Jonathan D. and Immer, Katharina and Klessen, Ralf S. and Longmore, Steven N. and Mills, Elisabeth A. C. and Molinari, Sergio and Smith, Rowan and Sormani, Mattia C. and Tress, Robin G. and Zhang, Qizhou},
title = {3D CMZ. II. Hierarchical Structure Analysis of the Central Molecular Zone},
journal = {The Astrophysical Journal}
}

@article{Battersby_2025_3DCMZI,
doi = {10.3847/1538-4357/adb5f0},
url = {https://dx.doi.org/10.3847/1538-4357/adb5f0},
year = {2025},
month = {may},
publisher = {The American Astronomical Society},
volume = {984},
number = {2},
pages = {156},
author = {Battersby, Cara and Walker, Daniel L. and Barnes, Ashley and Ginsburg, Adam and Lipman, Dani and Alboslani, Danya and Hatchfield, H Perry and Bally, John and Glover, Simon C. O. and Henshaw, Jonathan D. and Immer, Katharina and Klessen, Ralf S. and Longmore, Steven N. and Mills, Elisabeth A. C. and Molinari, Sergio and Smith, Rowan and Sormani, Mattia C. and Tress, Robin G. and Zhang, Qizhou},
title = {3D CMZ. I. Central Molecular Zone Overview},
journal = {The Astrophysical Journal}
}

@article{Lee2008,
doi = {10.1086/524128},
url = {https://dx.doi.org/10.1086/524128},
year = {2008},
month = {feb},
publisher = {},
volume = {674},
number = {1},
pages = {247},
author = {Sungho Lee and Soojong Pak and Minho Choi and Christopher J. Davis and T. R. Geballe and Robeson M. Herrnstein and Paul T. P. Ho and Y. C. Minh and Sang-Gak Lee},
title = {Three-Dimensional Observations of H2 Emission around Sgr A East. I. Structure in the Central 10 pc of Our Galaxy},
journal = {The Astrophysical Journal}
}

@ARTICLE{Karlsson2003,
       author = {{Karlsson}, R. and {Sjouwerman}, L.~O. and {Sandqvist}, Aa. and {Whiteoak}, J.~B.},
        title = "{18-cm VLA observations of OH towards the Galactic Centre. Absorption and emission in the four ground-state OH lines}",
      journal = {\aap},
         year = 2003,
        month = jun,
       volume = {403},
        pages = {1011-1021},
          doi = {10.1051/0004-6361:20030309},
       adsurl = {https://ui.adsabs.harvard.edu/abs/2003A&A...403.1011K}
}

@ARTICLE{Ryu2013,
       author = {{Ryu}, Syukyo Gando and {Nobukawa}, Masayoshi and {Nakashima}, Shinya and {Tsuru}, Takeshi Go and {Koyama}, Katsuji and {Uchiyama}, Hideki},
        title = "{X-Ray Echo from the Sagittarius C Complex and 500-year Activity History of Sagittarius A*}",
      journal = {\pasj},
         year = 2013,
        month = apr,
       volume = {65},
       number = {2},
          eid = {33},
        pages = {33},
          doi = {10.1093/pasj/65.2.33},
archivePrefix = {arXiv},
       eprint = {1211.4529},
 primaryClass = {astro-ph.GA},
       adsurl = {https://ui.adsabs.harvard.edu/abs/2013PASJ...65...33R}
}

@article{Tsuboi2018,
	adsurl = {https://ui.adsabs.harvard.edu/abs/2018PASJ...70...85T},
	archiveprefix = {arXiv},
	author = {{Tsuboi}, Masato and {Kitamura}, Yoshimi and {Uehara}, Kenta and {Tsutsumi}, Takahiro and {Miyawaki}, Ryosuke and {Miyoshi}, Makoto and {Miyazaki}, Atsushi},
	doi = {10.1093/pasj/psy080},
	eid = {85},
	eprint = {1806.10246},
	journal = {\pasj},
	month = oct,
	number = {5},
	pages = {85},
	primaryclass = {astro-ph.GA},
	title = {{ALMA view of the circumnuclear disk of the Galactic Center: tidally disrupted molecular clouds falling to the Galactic Center}},
	volume = {70},
	year = 2018}

@article{HerrnsteinHo2005,
	adsurl = {https://ui.adsabs.harvard.edu/abs/2005ApJ...620..287H},
	archiveprefix = {arXiv},
	author = {{Herrnstein}, Robeson M. and {Ho}, Paul T.~P.},
	doi = {10.1086/426047},
	eprint = {astro-ph/0409271},
	journal = {\apj},
	month = feb,
	number = {1},
	pages = {287-307},
	primaryclass = {astro-ph},
	title = {{The Nature of the Molecular Environment within 5 Parsecs of the Galactic Center}},
	volume = {620},
	year = 2005,
	}

@article{Genzel1990,
	adsurl = {https://ui.adsabs.harvard.edu/abs/1990ApJ...356..160G},
	author = {{Genzel}, R. and {Stacey}, G.~J. and {Harris}, A.~I. and {Townes}, C.~H. and {Geis}, N. and {Graf}, U.~U. and {Poglitsch}, A. and {Stutzki}, J.},
	doi = {10.1086/168827},
	journal = {\apj},
	month = jun,
	pages = {160},
	title = {{Far-Infrared, Submillimeter, and Millimeter Spectroscopy of the Galactic Center: Radio Arc and +20/+50 Kilometer per Second Clouds}},
	volume = {356},
	year = 1990}

@ARTICLE{Yan2017,
       author = {{Yan}, Qing-Zeng and {Walsh}, A.~J. and {Dawson}, J.~R. and {Macquart}, J.~P. and {Blackwell}, R. and {Burton}, M.~G. and {Rowell}, G.~P. and {Zhang}, Bo and {Xu}, Ye and {Tang}, Zheng-Hong and {Hancock}, P.~J.},
        title = "{Towards a three-dimensional distribution of the molecular clouds in the Galactic Centre}",
      journal = {\mnras},
         year = 2017,
        month = nov,
       volume = {471},
       number = {3},
        pages = {2523-2536},
          doi = {10.1093/mnras/stx1724},
archivePrefix = {arXiv},
       eprint = {1707.02378},
 primaryClass = {astro-ph.GA},
       adsurl = {https://ui.adsabs.harvard.edu/abs/2017MNRAS.471.2523Y}
}

@ARTICLE{Chaves-Velasquez2025,
       author = {{Chaves-Velasquez}, Leonardo and {G{\'o}mez}, Gilberto C. and {P{\'e}rez-Villegas}, {\'A}ngeles},
        title = "{Gas dynamics in the central molecular zone and its connection with the galactic bar}",
      journal = {\pasa},
         year = 2025,
        month = jan,
       volume = {42},
          eid = {e014},
        pages = {e014},
          doi = {10.1017/pasa.2024.130},
archivePrefix = {arXiv},
       eprint = {2411.05684},
 primaryClass = {astro-ph.GA},
       adsurl = {https://ui.adsabs.harvard.edu/abs/2025PASA...42...14C}
}

@ARTICLE{Kumar2023,
       author = {{Kumar}, Kaushal},
        title = "{Exploring Optimization Techniques for Parameter Estimation in Nonlinear System Modeling}",
      journal = {arXiv e-prints},
         year = 2023,
        month = apr,
          eid = {arXiv:2305.00351},
        pages = {arXiv:2305.00351},
          doi = {10.48550/arXiv.2305.00351},
archivePrefix = {arXiv},
       eprint = {2305.00351},
 primaryClass = {math.OC},
       adsurl = {https://ui.adsabs.harvard.edu/abs/2023arXiv230500351K}
}

@software{LMFIT,
  author       = {Newville, Matthew and
                  Otten, Renee and
                  Nelson, Andrew and
                  Stensitzki, Till and
                  Ingargiola, Antonino and
                  Allan, Daniel and
                  Fox, Austin and
                  Carter, Faustin and
                  Rawlik, Michal},
  title        = {LMFIT: Non-Linear Least-Squares Minimization and
                   Curve-Fitting for Python
                  },
  month        = mar,
  year         = 2025,
  publisher    = {Zenodo},
  version      = {1.3.3},
  doi          = {10.5281/zenodo.15014437},
  url          = {https://doi.org/10.5281/zenodo.15014437},
  swhid        = {swh:1:dir:6dbda1361832412880d315cbf608ba498c177d82
                   ;origin=https://doi.org/10.5281/zenodo.598352;visi
                   t=swh:1:snp:0d236d4d8ff7f7248600297b145d047734607d
                   14;anchor=swh:1:rel:5f2a70f7c390d9b78cb661b349b044
                   753cf7e89e;path=lmfit-lmfit-py-f97ddf7
                  },
}

@article{Lipman_2025,
doi = {10.3847/1538-4357/adb5ee},
url = {https://dx.doi.org/10.3847/1538-4357/adb5ee},
year = {2025},
month = {may},
publisher = {The American Astronomical Society},
volume = {984},
number = {2},
pages = {159},
author = {Lipman, Dani and Battersby, Cara and Walker, Daniel L. and Sormani, Mattia C. and Bally, John and Barnes, Ashley and Ginsburg, Adam and Glover, Simon C. O. and Henshaw, Jonathan D. and Hatchfield, H Perry and Immer, Katharina and Klessen, Ralf S. and Longmore, Steven N. and Mills, Elisabeth A. C. and Smith, Rowan and Tress, R. G. and Alboslani, Danya and Zhang, Qizhou},
title = {3D CMZ. IV. Distinguishing Near versus Far Distances in the Galactic Center Using Spitzer and Herschel},
journal = {The Astrophysical Journal}
}

@ARTICLE{Sunyaev_and_Churazov1998,
       author = {{Sunyaev}, R. and {Churazov}, E.},
        title = "{Equivalent width, shape and proper motion of the iron fluorescent line emission from molecular clouds as an indicator of the illuminating source X-ray flux history}",
      journal = {\mnras},
         year = 1998,
        month = jul,
       volume = {297},
       number = {4},
        pages = {1279-1291},
          doi = {10.1046/j.1365-8711.1998.01684.x},
archivePrefix = {arXiv},
       eprint = {astro-ph/9805038},
 primaryClass = {astro-ph},
       adsurl = {https://ui.adsabs.harvard.edu/abs/1998MNRAS.297.1279S}
}

@ARTICLE{Stel2025,
       author = {{Stel}, Giovanni and {Ponti}, Gabriele and {Haardt}, Francesco and {Sormani}, Mattia},
        title = "{25 years of XMM-Newton observations of the Sgr A complex: 3D distribution and internal structure of the clouds}",
      journal = {\aap},
         year = 2025,
        month = mar,
       volume = {695},
          eid = {A52},
        pages = {A52},
          doi = {10.1051/0004-6361/202451359},
archivePrefix = {arXiv},
       eprint = {2501.09737},
 primaryClass = {astro-ph.GA},
       adsurl = {https://ui.adsabs.harvard.edu/abs/2025A&A...695A..52S}
}

@ARTICLE{Marin2023,
       author = {{Marin}, Fr{\'e}d{\'e}ric and {Churazov}, Eugene and {Khabibullin}, Ildar and {Ferrazzoli}, Riccardo and {Di Gesu}, Laura and {Barnouin}, Thibault and {Di Marco}, Alessandro and {Middei}, Riccardo and {Vikhlinin}, Alexey and {Costa}, Enrico and {Soffitta}, Paolo and {Muleri}, Fabio and {Sunyaev}, Rashid and {Forman}, William and {Kraft}, Ralph and {Bianchi}, Stefano and {Donnarumma}, Immacolata and {Petrucci}, Pierre-Olivier and {Enoto}, Teruaki and {Agudo}, Iv{\'a}n and {Antonelli}, Lucio A. and {Bachetti}, Matteo and {Baldini}, Luca and {Baumgartner}, Wayne H. and {Bellazzini}, Ronaldo and {Bongiorno}, Stephen D. and {Bonino}, Raffaella and {Brez}, Alessandro and {Bucciantini}, Niccol{\`o} and {Capitanio}, Fiamma and {Castellano}, Simone and {Cavazzuti}, Elisabetta and {Chen}, Chien-Ting and {Ciprini}, Stefano and {De Rosa}, Alessandra and {Del Monte}, Ettore and {Di Lalla}, Niccol{\`o} and {Doroshenko}, Victor and {Dov{\v{c}}iak}, Michal and {Ehlert}, Steven R. and {Evangelista}, Yuri and {Fabiani}, Sergio and {Garcia}, Javier A. and {Gunji}, Shuichi and {Hayashida}, Kiyoshi and {Heyl}, Jeremy and {Ingram}, Adam and {Iwakiri}, Wataru and {Jorstad}, Svetlana G. and {Kaaret}, Philip and {Karas}, Vladimir and {Kitaguchi}, Takao and {Kolodziejczak}, Jeffery J. and {Krawczynski}, Henric and {La Monaca}, Fabio and {Latronico}, Luca and {Liodakis}, Ioannis and {Maldera}, Simone and {Manfreda}, Alberto and {Marinucci}, Andrea and {Marscher}, Alan P. and {Marshall}, Herman L. and {Massaro}, Francesco and {Matt}, Giorgio and {Mitsuishi}, Ikuyuki and {Mizuno}, Tsunefumi and {Negro}, Michela and {Ng}, C. -Y. and {O'Dell}, Stephen L. and {Omodei}, Nicola and {Oppedisano}, Chiara and {Papitto}, Alessandro and {Pavlov}, George G. and {Peirson}, Abel L. and {Perri}, Matteo and {Pesce-Rollins}, Melissa and {Pilia}, Maura and {Possenti}, Andrea and {Poutanen}, Juri and {Puccetti}, Simonetta and {Ramsey}, Brian D. and {Rankin}, John and {Ratheesh}, Ajay and {Roberts}, Oliver J. and {Romani}, Roger W. and {Sgr{\`o}}, Carmelo and {Slane}, Patrick and {Spandre}, Gloria and {Swartz}, Doug and {Tamagawa}, Toru and {Tavecchio}, Fabrizio and {Taverna}, Roberto and {Tawara}, Yuzuru and {Tennant}, Allyn F. and {Thomas}, Nicholas E. and {Tombesi}, Francesco and {Trois}, Alessio and {Tsygankov}, Sergey S. and {Turolla}, Roberto and {Vink}, Jacco and {Weisskopf}, Martin C. and {Wu}, Kinwah and {Xie}, Fei and {Zane}, Silvia},
        title = "{X-ray polarization evidence for a 200-year-old flare of Sgr A$^{*}$}",
      journal = {\nat},
         year = 2023,
        month = jul,
       volume = {619},
       number = {7968},
        pages = {41-45},
          doi = {10.1038/s41586-023-06064-x},
archivePrefix = {arXiv},
       eprint = {2304.06967},
 primaryClass = {astro-ph.HE},
       adsurl = {https://ui.adsabs.harvard.edu/abs/2023Natur.619...41M}
}

@ARTICLE{Ginsburg2015,
       author = {{Ginsburg}, Adam and {Walsh}, Andrew and {Henkel}, Christian and {Jones}, Paul A. and {Cunningham}, Maria and {Kauffmann}, Jens and {Pillai}, Thushara and {Mills}, Elisabeth A.~C. and {Ott}, Juergen and {Kruijssen}, J.~M. Diederik and {Menten}, Karl M. and {Battersby}, Cara and {Rathborne}, Jill and {Contreras}, Yanett and {Longmore}, Steven and {Walker}, Daniel and {Dawson}, Joanne and {Lopez}, John A.~P.},
        title = "{High-mass star-forming cloud G0.38+0.04 in the Galactic center dust ridge contains H$_{2}$CO and SiO masers}",
      journal = {\aap},
         year = 2015,
        month = dec,
       volume = {584},
          eid = {L7},
        pages = {L7},
          doi = {10.1051/0004-6361/201527452},
archivePrefix = {arXiv},
       eprint = {1510.06401},
 primaryClass = {astro-ph.GA},
       adsurl = {https://ui.adsabs.harvard.edu/abs/2015A&A...584L...7G}
}
\bibliographystyle{aasjournal}

\end{document}